\documentclass[iop,apj]{emulateapj}
\usepackage{apjfonts}
\usepackage{multirow}
\usepackage{natbib}

\def\aa{{A\&A}}

\def\aj{{AJ}}

\def\amin{$^\prime$}
\def\annrev{{ARA\&A}}
\def\apj{{ApJ}}
\def\apjl{{ApJL}}
\def\apjs{{ApJS}}
\def\asec{$^{\prime\prime}$}

\def\farcm{\hbox{$.\mkern-4mu^\prime$}}
\def\farcs{\hbox{$.\mkern-4mu^{\prime\prime}$}}

\def\hal{H$\alpha$}
\def\hb{H$\beta$}

\def\hst{{\it HST}}
\def\kms{km s$^{-1}$}

\def\lax{{$\mathrel{\hbox{\rlap{\hbox{\lower4pt\hbox{$\sim$}}}\hbox{$<$}}}$}}
\def\gax{{$\mathrel{\hbox{\rlap{\hbox{\lower4pt\hbox{$\sim$}}}\hbox{$>$}}}$}}
\def\simlt{\lower.5ex\hbox{$\; \buildrel < \over \sim \;$}}
\def\simgt{\lower.5ex\hbox{$\; \buildrel > \over \sim \;$}}
\def\lum{erg s$^{-1}$}
\def\mbh{{$M_{\rm BH}$}}

\def\mnras{{MNRAS}}

\def\percm2{cm$^{-2}$}

\def\solmass{$M_\odot$}

\def\edd{$L_{{\rm bol}}$/{$L_{{\rm Edd}}$}}
\def\lbul{$L_{\rm bul}$}
\def\ser{S\'{e}rsic}
\def\hnr{$L_{\rm host}/L_{\rm nuc}$}

\def\rbulge{$M_{R,{\rm bul}}$}

\def\rnuc{$M_{R,{\rm nuc}}$}

\def\rhost{$M_{R,{\rm host}}$}

\def\btot{$B/T$}
\def\lstar{$L_{R}^{*}$}

\shorttitle{Host Galaxies of Type 1 AGNs}
\shortauthors{KIM et al.}

\begin{document}

\title{Stellar Photometric Structures of the Host Galaxies of Nearby Type 1 Active Galactic Nuclei\altaffilmark{1}}

\author{Minjin Kim\altaffilmark{2,3}, Luis C. Ho\altaffilmark{4,5}, Chien Y.
Peng\altaffilmark{6}, Aaron J. Barth\altaffilmark{7},
and Myungshin Im\altaffilmark{8}}

\altaffiltext{1}{Based on observations made with the NASA/ESA  {\it Hubble
Space Telescope}, obtained from the Data Archive at the Space Telescope
Science Institute, which is operated by the Association of Universities for
Research in Astronomy (AURA), Inc., under NASA contract NAS5-26555. These
data are associated with program AR-12133 and AR-12818.}

\altaffiltext{2}{Korea Astronomy and Space Science Institute, 
Daejeon 305-348, Korea; mkim@kasi.re.kr}

\altaffiltext{3}{University of Science and Technology, Daejeon 305-350, Korea}

\altaffiltext{4}{Kavli Institute for Astronomy and Astrophysics, Peking 
University, Beijing 100871, China; lho.pku@gmail.com}

\altaffiltext{5}{Department of Astronomy, School of Physics, Peking University, Beijing 100871, China}

\altaffiltext{6}{Giant Magellan Telescope Corporation,
251 S. Lake Ave., Suite 300, Pasadena, CA 91101; peng@gmto.org}

\altaffiltext{7}{Department of Physics and Astronomy, University of
California at Irvine, 4129 Frederick Reines Hall, Irvine, CA 92697-4575;
barth@uci.edu}

\altaffiltext{8}{Department of Physics and Astronomy, Frontier Physics
Research Division (FPRD), Seoul National
University, Seoul, Korea; mim@astro.snu.ac.kr}

\begin{abstract}
We present detailed image analysis of rest-frame optical images of 235 
low-redshift ($z$ \lax\ 0.35) type 1 active galactic nuclei (AGNs) observed 
with the {\it Hubble Space Telescope}.  The high-resolution images enable us 
to perform rigorous two-dimensional image modeling to decouple the luminous 
central point source from the host galaxy, which, when warranted, is further 
decomposed into its principal structural components (bulge, bar, and disk). 
In many cases, care must be taken to account for structural complexities such 
as spiral arms, tidal features, and overlapping or interacting companion 
galaxies.  We employ Fourier modes to characterize the degree of asymmetry of
the light distribution of the stars, as a quantitative measure of morphological
distortion due to interactions or mergers.  We examine the dependence of the 
physical parameters of the host galaxies on 
the properties of the AGNs, namely radio-loudness and the width of the 
broad emission lines. 
In accordance with previous studies, narrow-line (H$\beta$ FWHM $\leq 2000$ 
km~s$^{-1}$) type 1 AGNs, in contrast to their broad-line (H$\beta$ FWHM $> 
2000$ km~s$^{-1}$) counterparts, are preferentially hosted in later type, 
lower luminosity galaxies, which have a higher incidence of pseudo-bulges, 
are more frequently barred, and are less morphologically disturbed.  This 
suggests narrow-line type 1 AGNs experienced a more quiescent evolutionary 
history driven primarily by internal secular evolution instead of external 
dynamical perturbations.  The fraction of AGN hosts showing merger signatures 
is larger for more luminous sources.  Radio-loud AGNs generally preferentially 
live in earlier type (bulge-dominated), more massive hosts, although a 
minority of them appears to contain a significant disk component.  We do not 
find convincing evidence for enhanced merger signatures in the radio-loud 
population.
\end{abstract}


\section{Introduction} 

Active galactic nuclei (AGNs), being some of the most energetic sources in
the Universe, are powered by accretion onto supermassive black holes
(BHs) at the centers of galaxies. The discoveries of tight correlations
between BH mass and large-scale properties of their host galaxies (Kormendy \& 
Richstone 1995; Magorrian et al. 1998; Ferrarese \& Merritt 2000; Gebhardt 
et al. 2000) point to the possibility that the growth of BHs and the evolution 
of their surrounding host galaxy are closely related.  The underlying causes 
of this apparent coevolution, however, are still debated (Kormendy \& Ho 2013).
One hypothesis is that AGN-driven feedback may cause BHs and galaxies to grow 
together in lock-step (e.g., Kauffmann \& Haehnelt 2000; Hopkins et al. 2006).
If so, fundamental questions of interest to address are how BHs and their 
host galaxies relate and interact, and what produces the different types of 
AGNs.

Within this context, the detailed stellar photometric properties of AGN host 
galaxies help to establish when and why supermassive BHs are active or 
inactive, and potentially the feedback consequences that AGNs may have on the 
hosts.  While a significant number of imaging datasets of type 1 (broad-line) 
AGN host galaxies now exist at low to intermediate redshifts, obtained by 
various large-scale {\it Hubble Space Telescope} (\hst; e.g., Giavalisco et al. 
2004, Rix et al.  2004, Scoville et al. 2007, Koekemoer et al. 2011) and 
ground-based (e.g., Alam et al. 2015) surveys, the sample of well-studied 
type 1 AGN host galaxies is still small. This, in part, is a result of the 
fact that disentangling the luminous central AGN robustly still poses a 
technical challenge, even with the resolution of \hst\ (e.g., McLeod \& Rieke 
1995b; Bahcall et al. 1997; McLure et al. 1999; Dunlop et al. 2003).  Both the
high contrast between the bright AGN nucleus and the host and variations in the
\hst\ point-spread function (PSF) introduce systematic uncertainties in 
photometric measurements of host galaxies (Kim et al. 2008a). Thus, a detailed, 
quantitative study of a large sample of hosts of type 1 AGNs is still largely 
lacking.

Although the host galaxies of type 2 AGNs are easier to study because they are 
not overwhelmed by the luminous central point source present in type 1 AGNs,
key physical properties, such as the BH mass or even intrinsic luminosity, 
cannot be reliably measured in these systems.  By contrast, BH masses 
in type 1 AGNs can be estimated with reasonable 
accuracy from their broad 
emission lines ($\sim 0.3-0.5$ dex; e.g., Kaspi et al. 2000; Ho \& Kim 2015), 
and we can obtain their 
\begin{figure}[t!]
\centering
\includegraphics[width=0.47\textwidth]{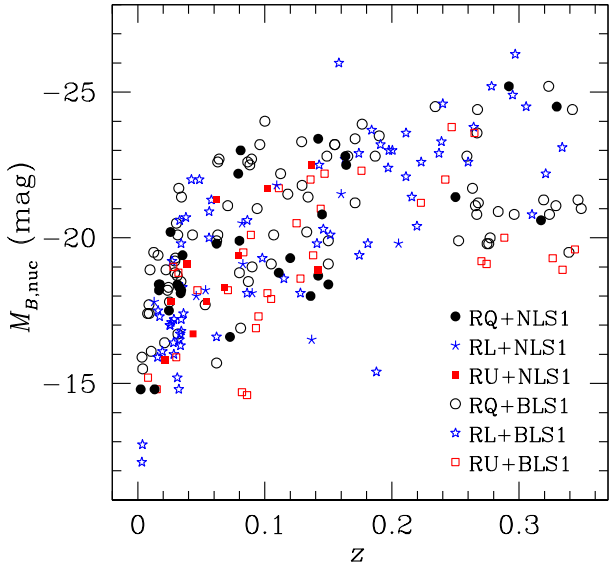}
\caption{
Redshift versus absolute magnitude of the nucleus in the $B$ band. The sample 
is divided by AGN type according to the width of the broad lines and the 
degree of radio-loudness.  Broad-line Seyfert 1s (BLS1) have FWHM $>$ 2000 
\kms; narrow-line Seyfert 1s (NLS1) have FWHM $\leq$ 2000 \kms.  Radio-loud 
(RL) and radio-quiet (RQ) sources satisfy $R \ge  10$ and $R < 10$, 
respectively, while those without proper radio data are radio-undefined (RU).
}
\end{figure}
\vskip 0.1in
\noindent
luminosities (and hence Eddington ratios) rather straightforwardly.  
The AGN phenomenon and the properties of AGN host 
galaxies are closely related to BH mass and Eddington ratio. For example, 
radio-loudness and width of broad emission lines appear to be closely 
connected to BH mass, Eddington ratio, and the morphological types of the host 
galaxies (e.g., Pounds et al. 1995; Crenshaw et al. 2003; Best et al. 2005; 
Sikora et al. 2007). Thus, investigating a 
comprehensive sample of host galaxies of type 1 AGN is essential to understand 
the coevolution of BHs and their host galaxies.

In this paper, we analyze images of type 1 AGNs obtained with the \hst, using
the largest sample to date for this purpose.  We perform a careful 
decomposition of AGN host galaxies into their principal structural components 
(e.g., bulge, bar, disk), at the same time allowing for, if warranted, 
complexities due to spiral arms, tidal arms, and other distortions due to 
tidal interactions and mergers. We derive detailed structural parameters 
(e.g., surface brightness, size, profile shape), bulge-to-total light ratio, 
and we quantitatively measure asymmetries using Fourier modes.

A number of past studies have utilized \hst\ images to investigate sizable 
numbers of nearby type 1 AGNs (e.g., Schade et al. 2000; Dunlop et al. 2003; 
Hamilton et al. 2008). Most of these efforts have focused only on specific 
categories of sources (e.g., luminous QSOs and X-ray selected AGNs). Here we 
place no such restrictions, aiming to study as wide a range of properties 
as possible, only requiring that the sample selection be restricted to 
type 1 AGNs so that BH masses 
can be estimated.  In addition to sample size and broad representation of 
AGN properties, our study utilizes a uniform method of analysis, pays close 
attention to the treatment of the PSF, and, for the first time, systematically 
treats the many structural complexities inherent to many AGN hosts galaxies.
Our large database has many applications for a wide variety of 
statistical studies of AGN host galaxies.  Here we focus on three issues. 

It is well-known that radio-loud AGNs are preferentially hosted by massive,
early-type galaxies (e.g., Matthews et al. 1964; Best et al. 2005; Floyd
et al. 2010). However, this finding has not been confirmed with a large sample
of type 1 AGNs (Dunlop et al. 2003). Meanwhile, it is still under debate 
whether radio-loudness is related to galaxy merging or tidal interactions
(Best et al.  2005; but see Heckman 1983; Ramos Almeida et al. 2012).  We 
revisit these issues by investigating how AGN radio-loudness correlates with 
host galaxy properties.
 
There is mounting evidence that the host galaxies of narrow-line Seyfert 1 
(NLS1; FWHM (\hb) $\leq 2000$ \kms) sources differ statistically from those of the
broad-line type 1 (BLS1; FWHM (\hb) $> 2000$ \kms) class, in the sense that 
NLS1s are more likely to be associated with barred galaxies and have more 
prominent star formation activity (e.g., Crenshaw et al. 2003; Deo et al. 
2006; Sani et al. 2010). In addition, the fraction of pseudo-bulges\footnote{
Kormendy \& Kennicutt (2004) describe pseudo-bulges as the bulge-like central 
concentration of stars in disk galaxies formed through internal secular
evolution, to be contrasted with classical bulges, which experienced major 
mergers, similar to elliptical galaxies.} has been reported to be 
significantly higher in NLS1s compared to BLS1s (Orban de Xivry et al. 2011). 
This might indicate that the AGN triggering mechanism of NLS1s, perhaps 
reflecting the predominance of secular evolution of their host galaxies, is 
distinctive from that of BLS1s. 
This work adopts the cosmological parameters: $H_0 = 100 h = 67.8$ km s$^{-1}$ 
Mpc$^{-1}$, $\Omega_m = 0.308$, and $\Omega_{\Lambda} = 0.692$ (Planck 
Collaboration et al. 2016). All magnitudes are in the Vega magnitude system 
(Bessel 2005).

%

\begin{figure*}[t]
\centering
\includegraphics[width=0.95\textwidth]{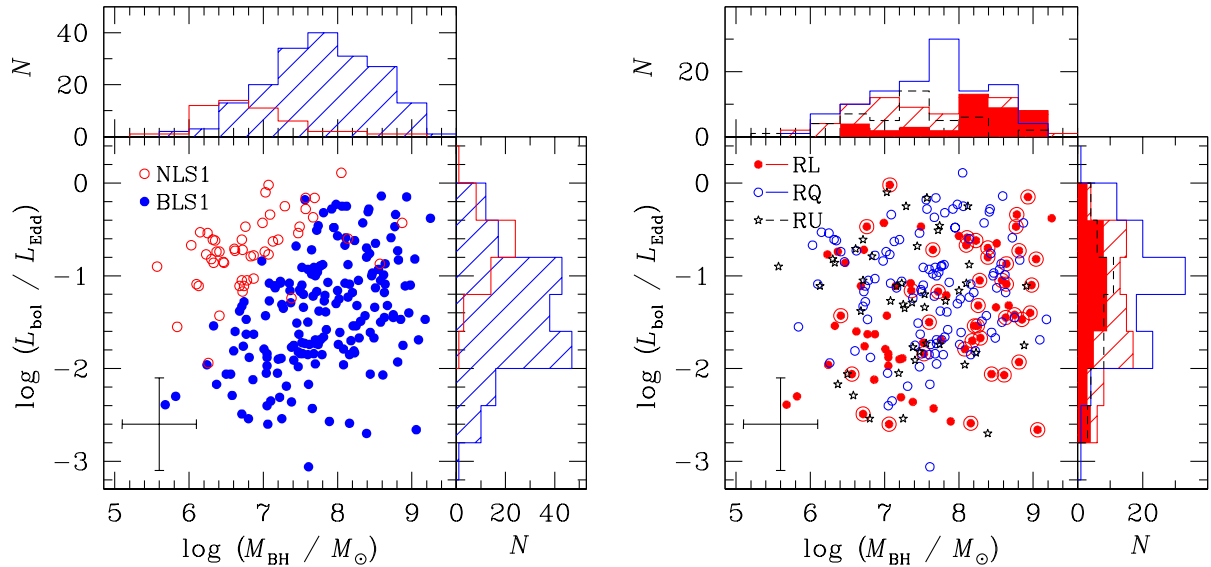}
\caption{
Distributions of \mbh\ and \edd\ for different AGN types. The left panel 
separates NLS1s from BLS1s, and the right panel divides the sample according
to radio-loudness.  The error bars at the bottom left corner show the typical 
uncertainties. The methods to estimate \mbh\ and \edd\ are described in 
\S{3.3}. While radio-loud objects ($R>10$) are represented by large red dots 
and hatched histogram, extreme radio-loud objects ($R>100$) are denoted by 
large circled red dots and red-filled histogram. 
}
\end{figure*}

\section{Data Selection}

The ultimate goal of our series of papers is to understand how the properties
of nearby AGNs are related to the structural and morphological characteristics 
of their host galaxies (Kim et al. 2007, 2008a, 2008b), using as the 
primary tool high-resolution \hst\ images.  For that purpose, we choose a 
sample of type 1 AGNs that has usable archival \hst\ images and 
sufficient spectroscopic information to estimate BH masses ($M_{\rm BH}$; 
Section 3.3).

We began by searching for all objects that are known to be AGNs of any type in 
the \hst\ archives.  As of June 2012, we found $\sim 1050$ candidates.  After a 
careful literature search, we discarded objects that are not known to have 
broad emission lines (i.e. type 2 AGNs or narrow-line radio galaxies; 
$\sim 450$ objects) or that have redshifts $z>0.35$ ($\sim 370$ objects).  We 
limit the sample to $z < 0.35$ to avoid complications from cosmic evolution of 
the BH-host relations, which has been reported above this redshift (Treu 
et al. 2007; Bennert et al. 2010). The spatial resolution of the \hst\ 
images corresponds to $\sim0.35$ kpc at this redshift limit ($z=0.35$), which 
is sufficient to resolve the bulges of host galaxies. 
These cuts resulted in an initial sample 
of $\sim 350$ nearby type 1 AGNs.  We then inspected the list to retain only 
those objects that have reliable spectroscopic data, either from the 
literature or from our own observations (Ho \& Kim 2009).

Of most relevance for our goal of estimating $M_{\rm BH}$ is the availability of
a robust measurement of the velocity widths of the broad emission 
lines\footnote{Because broad \hb\ and \hal\ are often blended with other 
narrow emission lines and Fe II multiplets, accurate measurement of their line 
widths requires spectra of relatively high S/N. The typical uncertainty of the 
FWHM measurements is $\leq 10\%$; for 6\% of the sample, the uncertainty is up 
to 20\%.} This
further reduced the sample to $\sim 270$ objects.  To facilitate comparison of
our results with previous studies, we restrict our attention to optical images
taken as closely as possible to the $V$, $R$, or $I$ bands\footnote{Roughly 40
objects have \hst\ images in multiple optical filters. For this study we only
consider the deepest image.  A forthcoming paper will use the multiple filters
to study the colors of the host galaxies.}. Since these AGNs are at low
redshifts, $k$-corrections will be minimal.  We choose all suitably deep 
Wide-Field Planetary Camera 2 (WFPC2) and Advanced Camera for Surveys (ACS) 
images, guided by the signal-to-noise ratio (S/N) requirements given in Kim 
et al. (2008a). 35 objects are rejected because they only have UV
or B$-$band images, or because their S/N is smaller than 250.\footnote{
Kim et al. (2008a) showed that accurate measurement of the host brightness 
requires ${\rm S/N_{nuc}} > 250$ especially if the host galaxy is fainter than
the nucleus.}

The final sample, summarized in Table~1 and Figures 1--2, consists of 235 
objects.  The sample contains both quasars and lower-luminosity Seyfert 1 
galaxies.  
We note that the sample is selected in a heterogeneous manner. Thus, it 
is worthwhile to briefly summarize the characteristics of the objects. The \hst\
images of the final sample were obtained through 28 different \hst\ programs. 
While most of the objects are identified as AGNs from optical color selection
(e.g., Bahcall et al. 1994; Letawe et al. 2008; Floyd et al. 2004), 
$\sim 25\%$ of the sample was selected through X-ray surveys 
(e.g., Schade et al. 2000). 
Also, the fraction of radio-loud AGNs in our sample ($\sim33\%$) is 
significantly higher than that of other typical AGN samples ($\sim 10-20\%$;
e.g., Kellermann et al. 1989; Ivezi\'{c} et al. 2002).  
This is a consequence of HST programs specifically devoted to
radio-selected AGNs (e.g., McLure et al. 1999; Dunlop et al. 2003; de Koff
et al. 1996). Approximately $20\%$ of the sample are NLS1s, comparable to
optically selected AGN samples ($\sim15-20\%$; Zhou et al. 2006).
Finally, a significant fraction ($\sim 20\%$) of the sample comes from the
large \hst\ survey of nearby AGNs by Malkan et al. (1998).
For the sake of completeness, the objects analyzed in Kim et al. 
(2008b) are also included (45 objects).

The majority of the sample ($\sim 70\%$) are observed with WFPC2. WFPC2 is 
composed of three wide-field cameras (WF) and one planetary camera (PC). The 
field-of-view is 1\farcm3 with a pixel scale of 0\farcs1 for the WF and 
34\asec\ with with a pixel scale of 0\farcs046 for the PC.  Among the 167 
sources with WFPC2 data, we use PC images for 80, WF images for 40, and full 
WFPC2 mosaic images for 47.  The rest of the sample, except for one object, 
was observed with ACS, which comprises two optical cameras: 39 used the High 
Resolution Camera (HRC) and 28 used the Wide Field Camera (WFC). The HRC has a 
high angular resolution (0\farcs027 pixel$^{-1}$) but a narrow field-of-view 
(29\asec$\times$26\asec), while the WFC has relatively high throughput and 
covers a wide field-of-view of $\sim$202\asec\ with $\sim$0\farcs049 pixels.  
Only one object (Mrk 1048) was observed with Wide-Field Planetary Camera 1 
(WF/PC), one of the first-generation instruments that was replaced with WFPC2 
in 1993. It contains four WF chips and four PC chips, whose characteristics 
are similar to those on WFPC2. 

The sample has been observed with various broad-band and medium-band optical 
filters, such as F547M, F550M, F555W, F606W, F625W, F675W, F702W, F785LP, 
F791W, and F814W, which are roughly equivalent to conventional $V$, $R$, and 
$I$ filters.  A total of 124, 45, and 66 targets were observed in the $V$-like 
(F547M, F550M, F555W, F606W, and F625W), $R$-like (F675W and F702W), and 
$I$-like (F785LP, F791W, and F814W) filters, respectively.  The exposure time 
varies between $\sim 230$~s and $10,000$~s, with a median value of $\sim 
600$~s. These depths reach a $3\sigma$ surface brightness limit of 24--28 mag 
arcsec$^{-2}$ over an area of 1 arcsec$^{2}$.  The S/N within such an area in 
the center ranges from $\sim$250 to a few thousand.

\begin{figure*}[htp]
\centering
\includegraphics[width=0.85\textwidth]{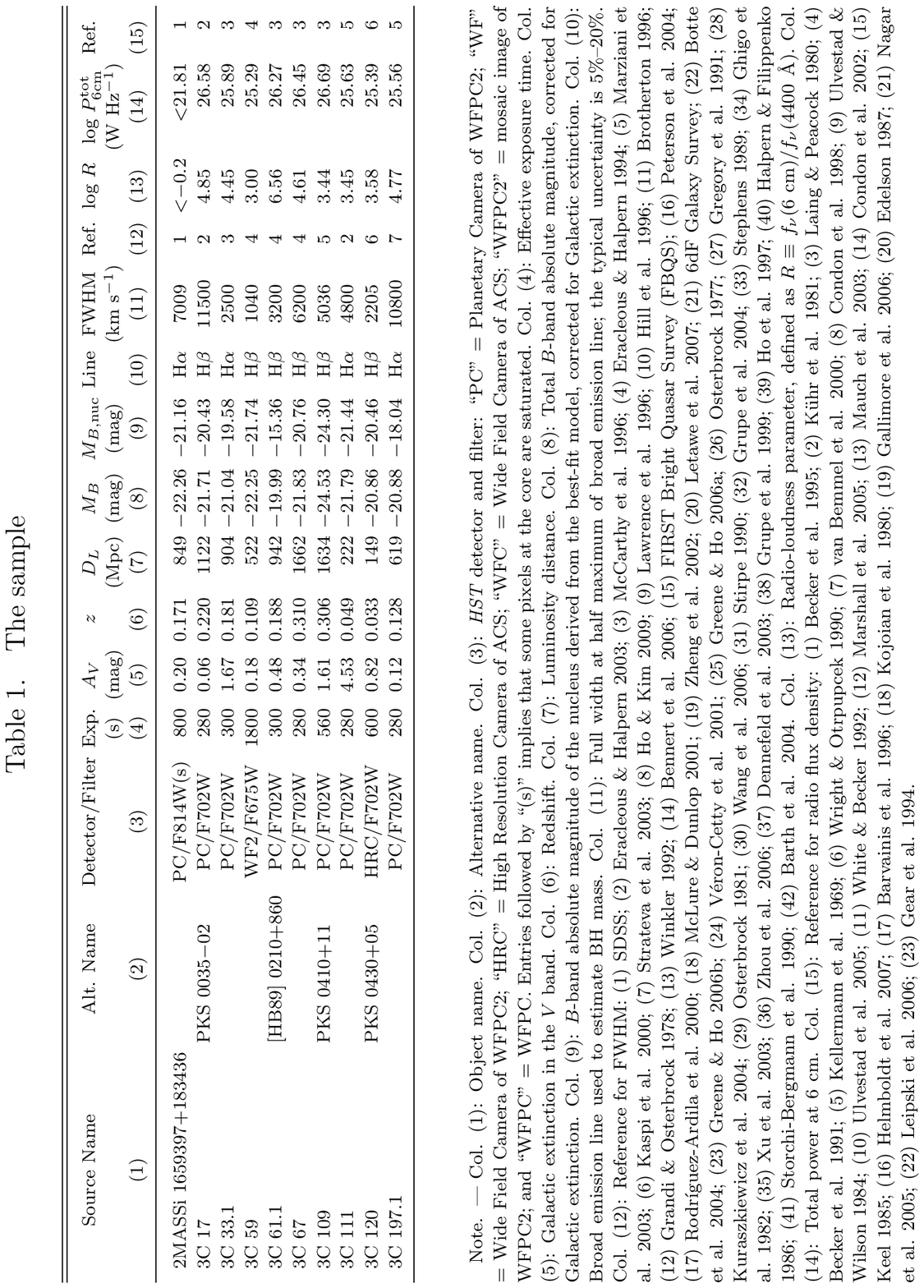}
\end{figure*}

As illustrated in Figure 2, the sample spans a wide range in \mbh, from 
$10^{5.6}$ to $10^{9.2}$ \solmass, and \edd, from 0.001 to 1 
(see \S{5.1} and \S{5.2}).  We assemble 
measurements of FWHM for the broad component of either \hb\ or \hal, taken 
either from literature sources we deem to be sufficiently reliable, or 
directly measured by ourselves from our own optical spectra.  
Following common 
practice, we classify the objects according to the FWHM of their broad 
emission lines: 184 objects qualify as BLS1s and 51 as NLS1s.  We further 
consider the radio brightness of the objects, using published radio flux 
density measurements from the literature.  As the literature data are quite 
heterogeneous in terms of angular resolution, which ranges from a few 
arcseconds to $\sim$6\amin, some care is required to assemble measurements 
that pertain only to the AGN, especially for more nearby Seyferts wherein 
\noindent
host 
galaxy contamination can be a serious issue (Ho \& Peng 2001).  Moreover, the 
published measurements were taken at various wavelengths.  We homogenize the 
literature data by converting all flux densities to 6 cm assuming a radio 
spectral index of $f_{\nu} \propto \nu^{-0.5}$ (Ho \& Ulvestad 2001).  We 
include upper limits from FIRST (Becker et al. 1995), if available. 

The monochromatic flux density of the nucleus at rest-frame 4400 \AA\ 
($f_{4400\,\AA}$) and $B$-band absolute magnitude ($M_{B,{\rm nuc}}$) are
derived directly from our work, using the decomposition of the \hst\ 
images presented below.  We apply a $k$-correction using the composite 
quasar spectrum of Vanden~Berk et al. (2001), and we correct for Galactic 
extinction using values of $A_V$ from Schlafly \& Finkbeiner (2011) and the 
extinction curve of Cardelli et al. (1989).

We combine the radio and extinction-corrected optical flux densities to 
calculate the radio-loudness parameter, 
$R\equiv f_\nu(6\,{\rm cm})/f_\nu({\rm 4400\,\AA})$ (Kellermann et al. 1989). 
Our sample has an unusually large fraction of radio-loud (RL; $R\geq10$) 
sources (33\%; 78 objects) compared to radio-quiet (RQ; $R < 10$) sources 
(48\%; 113 objects); the radio status of 44 objects (19\%) is unknown (RU) 
because either there is no measured upper limit available or the available 
upper limits are too high to determine the radio-loudness parameter 
($R_{\rm upper} > 10)$. The conventionally used radio-loudness threshold 
of $R=10$ might not be sufficiently restrictive.  If, as advocated by 
Chiaberge \& Marconi (2011), we adopt a higher value of $R=100$, we are left 
with 41 extreme radio-loud (RL2) sources. The spatial resolution of the 
radio data ranges from 5\asec\ to 6\amin.

\section{Data Reduction}

Our data reduction procedure depends on the \hst\ camera used.  Basic data 
processing was performed by the standard \hst\ pipeline.  If there are 
multiple images with similar exposure time, we combine them to increase the 
S/N and to facilitate cosmic ray rejection.  For cosmic ray removal in single 
images, we use the {\tt lacos\_im} task within IRAF\footnote{IRAF is 
distributed by the National Optical Astronomy Observatory, which is operated 
by the Association of Universities for Research in Astronomy (AURA) under 
cooperative agreement with the National Science Foundation.} (van Dokkum 2001).
Multiple dithered images taken with ACS are combined and corrected for 
distortion using the software {\tt MultiDrizzle} (Koekemoer et al. 2002).

If the full WFPC2 frame is necessary in order to cover the entire galaxy 
(e.g., ESO 031$-$G8), we combine the images of the four chips using the IRAF 
task {\tt wmosaic}. The mosaiced images introduce gaps between the CCD chips;
we mask out these gaps for the subsequent fits. 

Due to the high contrast ratio between the nucleus and the underlying host 
galaxy, a long exposure commonly results in saturation of the galaxy center.
If a short exposure is available in addition to the long exposure, we replace 
the saturated pixels and bleeding columns in the long exposure by scaling the 
corresponding pixels from the unsaturated image. If no short, unsaturated 
exposure is available, we simply mask the saturated pixels during the fit.  

Accurate sky subtraction is of great importance to obtain accurate photometric
measurements. We initially determine a rough estimate of the sky level using 
the IRAF task {\tt sky}.  For a more refined measurement, we perform aperture 
photometry of the source as a function of radius after subtracting the sky 
background and masking nearby sources.  If the background is subtracted 
correctly, the cumulative flux eventually reaches a constant value at large 
radii. However, if the sky is overestimated, the aperture flux at large radius 
will decrease; the opposite trend will occur if the sky is underestimated.  In 
order to find the best-fit sky value, we vary the sky manually until the curve 
of growth flattens at large radius.  For images in highly crowded fields, for 
which the above method is inapplicable, we solve for the sky value during the 
image-fitting process (Section 4.1). 

Our image decomposition requires a robust estimation of the PSF.  Whenever 
possible we use empirical PSFs derived from bright, 
unsaturated stars observed using the same filter, camera, and position within 
the camera field-of-view.  But such data are quite rare.  In practice most 
of our analysis (189 out of 235 sources) is done using synthetic PSFs 
generated by {\tt TinyTim} (Krist 1995),
which have been shown to work reasonably well for our application (Kim et al. 
2008a).  The empirical or synthetic PSFs are reduced in the same manner as the 
science images so as to preserve the native shape of the PSF.  For PSF 
stars observed with the ACS/WFC, we adopt IDL wrappers developed to account 
for the PSF changing during {\tt MultiDrizzle} (Rhodes et al. 2006, 2007).

\section{Image Decomposition}

Our main goal is to measure the luminosity and structural parameters of the 
bulges of the host galaxies.  To this end, a comprehensive global model must 
be constructed for the entire galaxy to properly disentangle the bulge from 
the active nucleus, which often can dominate the light, as well as other 
components that might be present, such as a disk and bar (Table 2).

The details of our technique for decomposing \hst\ images of AGN host galaxies
have been discussed in previous papers by our group.  Our method is based on 
two-dimensional (2-D) parametric fitting using the code {\tt GALFIT} (Peng et 
al. 2002).  Kim et al. (2008a) performed extensive simulations to quantify the 
systematic uncertainties in measurements of bulge parameters under conditions 
that closely mimic \hst\ images of AGN host galaxies.  Based on these controlled
experiments, we devised a strategy to minimize the error on the recovery of 
the photometric parameters of the bulge component.  In brief, the dominant 
uncertainty on the bulge luminosity highly depends on how well we can reduce 
the errors due to mismatch of the PSF, which mainly arise due to its variation 
in time and position within the detector.  Another important contribution
to PSF mismatch comes from the fact that \hst\ PSFs are undersampled.   We 
showed that one can mitigate the effects of an undersampled PSF by broadening 
both the science image and the PSF image to critical sampling.  We follow the
same strategy in this study.

We use an updated version (3.0) of {\tt GALFIT} (Peng et al. 2010). This code 
fits 2-D images with multi-component models, which include not only a nucleus 
and a bulge, but, especially in lower-mass systems, often also a disk and a 
bar.  These additional components are crucial for obtaining a proper 
decomposition of the host galaxies (Greene et al. 2010; Jiang et al.  2011; 
Gao \& Ho 2017).  The new version of the code has new capabilities to fit 
non-axisymmetric features such as spiral arms, tidal features, and distortions 
due to tidal interactions.  These higher-order components, too, can affect the 
accuracy of the bulge measurements.

As described in Kim et al. (2008b), we fit the nucleus with a point source 
model represented by the PSF and the host galaxy with the \ser\ (1968) profile

\begin{equation}
\mu(r) = \mu_e e^{-\kappa[(r/R_e)^{1/n}-1]},
\end{equation}

\noindent
where $\mu(r)$ is the surface brightness at radius $r$, $R_e$ is the effective 
radius (within which half of the total light is contained), $\mu_e$ is the 
effective brightness at $R_e$, $n$ is the \ser\ index, and $\kappa$ is a 
constant coupled to $n$.  Depending on the complexity of each source, we fit 
it with one or more of the following models:
(1) a single \ser\ component with $n$ free;
(2) a single component with $n=4$;
(3) a single component with $n=1$;
(4) two components with $n=1$ for the disk and $n=4$ for the bulge;
(5) two components with $n=1$ for the disk and $n=$ free for the bulge; and
(6) two components with $n$ free for the disk and the bulge.
We always begin with the simplest single-component models and gradually 
increase in complexity if the residuals indicate that the source is more 
complex.  

We allow $n$ to be fit as a free parameter in order to account for the 
possibility of pseudo-bulges (Kormendy \& Kennicutt 2004), but in practice the
solution is not always trustworthy because it can yield unphysical parameter 
values (e.g., $n\ge10$, unrealistically small $R_e$).  This often arises when 
the central nucleus is substantially brighter than the underlying galaxy, and 
small PSF mismatches in the center are being compensated by the free \ser\ 
component.  Under these conditions, it is often better to fix the value of $n$ 
of the bulge component to minimize systematic uncertainty on the bulge 
luminosity (\lbul).  Our tests suggest that values of $n$ in the range 2--5 
have only a small effect on \lbul, and so for concreteness we adopt $n = 4$ in 
these cases.  Examples of fits with a single component for the host are shown 
in Figure 3.

We model the host with two components---bulge plus a disk---if there is a clear 
indication of a disk or extended features in the original image or in the 
residual image. The choice between a single-component and a double-component 
fit is not always clear.  We include an extra component only if it 
significantly improves the fit compared to a single component 
judged from the visual inspection of the residual image, or if it is 
clearly visible in the residual image (Figure 4). 

The disk component, even if present, may be undetectable if the images are 
too shallow.  Another complication arises from the fact that the fits are not 
always unique.  While a single bulge may adequately fit the host, in some 
cases we cannot rule out the possibility that a disk component can also be 
accommodated.  Following the methodology of Kim et al. (2008b), we derive 
upper limits for a disk component in objects for which none is detected or 
required by simulating a disk on the science image.  We constrain the 
brightness and scale length $h$ of an exponential disk using the known 
relation between bulge-to-total light ratio ($B/T$) and the ratio of bulge 
$R_e$ to $h$ observed in nearby early-type disk galaxies (de Jong et al. 
2004).  We generate artificial images comprised of a model bulge and disk with 
different $B/T$, add noise to match the S/N of the object in question, and 
then attempt to recover the two components with our fitting procedure.  The 
upper limit on the disk contribution is set by the minimum disk brightness 
below which we fail to recover it.  

A significant fraction of the sample shows clear evidence of spiral arms 
(Figure 4) and bars (Figure 5). The latest version of {\tt GALFIT} has the 
ability to model these features (Peng et al. 2010).  The code creates spiral 
structure by a hyperbolic tangent rotation function plus high-order Fourier 
modes.  Bars are represented by a Gaussian-like \ser\ profile ($n=0.5$; 
Freeman 1966; de Jong 1996).  When both a bar and spiral arms are present 
simultaneously through visual inspection of the \hst\ images, 
and the two features appear to have clearly distinct profiles 
(e.g., different axis ratio, discontinuity in their boundary), we use a 
truncation function to create a composite model that has a truncated inner 
profile and a truncated outer profile.  The truncated profile is represented 
by:

\begin{equation}
\begin{array}{l}
f(x,y)=f_i(x,y)\, {\rm tanh}(x,y,r_{\rm break,in}, \Delta r_{\rm soft,in}) \\
\ \ \ \ \ \ \ \ \ \ +f_o(x,y)\, {\rm tanh}(x,y,r_{\rm break,out}, \Delta r_{\rm soft,out})
\end{array} 
\end{equation}

\noindent
where $f_i(x,y)$ and $f_o(x,y)$ represent the original light profile for the 
inner and outer regions, respectively, $r_{\rm break}$ is the break radius,
and $\Delta r_{\rm soft}$ is the softening length. The inner truncation 
function tapers the light profile in the core, whereas the outer truncation
function tapers the light profile in the wings.  At $r_{\rm break}$, the 
surface brightness of the truncated profile is 99\% of that in the original 
profile.  The surface brightness is 1\% of the original at 
$r_{\rm break,in}+\Delta r_{\rm soft,in}$ for the inner profile and at 
$r_{\rm break,out}-\Delta r_{\rm soft,out}$ for the outer profile.  We can also
fit an outer ring by using solely an outer profile with an inner truncation 
function (see, e.g., MS~1545.3+0305 in Figure 11).
 A host galaxy is identified to have a bar if the fitting result improves 
significantly with the addition of the bar component, judged from the 
visual inspection of the residual image, and the fitted bar is 
physically reliable ($b/a \leq 0.6 $).

\begin{figure*}[htp]
\centering
\includegraphics[width=0.95\textwidth]{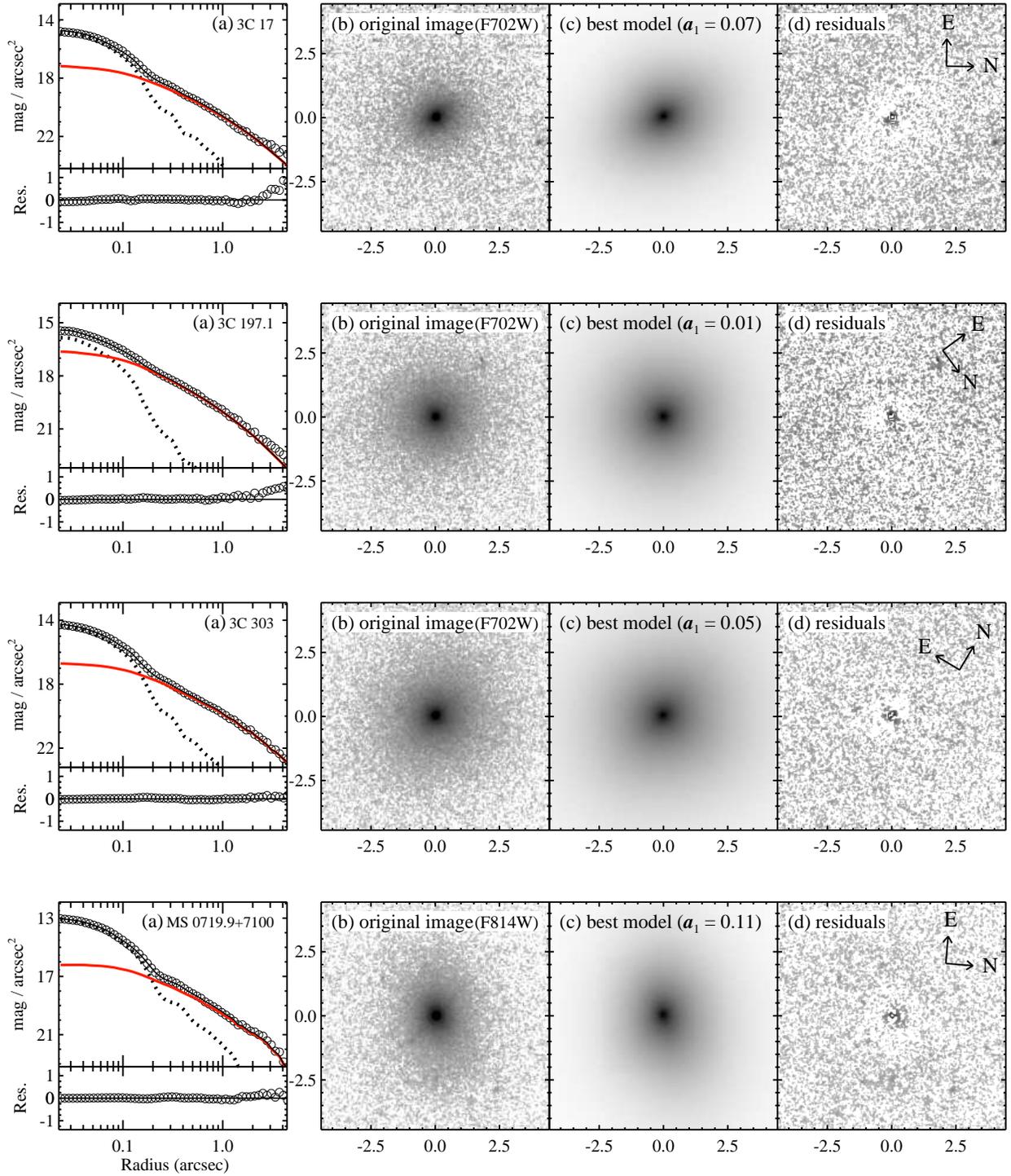}
\caption{
Examples of {\tt GALFIT} decomposition of elliptical AGN hosts.  ({\it a}) 
Azimuthally averaged profile, showing the original data ({\it open circles}), 
the best fit ({\it solid black line}), and the sub-components
[PSF ({\it dotted black line}) and bulge ({\it red solid line})]. 
The residuals are plotted on the bottom. Because the images are
already sky-subtracted, the outer region of the 1-D profiles are not smooth.
We present the 2-D image of ({\it b}) the original data, ({\it c}) the 
best-fit model for the host (the AGN is excluded to better highlight the 
host), with the amplitude of the first Fourier mode ($a_1$) labeled, and 
({\it d}) the residuals. The units of the images are in arcseconds. All images 
are displayed on an asinh stretch.}
\end{figure*}

\begin{figure*}[htp]
\centering
\includegraphics[width=0.95\textwidth]{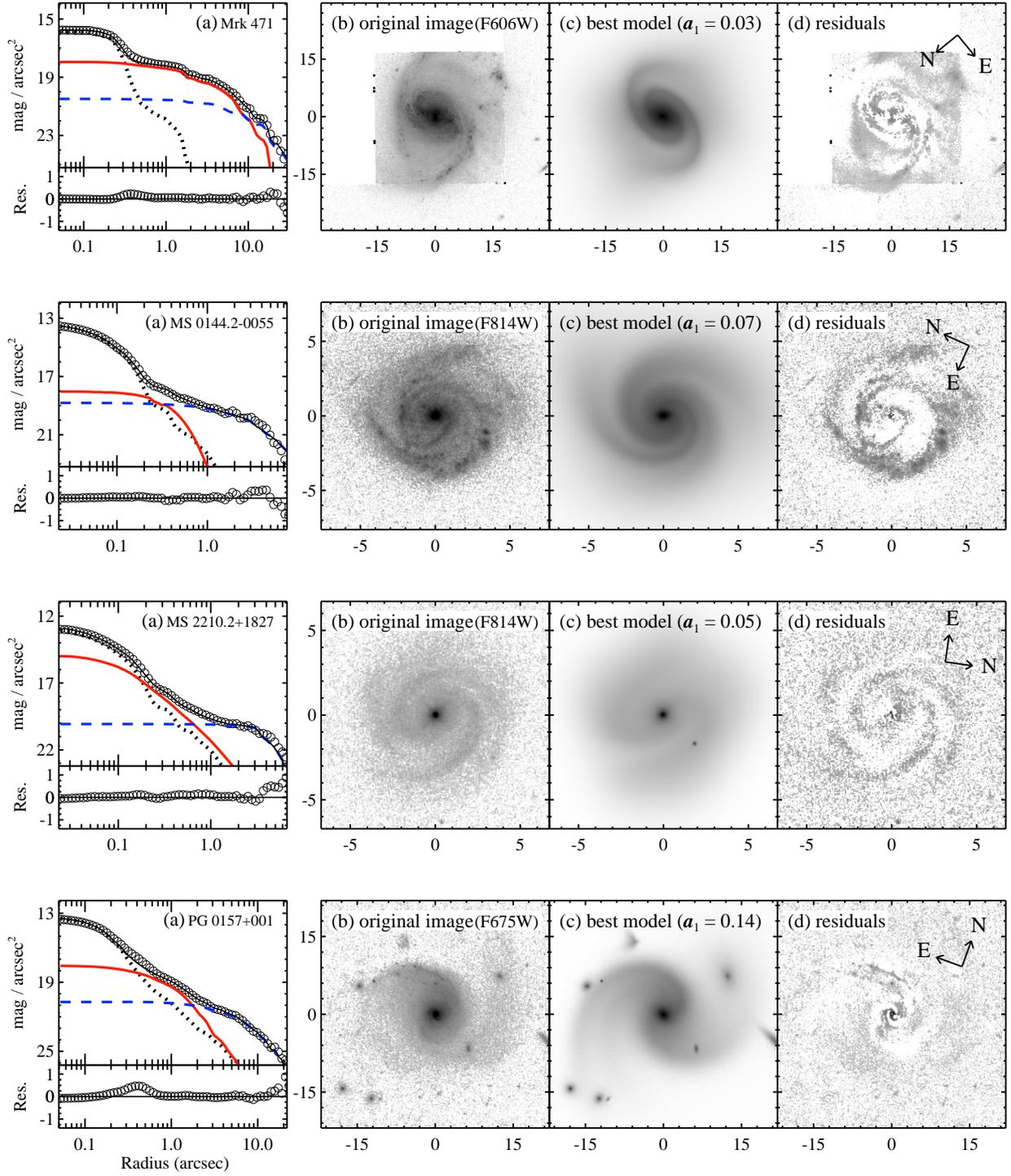}
\caption{
Examples of {\tt GALFIT} decomposition of unbarred disk galaxies.  Similar to 
Figure 3, except that the disk profiles are plotted as blue dashed lines.
}
\end{figure*}

\begin{figure*}[htp]
\centering
\includegraphics[width=0.95\textwidth]{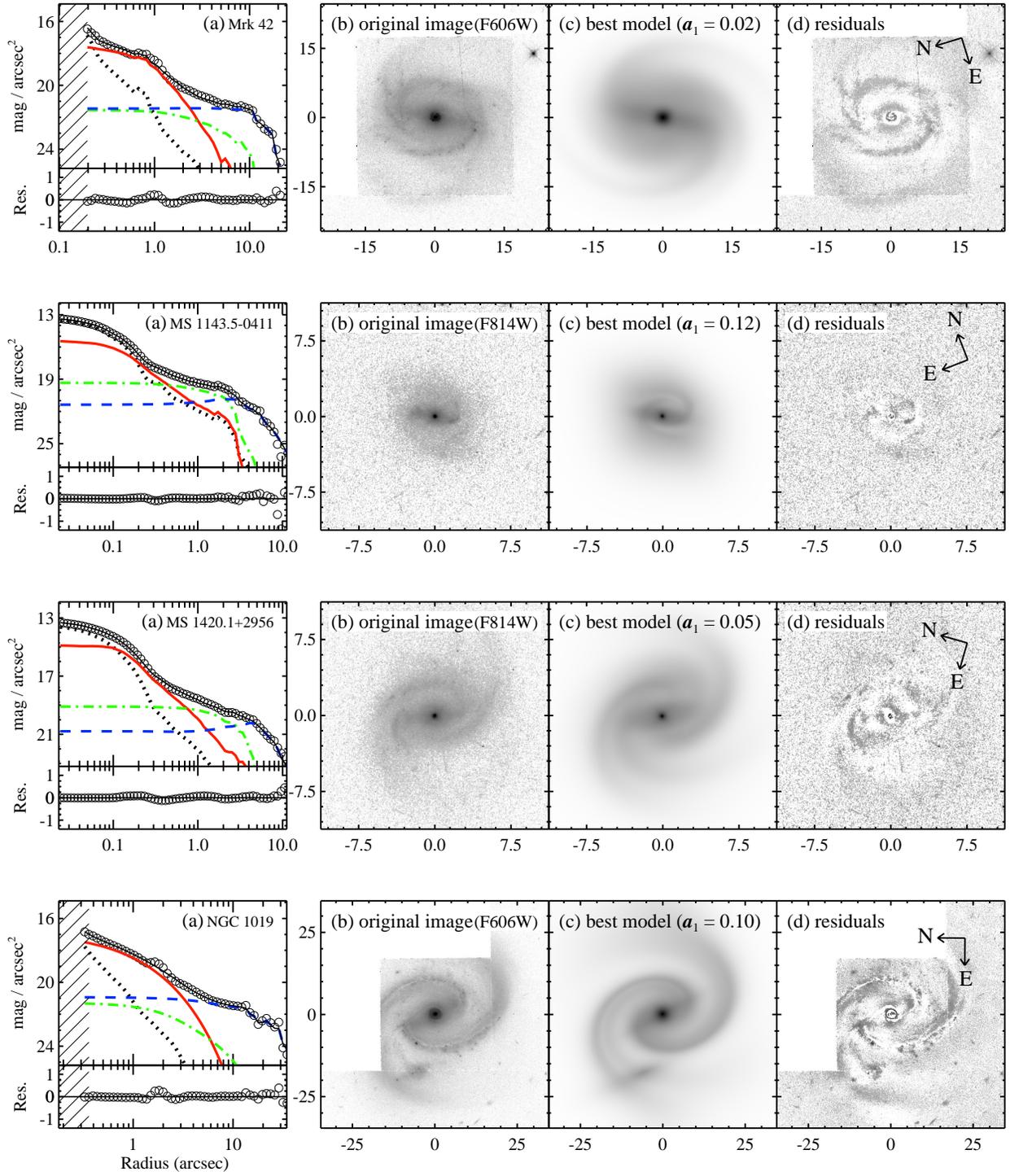}
\caption{
Examples of {\tt GALFIT} decomposition of barred disk galaxies. Similar to 
Figure 4, except that the bar profiles are plotted as green dot-dashed lines. 
The shaded region represents saturated pixels.
}
\end{figure*}

\begin{figure*}[htp]
\centering
\includegraphics[width=0.95\textwidth]{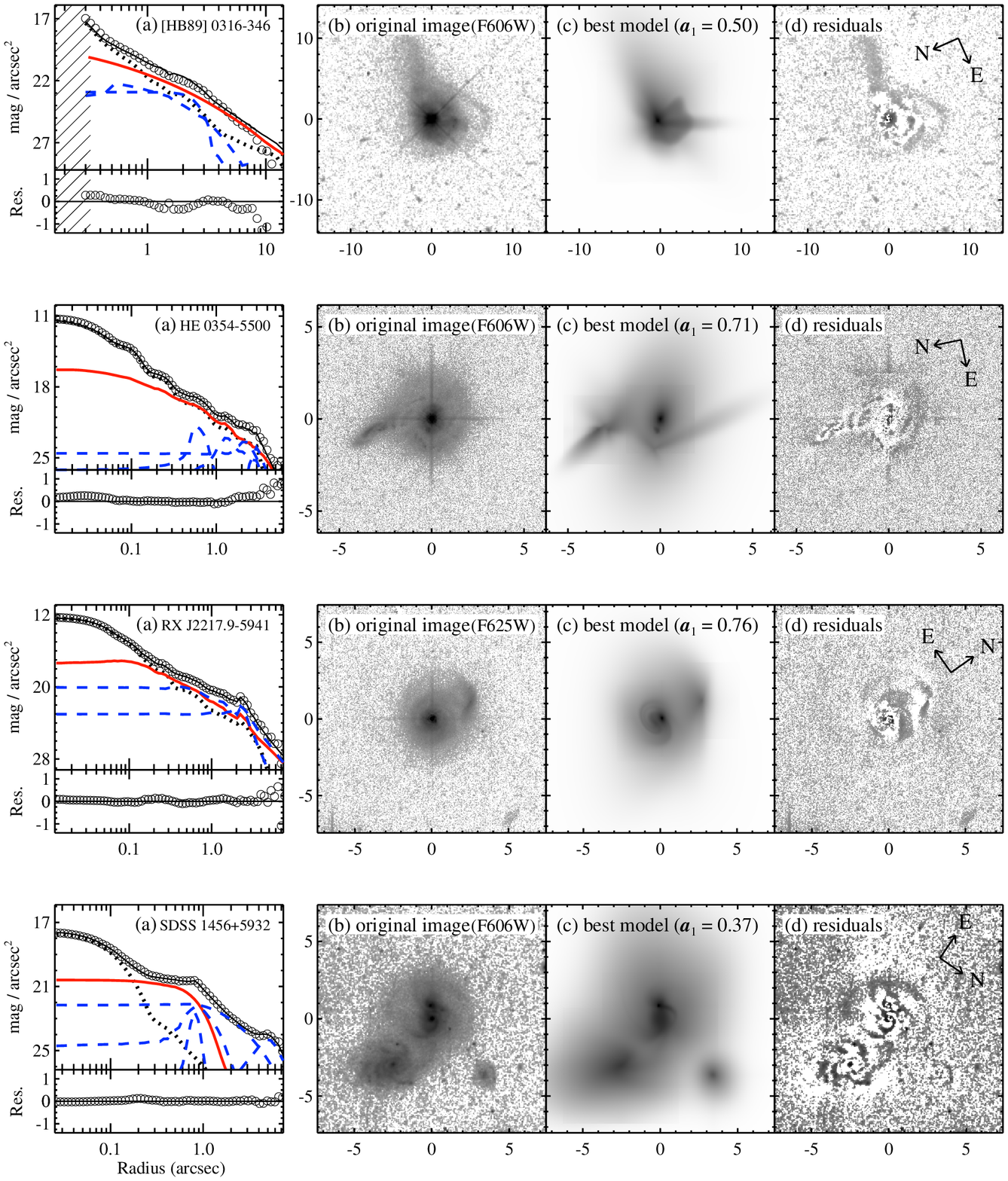}
\caption{
Examples of {\tt GALFIT} decomposition of interacting galaxies. Same as in 
Figure 5. 
}
\end{figure*}

As in Kim et al. (2008b), we use Fourier components to quantify the degree of 
asymmetry of the hosts.  The Fourier mode, which describes deviations of the 
original radial profile in the azimuthal direction, is defined as

\begin{equation}
r(x,y) = r_0(x,y)\, \left(1 + \sum_{m=1}^{N} a_m {\rm cos}[m (\theta + 
\pi_m)] \right),
\end{equation} 

\noindent
where $r_0(x,y)$ is the shape of the original ellipse in the radial coordinate,
$a_m$ is the amplitude of mode $m$, $\theta$ is the position angle of the 
ellipse, and $\pi_m$ is the relative alignment of mode $m$ relative to $\theta$.
The first ($m=1$) mode represents lopsidedness (how bright one side is 
compared to the other side) in the galaxy. Thus, the amplitude of the $m=1$ 
mode can be used to gauge the degree of asymmetry of the galaxy (Peng et al. 
2010).  In this study, we employ $a_1$ to select host galaxies that show 
morphological signs of interaction. Apart from this quantitative analysis, we 
also look for signatures of interaction and companion galaxies through visual 
inspection. Figure 6 shows fits of some of the most complex interacting 
galaxies in our sample. Note that Fourier components to measure 
lopsidedness are distinctive from those used to model spiral arms.
We also fit companions for 92 objects. We normally use a minimum
number (usually just one) of \ser\ component to model companion galaxies.
Foreground stars are fit with the PSF. 

Apart from {\tt GALFIT}-based parametric fits, we also obtain nonparametric 
measurements of the host galaxy luminosity by performing aperture photometry on 
the images after subtracting the active nucleus (Greene et al. 2008; Kim et 
al. 2008b).  By comparing this measurement of the host luminosity with that 
based on parametric fits, we can gauge the reliability of the fits 
independently.  We remove the AGN either by using the nucleus component
derived from the best-fit {\tt GALFIT} model or by scaling the PSF to match 
the brightness of the central pixel.  In general we find very good agreement 
between the host magnitudes derived from aperture photometry ($m_{\rm aper}$) 
and those obtained from model fitting ($m_{\rm fit}$).  The only exceptions are
cases in which close companion galaxies are present (e.g., [HB89]~1549+203), 
saturation affects the core, or the image does not cover the entire galaxy 
(e.g., Mrk 40).  Excluding such outliers, $\langle m_{\rm fit}-m_{\rm aper} 
\rangle =0.03\pm0.17$ mag.

\section{Derived Properties}

\subsection{AGN and Host Galaxy Luminosities}
In order to determine the luminosity of the nucleus and the host in the
rest-frame $R$ band, we need to convert the observed \hst\ filters to $R$
and apply a $k$-correction. Both are performed using our own IDL routines and
the response functions supplied by the Space Telescope Science Data Analysis
Software SYNPHOT package.
For the nucleus, we employ the quasar template spectrum from the Sloan Digital
Sky Survey (SDSS; Vanden~Berk et al. 2001), while for the host we use galaxy
template spectra from Calzetti et al. (1994) and Kinney et al. (1996).  We
assign a Hubble type to each host based on the $B/T$ derived from our
decomposition, using the relation between the two quantities given in Simien
\& de Vaucouleurs (1986); we apply the same correction to the bulge and disk
component.  Table 3 lists the final luminosities of the nucleus, expressed 
as absolute magnitudes in the $R$ band, $M_{R,{\rm nuc}}$,  and as a 
monochromatic luminosity at 5100 \AA,  $\lambda L_{5100}^{\rm image}$.  
The $R$-band luminosities are given for the bulge alone and for the entire 
host galaxy.

\begin{figure*}[htp]
\centering
\includegraphics[width=0.60\textwidth]{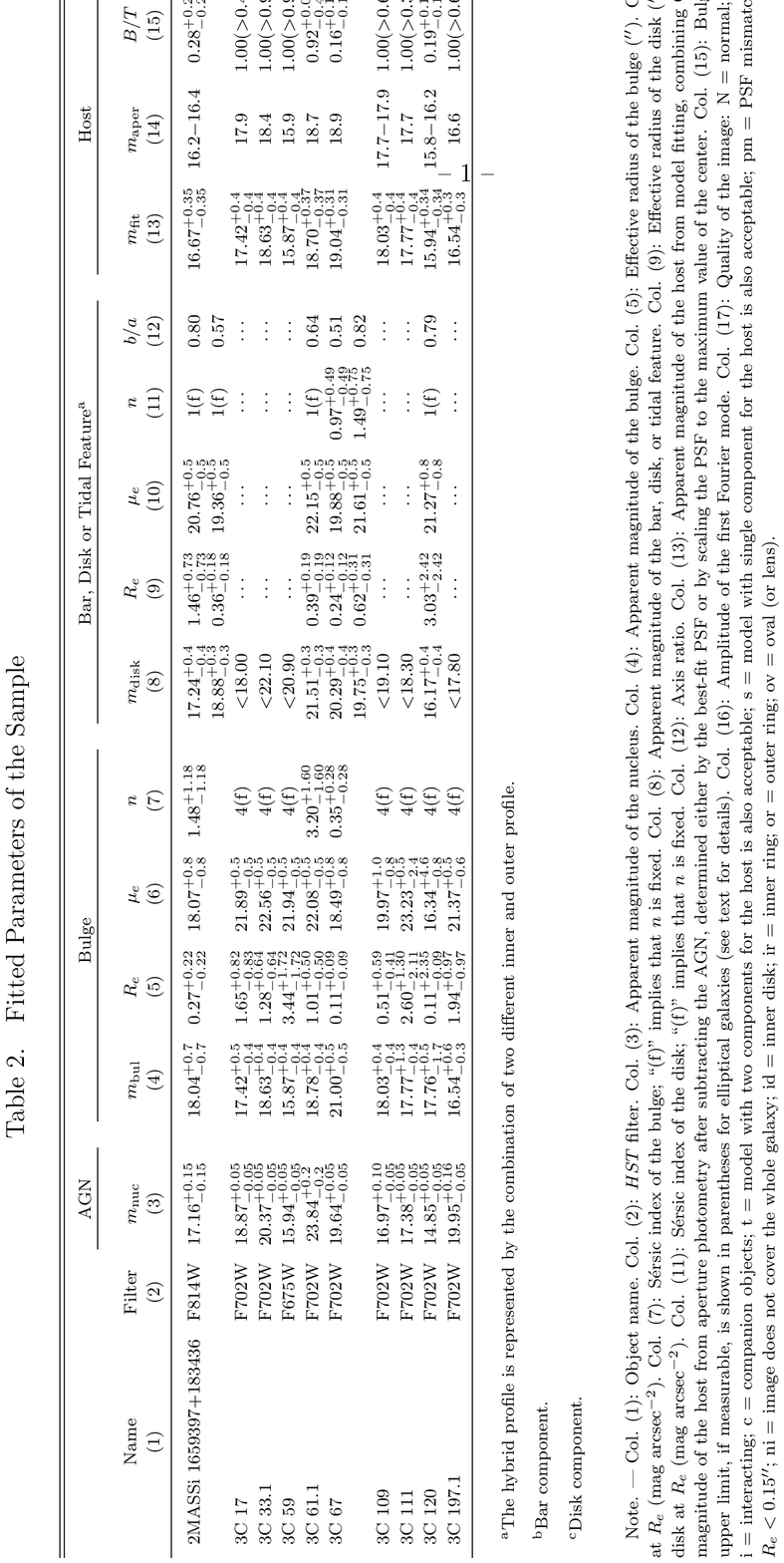}
\end{figure*}

\begin{figure*}[htp]
\centering
\includegraphics[width=0.9\textwidth]{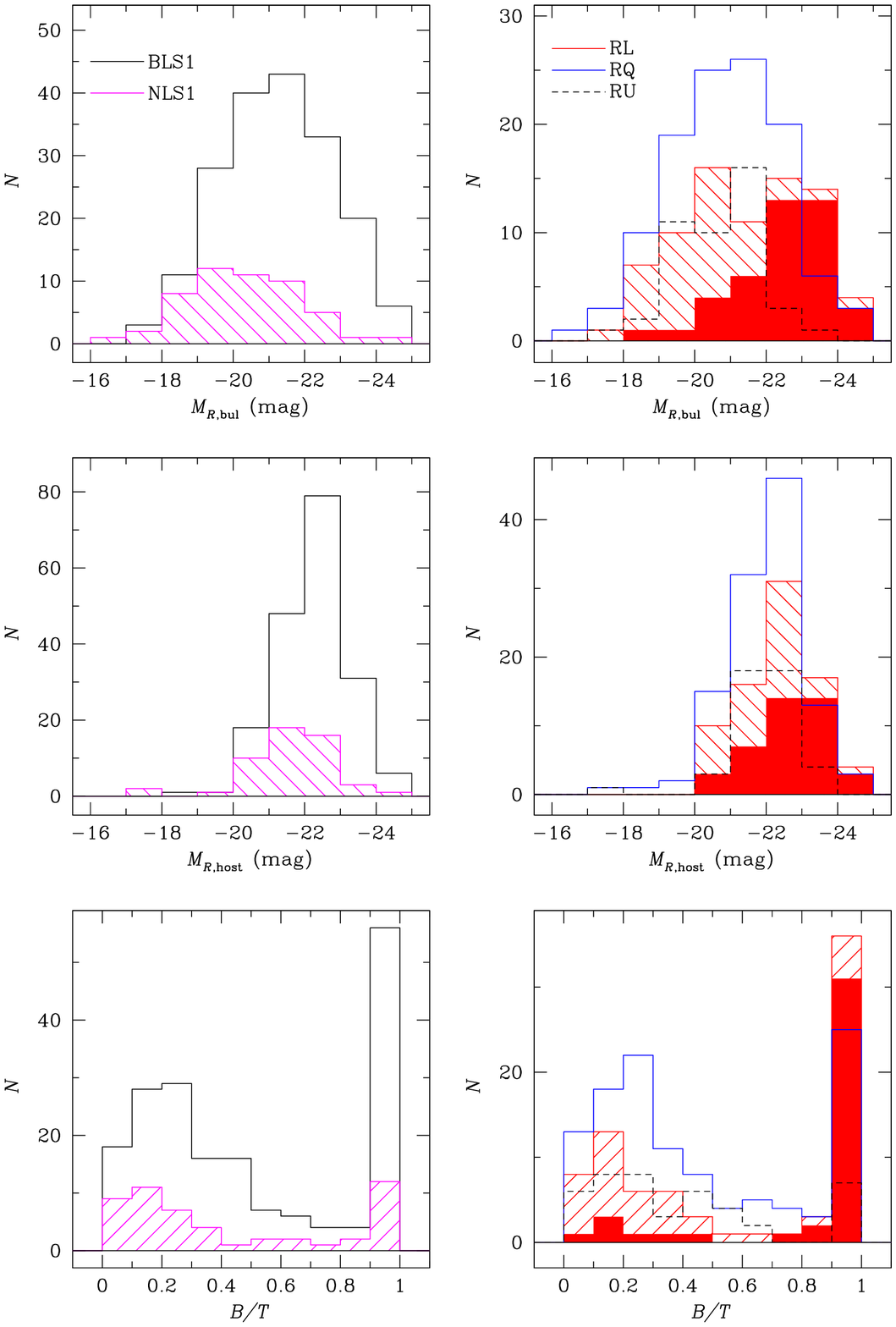}
\caption{
Distributions of $R$-band absolute magnitude for the bulge (top), absolute
magnitude for the host (middle), and bulge-to-total light ratio (bottom) for
different types of AGNs.  We divide the sample by broad-line type on the left
and radio-loudness on the right.  As in Figure 2, extreme radio-loud objects
are denoted by red-filled histogram.
}
\end{figure*}

\subsection{Black Hole Masses}

The subsequent analysis in this and in our subsequent papers makes use of BH 
masses.  We estimate virial BH masses following \mbh\ 
$= f v^2 R/G$, where $R$ is the size of the broad-line region (BLR) obtained 
from an empirical correlation between $R$ and AGN luminosity (Kaspi et al. 
2000; Bentz et al. 2013), $v$ is the virial velocity of the BLR inferred 
from the width of broad emission lines (commonly parameterized as the FWHM or 
as the line dispersion $\sigma_{\rm line}$), and the normalization coefficient 
$f$ absorbs the poorly understood structure and kinematics of the BLR.  In 
practice, $f$ is calibrated using local type 1 AGNs that have $R$ derived from 
reverberation mapping and measurements of bulge stellar velocity dispersion 
$\sigma_*$, by scaling their virial \mbh\ to the \mbh $-\sigma_*$ relation of 
inactive galaxies. The calibration by Woo et al. (2010), referenced to 
the \mbh $-\sigma_*$ relation of G\"ultekin et al. (2009), gives $f=1.31$ for 
$v$ set to FWHM of broad H$\beta$.  This normalization, in conjunction with 
the BLR size-luminosity relation of Bentz et al. (2013), yields (Park et al. 
2012):

\begin{equation}
\begin{array}{l}
{\rm log} (M_{\rm BH}/M_\odot)  =  6.97 + 0.518\, {\rm log}\left 
(\frac{L_{5100}} {10^{44} \ {\rm erg\  s}^{-1}} \right) \\
\ \ \ \ \ \ \ \ \ \ \ \ \ \ \ \ \ \ \ \ \ \ \ \ \ \ \ \
+ 1.734\, {\rm log}\left (\frac{{\rm FWHM}}{10^3 \ {\rm km\ s}^{-1}} \right).
\end{array}
\end{equation}  

Most of the line widths come from the literature.  Whenever possible, we use 
our own measurements derived from high-quality Magellan spectra (Ho \& Kim 
2009).  For a small fraction of the sample, we located matching archival 
spectra from SDSS (Alam et al. 2015), the FIRST Bright Quasar Survey 
(White et al. 2000), and the 6dF Galaxy Survey (Jones et al. 2004). We 
reanalyzed these archival data, performing spectral decomposition following 
the methodology of Kim et al. (2006) and Ho \& Kim (2009), which includes 
fitting the AGN continuum, the stellar continuum, Fe emission, and deblending 
of narrow and broad emission lines.   Note that in the 
derivation of \mbh\ we do not use the continuum luminosity ($L_{5100}$) from 
the spectral analysis but instead the nucleus luminosity derived from our 
photometric decomposition, which we believe to be more robust against
host galaxy contamination.  And while most of the line widths pertain to 
H$\beta$, a minority come from H$\alpha$. We adopt H$\alpha$ instead of 
\hb\ when a FWHM measurement for the latter is unavailable
or unreliable, mainly because Fe II multiplets were not included in the 
spectral fitting or \hb\ is heavily affected by the young stellar population.

Both lines share very similar 
kinematics.  In general, however, H$\alpha$ tends to be slightly narrower than 
H$\beta$; we convert the FWHM of H$\alpha$ to H$\beta$ using the statistical 
relation between the two line widths given by Greene \& Ho (2005). Note that,
for 37 objects, BH masses come from the reverberation mapping. 

The dominant uncertainty for BH virial masses comes from the current
uncertainty on the virial coefficient $f$, which is $\sim 0.4$ dex (Onken et 
al. 2004; Collin et al. 2006; Woo et al. 2010; Ho \& Kim 2014).  
Variability introduces an additional source of uncertainty, since the \hst\ 
images, from which we extract the AGN luminosity, and the spectra, which 
provide the line widths, 
were not observed simultaneously.  However, variability affects BH virial mass 
estimates only at the level of $\sim 0.1$ dex (Wilhite et~al. 2006; Denney 
et~al. 2009; Park et al. 2012).  We conservatively and uniformly assign an 
overall error budget of 0.5 dex to \mbh.  This is in a good agreement with the 
uncertainties quoted in previous studies (e.g., Vestergaard \& Peterson 2006; 
Park et al. 2012). 

We estimate bolometric luminosities using the continuum luminosity at 
5100 \AA\ measured from the imaging analysis and a conversion factor 
$L_{\rm bol} = 9.8 L_{\rm 5100}$ 
(McLure \& Dunlop 2004).  Our sample spans a relatively wide range of 
Eddington ratios\footnote{The Eddington luminosity is defined as 
$L_{\rm Edd}\,=\,1.26 \times 10^{38} \left(M_{\rm BH}/ M_{\odot}\right)$ 
\lum .}, from $\sim 0.001$ to 1, with $\langle$\edd$\rangle \approx 0.1$
(Figure 2).

\subsection{Bulge Type Classification}
We divide the sample of late-type galaxies into two classes using 
the \ser\ index and bulge-to-total light ratio [classical bulges 
($n > 2$ and $B/T > 0.2$) and pseudo-bulges ($n \le 2$ or $B/T \le 0.2$; 
Fisher \& Drory 2008; Gadotti 2009).
However, for some objects, there are ambiguities involved for the classification. 
We classify the bulge-type of host galaxies as ``intermediate'' (1) if the host is 
highly disturbed, which makes us difficult to constrain the \ser\ index and $B/T$ reliably
(e.g., 3C 57), (2) if it is hard to choose the best single model to describe the 
host profile because either models (with a classical bulge or a pseudo-bulge) 
work equally well for the fit (e.g., MS 0111.9$-$0132, MS 1455.7+2121, etc.), 
(3) if there is severe discrepancy between our fitting result and that
from the literature (e.g., Mrk 509, Mrk 609, etc.). 

\subsection{Uncertainties on Host Galaxy Parameters}

The dominant sources of uncertainty for the host galaxy measurements arise 
from systematic rather than statistical errors.  These have been extensively 
explored by Kim et al. (2008a) through controlled tests using simulated 
images, which take into into account factors such as PSF variation, S/N, and 
the brightness contrast ratio between AGN and the host.  We find that the most 
important factor affecting the final uncertainty is the luminosity ratio of 
the bulge to the nucleus (\hnr). If \hnr\ $\ge 0.2$, the bulge luminosity 
typically has an uncertainty of 0.3 mag, but it increases to 0.4 mag if 
\hnr$< 0.2$.  The error budget increases by $\sim 0.1$ mag if a disk component 
is present, and by $\sim 0.3$ mag if the core of the image is saturated.  For 
the relatively bright AGNs in our sample, the luminosity of the nucleus is quite
well determined, with errors \lax$0.05$ mag, degrading to $\sim 0.1$ mag only 
when the nucleus is much fainter than the host (\hnr$\ge 50$). If the best 
model for the host is not unique, we incorporate measurements from the 
second-best model into the error budget (e.g., 3C 17).

As $k$-corrections need to be applied to the host luminosities (Section 5.1),
the final $R$-band bulge luminosities are sensitive both to the choice of 
galaxy template spectra and the original \hst\ filter in question.  We study 
these effects by experimenting with template spectra of different galaxy types 
and calculate the scatter in the inferred output bulge luminosities.  We choose
a range of Hubble types that reasonably bracket the likely morphological type 
of the host galaxy in question.  Not surprisingly, the uncertainty is larger 
for bluer filters and for more distant objects since we will be sampling the 
rest-frame light further away from the central wavelength of the $R-$band 
filter, but the typical error budget is negligible ($\leq 0.05$ mag).  The 
bulge luminosity can be systematically underestimated at most by 0.1 mag, if 
the filter is bluer than F606W and the redshift is larger than 0.3.  There is, 
however, one important caveat.  If the stellar population of the host is very 
young (e.g., that of an Sc or starburst galaxy), the final luminosity can be 
systematically overestimated by as much as 0.8 mag at $z\approx 0.3$ since we 
assume the stellar population to be old (e.g., that of an elliptical). 

Our $k$-corrections for the nucleus luminosity assume a fixed quasar composite
spectrum.  The continuum shape and emission-line spectra of individual 
optically selected quasars typical vary by $\sim 20\%$ around the mean
(Vanden~Berk et al. 2001).  Monte Carlo simulations indicate that this level
of spectral variation introduces variations of less than 0.03 mag for 
$M_{R, \rm nuc}$ and 0.1 dex for $L_{5100}$, respectively.
\begin{figure*}[ht]
\centering
\includegraphics[width=0.95\textwidth]{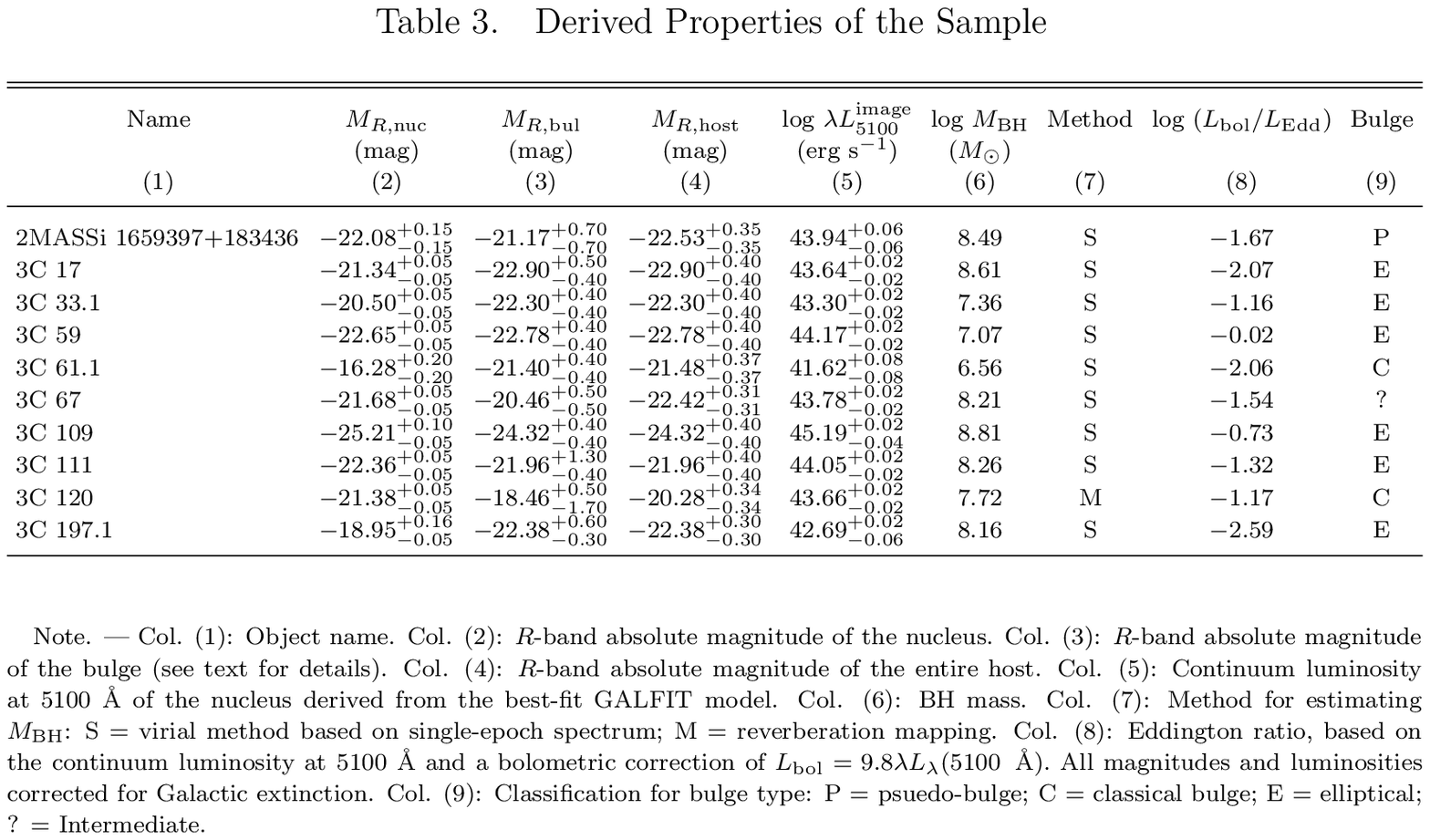}
\end{figure*}

\begin{figure*}[ht]
\centering
\includegraphics[width=0.95\textwidth]{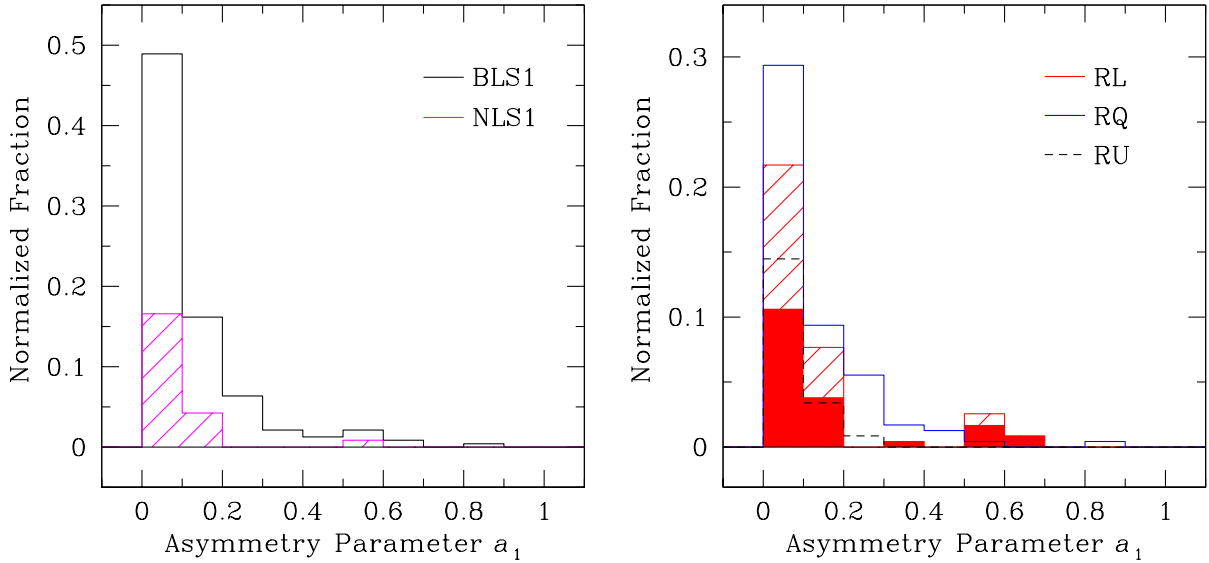}
\caption{
Distributions of the asymmetry parameter $a_1$ for different types of AGNs.
We divide the sample into broad line type on the left and radio-loudness on
the right. In the right panel, extreme radio-loud objects
are denoted by red-filled histogram.}
\end{figure*}
\vskip 0.3cm

Although our objects are relatively nearby, they can still be affected by 
cosmological evolution of the stellar mass-to-light ratio ($M_*/L$).  More 
distant galaxies have younger stars and hence lower $M_*/L$.
In order to correct for this effect, we adopt the correction 
$d {\rm log}(M_*/L_B)/dz=-0.72$ 
taken from Treu et al. (2002).  According to the stellar population models of 
Bruzual \& Charlot (2003), this magnitude of evolution of $M_*/L_B$ is well 
described by a single stellar population with a formation redshift of 1.5 
[assuming solar metallicity and a Salpeter (1955) initial mass function]. The 
corresponding evolution in the $R$ band is $d {\rm log}(M_*/L_R)/dz \approx 
-0.29$. Given that $\langle z \rangle \approx 0.1$, the correction for the 
$R$-band magnitudes is negligibly small ($\sim$ 0.07 mag).  

Lastly, some of our objects have saturated AGN cores that we could not fully 
correct using short, unsaturated exposures.  Under these circumstances we 
masked out the few central saturated pixels during the fitting process.  Our 
simulations indicate that this procedure can increase the uncertainty of the 
bulge luminosity by $\sim$0.2 mag.

Determining errors on $R_e$ and $n$ is non-trivial, as their errors are highly 
coupled and, as mentioned in Section 4, $n$ can be difficult to fit reliably.
For moderately bright hosts (\hnr\ $\ge 0.2$) for which $n$ can be fit freely, 
our tests show that we can determined $R_e$ and $n$ to within 25\% and 15\%, 
respectively.  However, when the host is faint relative to the nucleus 
(\hnr\ \lax\ 0.2), the error on $R_e$ and $n$ can be as large as 0.5 dex.

\section{Statistical Trends}

Table 4 summarizes the mean properties of various subclasses of AGNs and their 
host galaxies.

\subsection{Radio-loudness}

There is increasing evidence that RL galaxies are somewhat more massive and 
more early in type than RQ galaxies (e.g., Smith et al. 1986; Best et al. 
2005; Floyd et al. 2010).  We test whether the trend with radio emission also 
holds for the host galaxies of type 1 AGNs.  The top and middle panels in 
Figure 7 show the luminosity distributions of the bulges and host galaxies 
for the different subsets of AGNs.  We find that RL AGNs tend to have 
more luminous bulges than RQ sources. The Anderson-Darling test 
(A-D test) rejects the null hypothesis that the samples are drawn from 
the same parent distributions with a probability of less than 3\%.  The 
difference between the median \rbulge\ of the two subsamples is 
$\sim 0.56$ mag, and for \rhost\ the difference is 
$\sim 0.31$ mag.  This is consistent with prior studies, which find 
that, on average, the host galaxies of RL quasars are brighter than those of 
their RQ counterparts (Dunlop et al. 2003).  The median \btot\ for the host 
galaxies of RL AGNs is significantly larger than that of RQ AGNs 
($0.81\pm0.19$ vs. $0.33\pm0.20$).  The difference in \btot\ is even larger for
the extreme radio-loud sources (RL2; median \btot$= 1.00$). 
These trends are broadly consistent with the findings of Floyd et al. (2010), 
who report that 3CR radio galaxies are preferentially bulge-dominated.  
In addition, the discrepancy between RL AGNs and RQ AGNs becomes more
significant if we classify RU AGNs as RQ AGNs. From the A-D test, the null 
hypothesis that the two samples come from the same population can 
be rejected with a probability of less than 1\% for the distributions of bulge 
luminosity, host luminosity, and \btot.

RL AGNs are traditionally associated with massive 
elliptical galaxies (e.g., Matthews et al. 1964). The median 
\rhost\ of the RL sources in our sample is $-22.4$ mag, 1.4 mag brighter than 
\lstar\ of local elliptical galaxies 
(Tempel et al. 2011). The extreme RL (RL2) population, with median \rhost\ = 
$-22.8$ mag, stands out even more. Therefore, we confirm that the majority of 
type 1 RL AGNs are hosted by massive galaxies. 
A significant number of RL sources in our sample 
(35 objects or $\sim45\%$ of the RL sources) seem to be hosted by rather 
late-type, disk-dominated ($B/T < 0.5$) galaxies\footnote{Some S0 galaxies can 
be disk-dominated, having $B/T$ as small as $0.3-0.4$ (e.g., Im et al. 2002, and references therein).}.
This is rarely seen among classical radio galaxies and RL quasars (e.g., 
Dunlop et al. 2003; Madrid et al.  2006), although some notable exceptions 
have been reported (Bagchi et al. 2014; Kotilainen et al. 2016).  The majority 
of these sources in our sample have moderate absolute ($P_{6 \rm cm} \leq 
10^{22}$ W Hz$^{-1}$) or relative ($R \leq 100$) radio powers.  As Ho \& Peng 
(2001) demonstrate, the radio-loudness parameter $R$ for AGNs of moderate 
luminosity can be severely underestimated if the host galaxy contamination to 
the optical light is not properly removed.  Were it not for the 
high-resolution of the \hst\ images and the careful separation of the nucleus 
from the host, which can contribute $\sim$4\%--99\% of the total optical 
light, some of the disk-dominated RL objects in this study also would 
traditionally have been classified as RQ.  The radio morphologies appear to be 
compact, with no obvious signs of extended jets or lobes, but the resolution 
of the available radio maps (5\asec\ to 6\amin) is too poor and heterogeneous 
to yield meaningful insights into source structure.  In our calculation of the 
$R$ parameter, we assume that all of the detected radio emission comes from 
the AGN; this may overestimate the radio power of the AGN because we have not
accounted for the contribution from the host galaxy (e.g., Lal \& Ho 2010).

Four of the objects with a disk-like component have sufficiently strong 
radio sources ($P_{6 \rm cm} > 10^{25}$ W Hz$^{-1}$; $R > 10^3$) that they 
are, without a doubt, genuinely RL.  Three of them (3C 67, 3C 120, and 
PG 2349$-$014) show clear evidence of tidal interaction.  This is consistent 
with other nearby radio galaxies, for which those with ``disky'' stellar 
component are mostly either merging or post-merging systems (Floyd et al. 
2008).  

Galaxy interactions are often implicated for triggering AGNs (e.g., Stockton
et al. 1982; Bennert et al. 2008).  However, it is unclear whether ongoing
mergers enhance radio emission in AGNs (Dunlop et al. 2003, Best et al. 2005;
but see Heckman 1983). In order to investigate possible influence of mergers 
or interactions on radio activity, we examine the degree of asymmetry of the 
hosts, as measured by the amplitude of the first Fourier mode ($a_1$) derived 
from our 2-D fits.  The right panel in Figure 8 shows that the majority of 
radio-loud AGN are hosted by undisturbed galaxies. If anything, fraction of 
hosts with $a_1 \ge 0.2$ is actually slightly higher for RQ AGNs compared to 
RL AGNs, exactly the opposite of what is expected. The extreme RL AGNs (RL2)
appears to be hosted by similar fraction of disturbed galaxies as RQ AGNs. 
Overall, there appears to be no clear casual connection between radio 
emission and ongoing merger.

%
\begin{figure*}[t]
\centering
\includegraphics[width=0.75\textwidth]{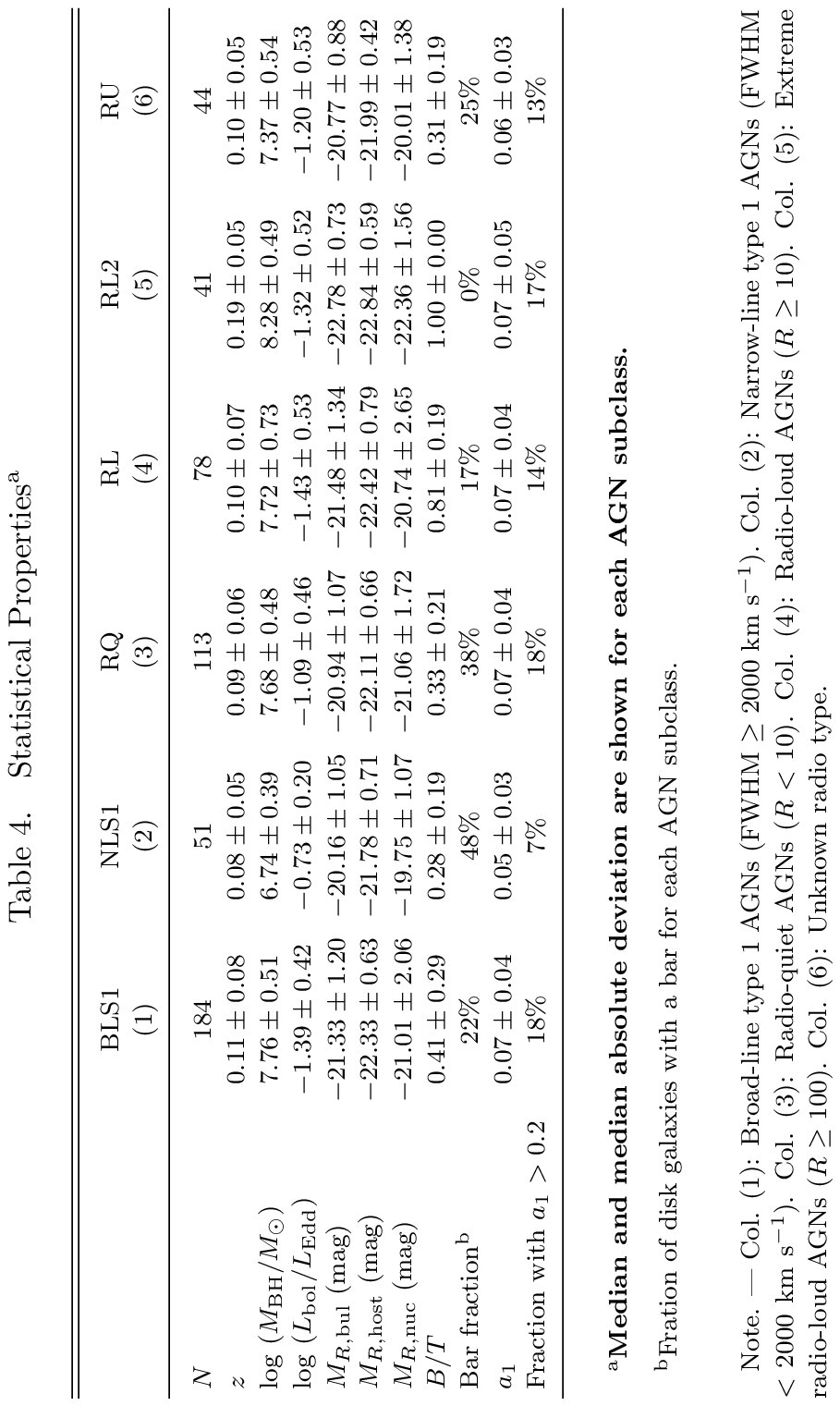}
\end{figure*}
\vskip 0.3cm

\subsection{NLS1: Does Secular Evolution Play an Important Role?}

NLS1s distinguish themselves from BLS1s not only in terms of their nuclear 
properties, but also in terms of the large-scale characteristics of their host 
galaxies (see Boller 2004 for a review).  While the original \hst\ 
observations analyzed in this study were not based on selection by host galaxy 
morphology, we find that BLS1 galaxies tend to have more luminous bulges and 
have earlier morphological types than NLS1s (Figure 7 and Table 4).  The 
difference ($\sim 1.2$ mag) in \rbulge\ between BLS1 and NLS1 hosts 
corroborates with the notion that the central BHs in the former is more 
massive than those in the latter. 
The difference is not as great in terms of total luminosity, but it is still 
significant (0.55 mag). 
The A-D test for both the distributions of bulge luminosity and total 
luminosity rejects the null hypothesis that the samples are 
drawn from the same parent distributions with a probability of less than 1\%.

Moreover, if NLS1s experience more elevated star 
formation activity compared to BLS1s (Sani et al. 2010), we expect NLS1s to 
have younger stellar populations and hence lower mass-to-light ratios than 
BLS1s, which would further enhance the mass difference between the two classes.

We find no significant difference in the overall distribution of global 
morphological asymmetry between NLS1s and BLS1s, but the fraction of disturbed 
hosts ($a_1 \ge 0.2$) evidently decreases for NLS1s.  Only 4 out of 51 objects 
($[$HB89$]$~1321+058, Mrk 896, PG1700+518, RX J2217.9-5941) show weak signs of 
interactions (Table 2).  Numerical simulations (Lotz et al. 2008, 2010) show 
that morphological disturbances last much longer in gas-rich mergers than in 
dry mergers.  Indeed, among nearby  ($z<0.1$) field galaxies, the merging 
fraction among spiral galaxies is higher than in elliptical galaxies (Darg et 
al. 2010).  Since NLS1s are preferentially hosted by late-type (low-$B/T$) 
galaxies, the apparent rarity of merger or tidal features for NLS1s is 
especially telling.  Darg et al. (2010) find that the merging fraction for 
nearby late-type galaxies is approximately $8\%-11$\%, essentially identical 
to the detection frequency of global asymmetry in our sample of NLS1s (8\%).  
Taken at face value, these results suggest that dynamical interactions play a 
minor role in triggering AGN activity for NLS1s.  Secular evolution seems to 
be more important for NLS1s (Orban de Xivry et al. 2011).  These conclusions 
need to be verified using a statistically more rigorous sample with better 
control on surface brightness depth.

If galaxy mergers play a less crucial role in the lifecycle of NLS1s, then 
pseudo-bulges should be more common in NLS1s than in BLS1s.  Using the \ser\ 
index ($n < 2$; Fischer \& Drory 2008) as a proxy for identifying 
pseudo-bulges, we find that the fraction of pseudo-bulges in disky NLS1s is 
$\sim 71\%$.  Gadotti (2009) warns, however, that the \ser\ index threshold 
may not be rigorous enough to cleanly distinguish pseudo-bulges from classical 
bulges.  We therefore impose an additional criterion, namely that, in addition 
to $n < 2$, the bulge fraction of the host must be $B/T < 0.2$.  This reduces 
the pseudo-bulge fraction in NLS1s to $\sim 43\%$, but it is still much larger 
than that in BLS1s ($\sim 28\%$).  As the fraction of pseudo-bulges may depend 
on galaxy mass (e.g., Gadotti 2009), we perform the same comparison in bins of
fixed host galaxy luminosity.  For 1 mag-wide bins within $-20 < M_{R,\rm host}
< -23$, the fraction of pseudo-bulges in NLS1s ($33\%-58\%$) always 
significantly exceeds that in BLS1s ($12\%-29\%$), by a factor of $\sim2$.

Several studies claim that the frequency of bars is significantly larger in 
the hosts of NLS1s than in BLS1s (Crenshaw et al. 2003; Deo et al. 2006; Ohta 
et al. 2007).  Among NLS1s hosted by spiral galaxies, the bar fraction is 
$\sim 48\%$ which is marginally lower than that in other studies of nearby 
(low-luminosity) Seyfert galaxies ($65\% - 73\%$).  By contrast, spiral 
galaxies with BLS1s have a bar fraction of only $\sim 22\%$, in excellent 
agreement with previous \hst\ studies ($\sim 25\%$; Crenshaw et al.  2003).  
Care must be exercised in comparing bar fractions between subsamples, because
the incidence of bars in the general local galaxy population depends strongly 
on galaxy color and morphology. Redder, more bulge-dominated disk galaxies 
tend to have a higher bar fraction (Masters et al. 2011).  Interestingly, the 
trend we observe in our AGN sample is precisely the opposite: the bar fraction 
is higher for the NLS1s than in BLS1s.  Our analysis shows that NLS1s are less 
bulge-dominated than BLS1s, and, if they experience increased levels of star 
formation (Sani et al. 2010), we also expect them to be bluer.  This implies 
that the observed higher bar fraction in NLS1 must be a genuine 
{\it enhancement}\ relative to the general galaxy population.  

The above four characteristics---more disk dominance, higher fraction of 
pseudo-bulges, higher bar frequency, and lower merger fraction---implicate 
secular processes as the main agent for BH growth in NLS1s.

Among our sample of 51 NLS1s, 12 (24\%) appear not to have a disk component.  
What kind of galaxies are these?  Given the low BH and stellar masses of 
NLS1s, one interesting possibility is that hosts may be spheroidals rather 
than genuine low-mass elliptical galaxies (Greene et al. 2008; Jiang et al. 
2011).  
Spheroidals, like pseudo-bulges, occupy a different locus on the 
Kormendy relation than do classical bulges and elliptical galaxies (Kormendy 
et al. 2009).  Figure 9 compares the host galaxies of the 12 apparently 
diskless NLS1s with a sample of inactive galaxies.  Apart from POX 52, whose 
host is a genuine spheroidal (Barth et al. 2004), the majority of the NLS1s 
largely follow the sequence of classical bulges and elliptical galaxies.
It appears that these diskless NLS1s are low-mass elliptical 
galaxies rather than spheroidals. One object (RX J1702.5+3247) with 
a small effective radius ($\sim$0\farcs07) and high surface brightness 
appears to be significantly off from the Kormendy relation of normal galaxies.
Given the fact that the size of the bulge is comparable to that of the PSF
and the residuals still show an extended feature, the host modelling can be
\begin{figure}[t!]
\centering
\includegraphics[width=0.47\textwidth]{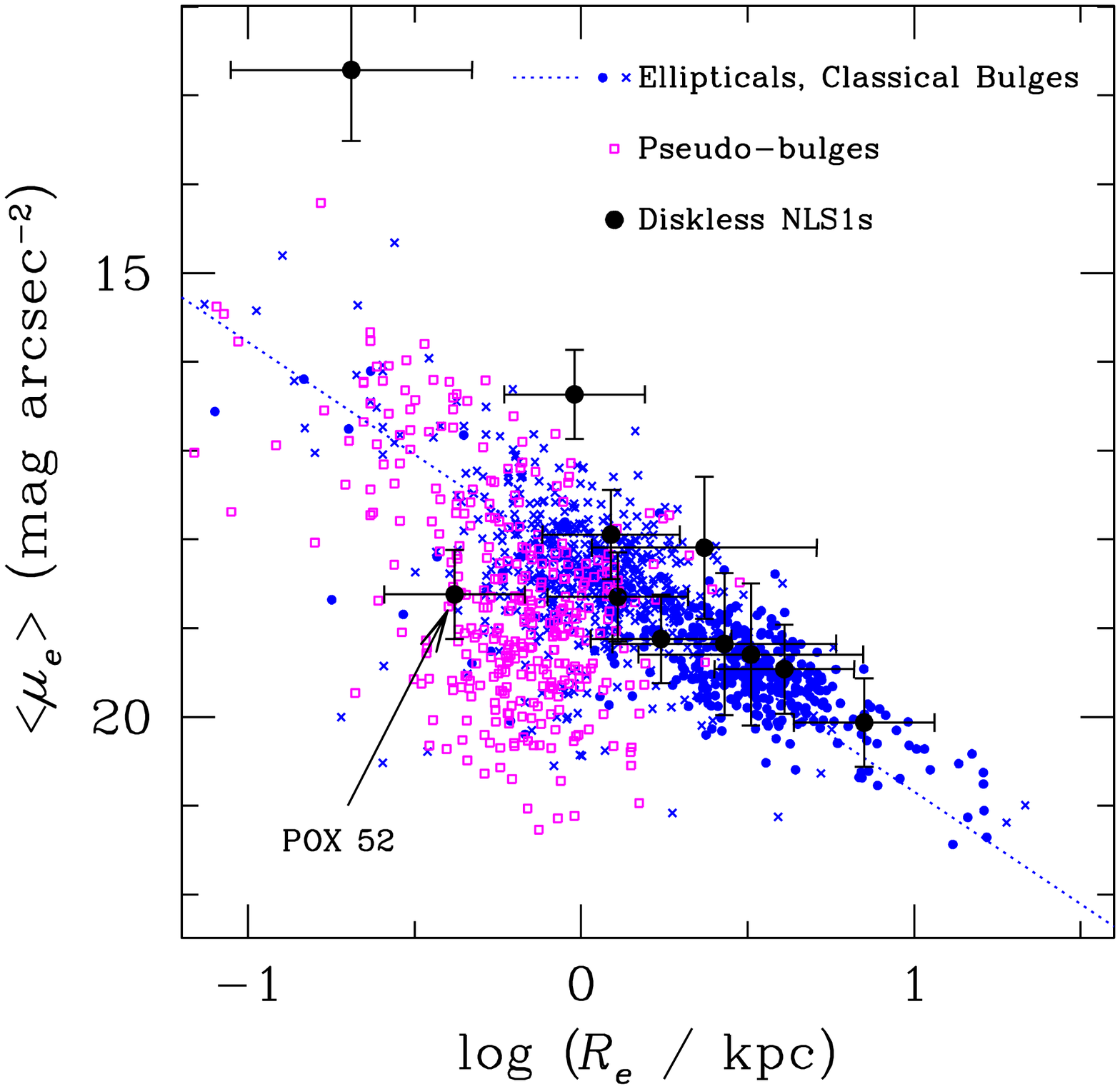}
\caption{
Correlation between the effective radius $R_e$ and the mean surface
brightness at $R_e$ of bulges and elliptical galaxies (the Kormendy relation),
comparing the subsample of 12 NLS1s whose host galaxy apparently lacks a disk
component (black filled dots) with inactive galaxies.
We show three samples for comparison: elliptical galaxies (blue circles) from
Bender et al. (1992), Gadotti (2009), and Kormendy et al. (2009); disk
galaxies with classical bulges (blue crosses) from Bender et al. (1992),
Fisher \& Drory (2008), Gadotti (2009), and Laurikainen et al. (2010); and
disk galaxies with pseudo-bulges (magenta squares) from Fisher \& Drory
(2008), Gadotti (2009), and Laurikainen et al. (2010).  The fit for the normal
elliptical galaxies and classical bulges is denoted by a blue dotted line.
}
\end{figure}
\vskip 0.2in
\noindent
heavily affected by the PSF mismatch. Therefore, the derived parameters for 
the host galaxy can be unreliable for RX J1702.5+3247.

\subsection{Merging Features}
Merger induced AGN activity is a long-standing issue 
that still remains unsolved. Letawe et al. (2010) showed
that the interaction frequency in nearby QSOs is marginally higher than that
in normal galaxies (See also Bennert et al. 2008; Ramos Almeida et al. 2012;
Hong et al. 2015). On the other hand, a lack of sign of interaction have been 
found in low-luminosity AGN and X-ray selected distant AGN (Schade et al. 2000;
Cisternas et al. 2011).
As discussed above, GALFIT allows us to quantify the degree of disturbance
using the Fourier mode components. While there can be several different
forms to quantify the degree of asymmetry using the Fourier components, Kim
et al. (2008b) showed that the amplitude of the first Fourier component
($a_1$) can be used as an indicator of interaction. We also find sign of
interaction by eye in the original image and residual image and classify
interacting hosts independently. There is a slight discrepancy between the 
two methods but the overall result is similar.

\begin{figure*}[ht]
\centering
\includegraphics[width=0.95\textwidth]{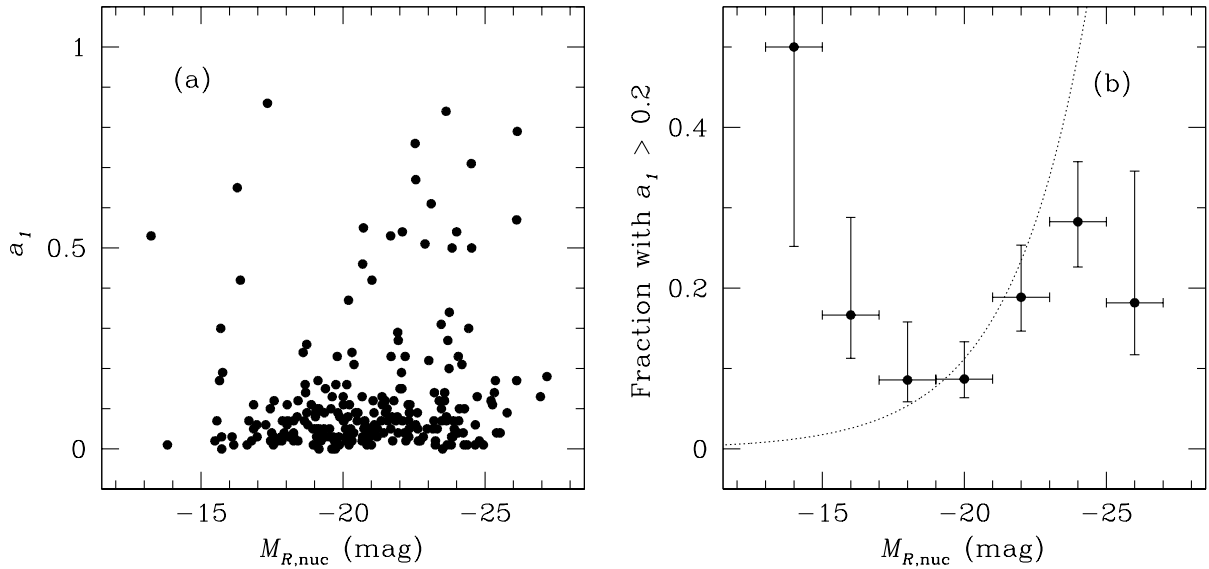}
\caption{
Dependence of sign of interaction on nuclear brightness. The left panel
shows the distribution of $a_1$ as a function of \rnuc.
In the right panel, black points show the fraction of interacting hosts
($a_1 > 0.2$) in bins of \rnuc.
Error bars in the vertical and horizontal axes represent
uncertainties of the fraction (Cameron 2011) and the size of the bin,
respectively.
The dotted line represents the
merger fractions of AGNs ($z \leq 3$) taken from Equation 2 of Treister et al.
(2012). The nuclear luminosity is converted to  bolometric luminosity using a
bolometric conversion of $L_{\rm bol} = 9.8 L_{\rm 5100}$.}
\end{figure*}

Using the amplitude of the first fourier mode ($a_1 > 0.2$), we find that 
39 objects ($\sim 17\%$) show signs of interaction while the visual 
inspection yields 40 merging hosts. Note that 12 hosts with $a_1 > 0.2$ 
are classified as non-interacting galaxies from the visual 
inspection. While dust lanes are present in the host for the majority of them, 
some of them are associated with close companion. It is unclear if 
there are merging features in the images. And 75 objects seem to 
have close companion in
the images. We compare $a_1$ to \rnuc\ in Figure 10a, which shows that, the vast
majority of merging hosts ($a_1 > 0.2$) have bright nucleus
(\rnuc $< -20 $ mag). This is consistent with the previous finding in which
sign of interaction is found more frequently in QSO hosts than in Seyfert
hosts (e.g., Treister et al. 2012). 
There are three outliers that have high values of $a_1$ ($> 0.5$) but 
a relatively low-luminosity nucleus (\rnuc $> -20 $ mag). While two of them 
(3C 61.1 and Mrk 40) clearly show 
merging features in the \hst\ images, one object (NGC 4143) appears to be
a non-interacting galaxy and its high value of $a_1$ mainly comes from the dust 
lanes in the host. 
Figure 10b shows the fraction 
of merging AGNs ($a_1 > 0.2$) as a function of the nucleus luminosity. For 
comparison, we overplot the merging fraction from Treister et al. (2012).  
Because the sample of Treister et al. (2012) contains more distant objects 
($z \leq 3$) than our sample, a direct comparison with our results is 
not straightforward. Nevertheless, we find that the trend in our sample is in 
good overall agreement with that from Treister et al. (2012). 
As a caveat, however, it is worthwhile
noting that our sample spans wide range of redshift and is observed under
different observing conditions (S/N, filters, and detectors).

\section{Summary}

We present detailed analysis of \hst\ optical images of a diverse sample of 
235 $z < 0.35$ type 1 AGNs for which reliable BH masses can be estimated from 
available optical spectroscopy, with the primary goal of investigating the 
structures of their host galaxies and their dependence on AGN properties.  
This represents the largest sample of low-redshift AGNs to date for which 
quantitative measurements of the host galaxies are derived from 
two-dimensional ({\tt GALFIT}) decomposition using high-resolution images. 
We perform a careful decomposition of the sources into their principal 
structural components (nucleus, bulge, bar, disk), accounting for the effects
of the PSF and simultaneously modeling, if necessary, spiral arms, tidal 
features, and other distortions due to tidal interactions and mergers.  
Whenever possible, we measure the luminosity profile of the bulge, classify 
its type (pseudo-bulge or classical bulge), and specify the overall morphology 
of the host galaxy in terms of its bulge-to-total light ratio.  Fourier modes 
are used to quantify the degree of asymmetry of the light distribution of the 
stars.

Our database will be used for a variety of forthcoming applications.  Here, we 
focus only on some issues relevant to the nature of two subclasses of AGNs, 
namely those with ``narrow'' (H$\beta$ FWHM $\leq 2000$ km~s$^{-1}$)) broad 
lines (so-called NLS1s) and radio-loud sources.  Our main results are:

\begin{itemize}

\item{Compared to broad-line (H$\beta$ FWHM $> 2000$ km~s$^{-1}$) type 1 AGNs, 
NLS1s have lower overall luminosity, lower bulge luminosity, lower 
bulge-to-total light ratio (later Hubble types), higher incidence of 
pseudo-bulges, higher frequency of bars, and fewer signs of morphological 
asymmetry.  This suggests that NLS1s have experienced fewer external dynamical 
perturbations and that their evolution is primarily driven by internal secular 
evolution.}

\item{The fraction of AGN hosts showing morphological signatures of tidal 
interactions and mergers is larger for more luminous AGNs.}

\item{The properties of the host galaxies vary systematically with the degree 
of radio-loudness of the AGN.  Radio-loud sources inhabit more 
bulge-dominated, more luminous hosts, but, contrary to earlier reports, in our 
sample they are not preferentially interacting or merger systems.}
\end{itemize}

\vskip 0.3in

We are grateful to an anonymous referee for constructive comments.
This work was supported by NASA grants HST-AR-10969.03-A and HST-GO-10428.11-A 
from the Space Telescope Science Institute (operated by AURA, Inc., under NASA 
contract NAS5-26555).  MK was supported by the National Research Foundation of 
Korea (NRF) grant funded by the Korea government (MSIP) (No. 2017R1C1B2002879).
MI was supported by the National Research Foundation of Korea (NRF) grant, 
No. 2017R1A3A3001362, funded by the Korea government (MSIP).
The work of LCH was supported by the National Key Program 
for Science and Technology Research and Development (2016YFA0400702) and the 
National Science Foundation of China (11303008, 11473002).  
Research by AJB is supported in part by NSF grant AST-1412693.
We made use of the 
NASA/IPAC Extragalactic Database (NED), which is operated by the Jet 
Propulsion Laboratory, California Institute of Technology, under contract with 
NASA.

\figurenum{11.1}
\begin{figure*}[ht]
\centering
\includegraphics[width=0.95\textwidth]{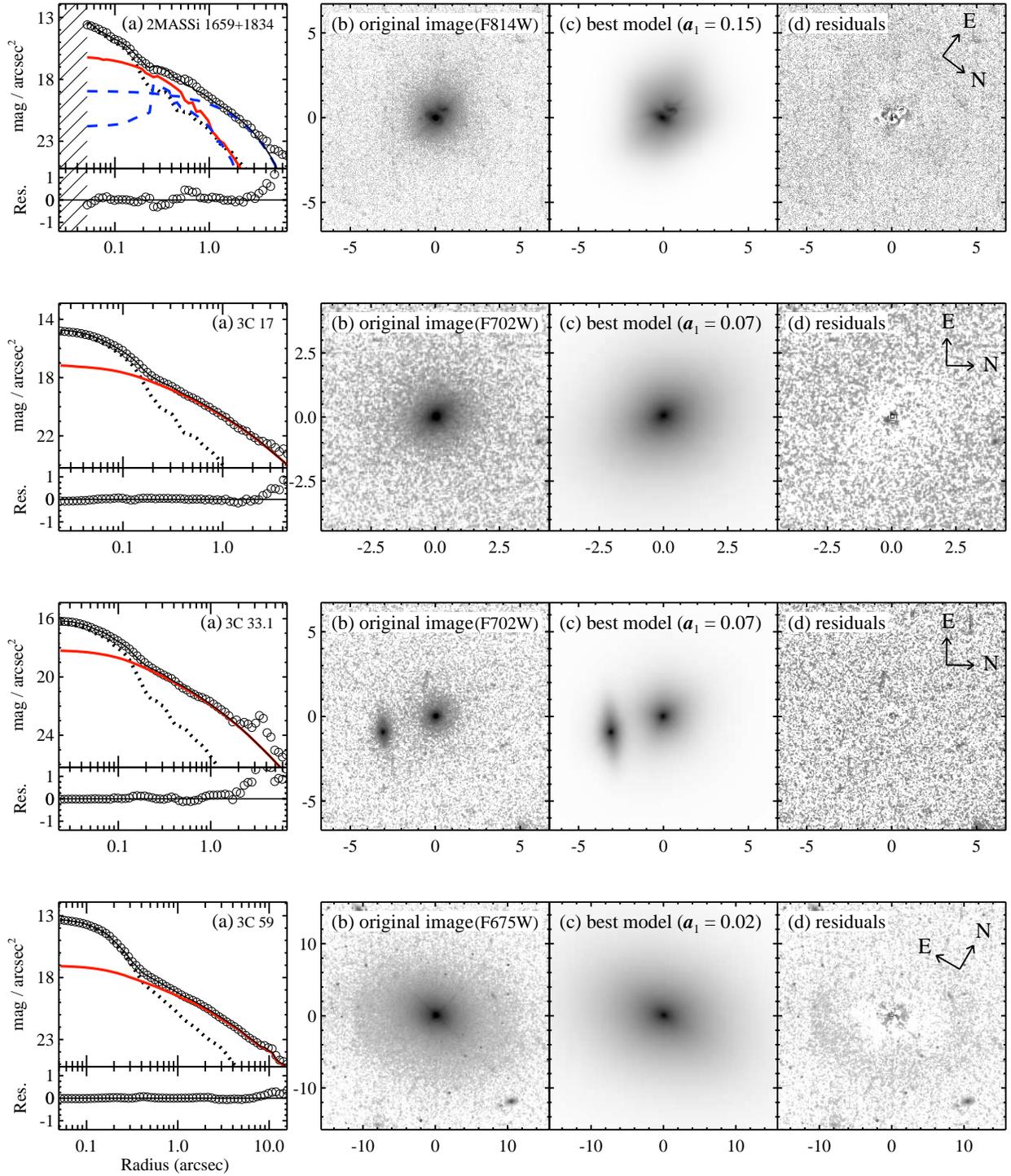}
\caption{
{\tt GALFIT} decomposition for our sample.
({\it a}) Azimuthally averaged profile, showing the original data
({\it open circles}), the best fit ({\it solid line}), and the sub-components
[PSF ({\it dotted black line}), bulge ({\it red solid line}), disk ({\it blue 
dashed line}), bar ({\it green dot-dashed lines})].  Shaded regions mark 
saturated pixels.  The residuals are plotted on the bottom. Because the images 
are already sky-subtracted, the outer region of the 1-D profiles are not 
smooth.  We present the 2-D image of ({\it b}) the original data, ({\it c}) 
the best-fit model for the host (the AGN is excluded to better highlight the 
host), with the amplitude of the first Fourier mode ($a_1$) labeled, and 
({\it d}) the residuals. The units of the images are in arcseconds. All images 
are displayed on an asinh stretch.  (The complete set of plots for Figure 11
is available in the online journal.)}
\end{figure*}
\clearpage

\appendix

\section{Notes on Individual Objects}

Comments on the fitting results for individual objects are given here.  The
corresponding fit results are located at the Figures indicated.  Note that the
luminosities from literature are corrected to our adopted cosmology.

\bigskip

{\it 2MASSi 1659397+183436} (Fig. 11.1) --- \ \
The image is slightly saturated at the center. It appears that the host galaxy
has a companion. The host is fit with a pseudo-bulge and a disk, while the
companion galaxy requires a disk with spiral arms.  Marble et al. (2003)
estimate that the nucleus-subtracted host galaxy has an $I$-band magnitude of
$-22.9$, which is $\sim 0.3$ mag fainter than our estimate. 

{\it 3C 17} (Fig. 11.1) --- \ \
While the host galaxy appears to be E-type, the fit with two components works
reasonably well. The aperture magnitude of the host inside 15 kpc from de Koff
et al. (1996) agrees with our estimates to within the uncertainties. Based on
ground-based and {\it HST}/NICMOS images, Zirbel (1996) and Donzelli et al.
(2007) also classify this object as an elliptical galaxy.  Floyd et al. (2008)
and Donzelli et al. (2007) fit the $H-$band NICMOS image and find $n\approx7.5$
and $2.7$, respectively.

{\it 3C 33.1} (Fig. 11.1) --- \ \
The host galaxy is fit with a single S\'ersic component. There is a companion
source which we fit with a bulge and a disk. Our measurement is in broad
agreement with the aperture photometry of de Koff et al. (1996). Floyd et al.
(2008) and Donzelli et al. (2007) derive $n\approx6.4$ and 2.3, respectively.

{\it 3C 61.1} (Fig. 11.2) --- \ \
There are three diffuse companions. The nucleus appears much fainter than the
host. The host galaxy is fit with a bulge and an off-center disturbed structure.
Donzelli et al. (2007) classify this object as an elliptical galaxy. The
measured $B/T=0.92$ also reveals that the host is an early-type galaxy.  Floyd
et al. (2008) and Donzelli et al. (2007) estimate $n\approx2.8$ and 3.5,
respectively, which agree well with our measurement. 

{\it 3C 67} (Fig. 11.2) --- \ \
The host galaxy is disturbed, leading to a rather ambiguous fit. 
We fit the host with a low-$n$ bulge and two disks with Fourier modes.
The aperture photometry from de Koff et al. (1996) closely matches 
our measurement.

{\it 3C 109} (Fig. 11.2) --- \ \
The host galaxy is fit with a single S\'ersic component.
De Koff et al. (1996) slightly overestimate the aperture magnitude because
they do not subtract the nucleus from the image.

{\it 3C 111} (Fig. 11.2) --- \ \
The host galaxy is fit with a single S\'ersic component. Martel et al. (1999)
argue that the host might be an elliptical-like galaxy. Donzelli et al. (2007)
estimate $n\approx5.8$. 

{\it 3C 120} (Fig. 11.3) --- \ \
The host galaxy is fit with a tiny bulge ($R_e \sim$0\farcs1) and a disturbed
disk with spiral arms. However, we also find that a fit with a single S\'ersic
component can be acceptable.  Using ground-based $K-$band data, Inskip et al.
(2010) also fit the host with two components, resulting in $B/T=0.88$. However,
they claim that the inner component ($R_e \sim 0.62$ kpc) and the larger
component ($R_e \sim 8.62$ kpc) can be regarded as as a disk and a bulge,
respectively, which is the opposite of our interpretation ($B/T=0.19$). By
contrast, Bennert et al. (2010) concluded that the host is an elliptical galaxy
(but see Bentz et al. 2009). The reason for this controversy is due to presence
of the disturbed disk-like feature. 


{\it 3C 197.1} (Fig. 11.3) --- \ \
While a single S\'ersic component gives the best fit, 
two components also works reasonably well. Donzelli et al. (2007)
classify this object as an elliptical galaxy and derive $n=4.4$.

{\it 3C 206} (Fig. 11.3) --- \ \
The central region is saturated and there is a tiny companion galaxy. 
Thus the fit could be a bit ambiguous. The host is fit either with a single 
bulge ($n = 1.6$; 18.12 mag) or with a tiny bulge ($R_e = 0.9$ pix; 17.95 mag) 
and a disk ($B/T = 0.55$). Interestingly, our estimation of the bulge 
luminosity is an order of magnitude smaller (fainter) than that 
($\approx16.9$ mag) from Hamilton et al. (2008). While the reason for the 
substantial difference is unclear, Hamilton et al. reported that their 
photometry method using the PSF-subtracted image systematically overestimates 
the host luminosity. In this particular object, companion galaxies might 
contribute to the discrepancy.

{\it 3C 219} (Fig. 11.3) --- \ \
There are more than 20 companion galaxies in the image. There seems to be 
a weak sign of interaction with companion galaxies. The host is well fit with
a single bulge but an additional extended component slightly helps. The bulge 
magnitude would be 0.3 mag fainter if we use two components model. McLure 
et al. (1999) and Donzelli et al. (2007) classified the host galaxy as an 
elliptical galaxy.

{\it 3C 223} (Fig. 11.4) --- \ \
The host is fit well with a single bulge ($n=3.22$; 16.78 mag). But, 
alternatively the fits with a single ($n=4$) or two component models also
reasonably work well. Thus the bulge luminosity can range 16.7 $-$ 17.7 mag.
Donzelli et al. (2007) fit the host with a single component with $n=3.5$),
which is in good agreement with our estimation.

{\it 3C 227} (Fig. 11.4) --- \ \
The image is saturated in the center. We fit the host with a single bulge 
($n=4$). The host is also classified as an elliptical ($n\approx4.3$) 
in Donzelli et al. (2007).

{\it 3C 277.1} (Fig. 11.4) --- \ \
The host is associated with an interacting galaxy and is fit with 
a bulge and a tidal tail. Out measurement of the host luminosity is well
matched with that of Hamilton et al. (2002).  

{\it 3C 287.1} (Fig. 11.4) --- \ \
The host is fit with a single bulge. There are several companion objects. 
But we find no clear sign of interaction. While Donzelli et al. (2007) 
claimed that this object has disk-like structure, Dunlop et al. (2003) fit
the host with a single component. It is possible that Donzelli et al.
misclassified the host due to the bright nucleus. The bulge luminosity from 
Dunlop et al. (2003) is in good agreement with that from our method.  


{\it 3C 303} (Fig. 11.5) --- \ \
While the host galaxy is well fit with a single bulge, the host model with two 
components works reasonably well. Donzelli et al. (2007) fit the host with 
a single component ($n=3.9$).

{\it 3C 332} (Fig. 11.5) --- \ \
The host galaxy is well-fit with a single bulge. Donzelli et al. (2007) claim
that the host is fit with a bulge and a disk. As in 3C 287.1, the fit from
Donzelli et al. can be affected by the bright nucleus. In addition, Floyd et al.
(2008) show that the \ser\ index of the host ranges from 1.7 to 4.5 depending on
the fitting  method (1D or 2D). We cannot completely rule out the existence of
the disk in this object.

{\it 3C 382} (Fig. 11.5) --- \ \
The image is heavily saturated at the center. The host galaxy only requires a
single bulge component to fit. Floyd et al. (2008) found that the host can be
fit with high \ser\ index ($n\approx5.8$).  

{\it 3C 390.3} (Fig. 11.5) --- \ \
There is a marginal PSF mismatch for this object. While a single bulge (17.09
mag) is sufficient for representing the host, the fit with two components
($m_{\rm bul} =17.45$ mag) works reasonable well, also.  The host is well-fit
with a \ser\ profile with $n\approx3.1$ in the \hst\ IR image (Floyd et al.
2008). Bennert et al. (2010) also described the host as an elliptical galaxy
(but see Bentz et al. 2009).

{\it 3C 445} (Fig. 11.6) --- \ \
The image is saturated in the center, thus the fit might be ambiguous.
The host is fit with a single bulge. Floyd et al. (2008) also claim that
the host might be an elliptical galaxy ($n=3.74$).

{\it ESO 031$-$G8} (Fig. 11.6) --- \ \
The image is saturated in the center. 
The fit is done with a bulge, a bar, and a disk with spiral arms. 
We use the hybrid method for the disk. 
The host is classified as a barred spiral galaxy from previous studies 
(Crenshaw et al. 2003; Malkan et al. 1998).

{\it ESO 362$-$G18} (Fig. 11.6) --- \ \
The image is saturated in the center. The fit is done with a mosaic image.  The
host galaxy is modeled with a pseudo-bulge ($n=1.07$) component and a disk
with spiral arms.  There might be a companion galaxy but it is not certain
whether the two galaxies are in merging process.

{\it ESO 383$-$G35} (Fig. 11.6) --- \ \
The image is saturated in the center in this WFPC2 mosaic image. The host galaxy
is classified as a S0 galaxy in Malkan et al.  (1998), and we fit it with a
bulge and a disk component.

{\it FBQS 125807.4+232921} (Fig. 11.7) --- \ \
The image is slightly saturated in the center. The host is fit with a single
bulge. Our measurement of the host luminosity ($M_R=\approx-22.1$) is in good 
agreement with that ($M_R=\approx-22.0$, converted to our assumed cosmology 
with $R-I=0.7$) from Marble et al. (2003). 

{\it Fairall 9} (Fig. 11.7) --- \ \
The host is fit with a bulge and a disk ($B/T=0.64$). Based on the ground-based
image, Kotilainen et al. (1993) obtain $B/T=0.83$. There is a good agreement
between our measurement and that ($B/T=0.52$) from Bentz et al. (2009) derived
with the same \hst\ image. 

{\it Fairall 1146} (Fig. 11.7) --- \ \
The WFC2 image is saturated in the center and the nucleus is very faint compared
to the host galaxy. The host appears to be a pseudo-bulge with \ser\ index of
$n=1.29$, a bar, and a disk with spiral arms.  While Crenshaw et al. (2003)
and Malkan et al.  (1998) conclude that the host is an unbarred spiral galaxy,
we find that a bar is required for a better fit.


{\it $[$HB89$]$ 0244+194} (Fig. 11.7) --- \ \
The host galaxy is well-fit with a single component, although the residual shows
an extended feature. McLure et al. (1999) identify the host as an elliptical
galaxy. 

{\it $[$HB89$]$ 0316$-$346} (Fig. 1.8) --- \ \
While the host galaxy is fit with a bulge and a highly disturbed component, the
galaxy is likely to be an early-type ($B/T=0.63$ and $n$(bulge)$=3.0$).
Based on the same \hst\ image, Bahcall et al. (1997) conclude that
$B/T=0.61-0.68$, consistent with our measurement. 

{\it $[$HB89$]$ 0759+651} (Fig. 11.8) --- \ \
The image is saturated in the center, thus the fit might be ambiguous.  Tidal
features are seen. The host galaxy  is fit with a bulge and two extended
components with azimuthal Fourier modes. The light of the host galaxy appears 
to be dominated by the bulge ($B/T=0.6$). Boyce et al. (1998) also claimed 
that the host might be a disk galaxy. 

{\it $[$HB89$]$ 1150+497} (Fig. 11.8) --- \ \
The host galaxy is fit with a bulge component, and a faint fine structure
($21.9$ mag; $B/T=0.98$). The residual image shows significant PSF mismatch,
thus the nucleus magnitude could be significantly underestimated. Floyd et al.
(2004) also described the host as an elliptical galaxy.


{\it $[$HB89$]$ 1321+058} (Fig. 11.8) --- \ \
The host galaxy is highly disturbed, thus the fit might be ambiguous.
The host is fit with a pseudo-bulge ($n=1.27$), and two disks with 
azimuthal Fourier modes. 

{\it $[$HB89$]$ 1402+436} (Fig. 11.9) --- \ \
The image is saturated in the center. We fit the host galaxy with a bulge and
extended disturbed components. But it appears to be a bulge-dominated galaxy
merging with a smaller galaxy (see also Hutchings \& Neff 1992).

{\it $[$HB89$]$ 1549+203} (Fig. 11.9) --- \ \
Using a ground-based image, Hutchings \& Neff (1992) find that the host
galaxy appears to a barred spiral. However, the \hst\ image shows that the
bar-like structure extending N-S is more likely to be spiral arms.  In contrast,
Dunlop et al. (2003) conclude that the fit is consistent with an elliptical
galaxy. But the discrepancy in the bulge luminosity estimates is relatively
small ($\approx0.4$ mag).  From our fit, we find that the host might have a
compact nucleus or tiny bulge ($R_e = 0.19$ pixels), and a disk with spiral 
arms.

{\it $[$HB89$]$ 1635+119} (Fig. 11.9) --- \ \
The host galaxy is well fit with a single bulge-like component, which is
consistent with findings from McLure et al. (1999).

{\it $[$HB89$]$ 1821+643} (Fig. 11.9) --- \ \
The contamination from surrounding objects is severe, thus we fit them ($\sim
20$ objects) simultaneously. A single bulge-like component suffices for fitting
the host galaxy, consistent with the result from Floyd et al. (2004).

{\it $[$HB89$]$ 2141+175} (Fig. 11.10) --- \ \
While the host galaxy is fit with a bulge component and an elongated disk-like
feature, it still appears to be dominated by the bulge ($B/T=0.83$). Previous
studies also conclude that it is an elliptical galaxy, although the image
reveals extended filaments (e.g., McLure et al. 1999; Hutchings et al. 1994). 

{\it $[$HB89$]$ 2201+315} (Fig. 11.10) --- \ \
The fit favors the host galaxy being E-type. Using the $H$-band image
with adaptive optics (AO), Guyon et al. (2006) also conclude that the host is
well fit with a single bulge component.

{\it $[$HB89$]$ 2215$-$037} (Fig. 11.10) --- \ \
The host appears to be an elliptical galaxy. Our estimate of the bulge 
luminosity is well matched to that from Dunlop et al. (2003). 

{\it $[$HB89$]$ 2344+184} (Fig. 11.10) --- \ \
While the host galaxy is well-fit using a bulge and a disk-like components
(see also Taylor et al. 1996; McLure et al. 1999), Hutchings \& Neff (1992) 
claim that it can be a barred spiral galaxy. 

{\it HE 0054$-$2239} (Fig. 11.11) --- \ \
The host galaxy is fit with a tiny bulge-like component ($R_e = 0$\farcs06) and
a disk with spiral arms.  Ohta et al.  (2007) find that the host is a barred
spiral galaxy, based on a ground-based image with 0\farcs5$-$1\farcs0 seeing.
However, we find no clear evidence for a bar.

{\it HE 0306$-$3301} (Fig. 11.11) --- \ \
The host is fit with a pseudo bulge ($n=0.24$) and a disk with spiral arms. 

{\it HE 0354$-$5500} (Fig. 11.11) --- \ \
This object appears to reside in a merging system. The host galaxy is highly 
disturbed and might have a disk with spiral arms. 

{\it HE 1043$-$1346} (Fig. 11.11) --- \ \
The host galaxy is fit with a bulge ($n=4$) component, an inner disk (with
spiral arms) and an additional outer disk with spiral arms. The bulge luminosity
can range from 18.6 to 19.3 mag depending on the model profile for the bulge. We
tried to add a bar but it appears to be unnecessary.

{\it HE 1110$-$1910} (Fig. 11.12) --- \ \
Whereas the best fit of the host galaxy using a single bulge component yields a
bulge luminosity of $17.52$ mag, a fit with two components (bulge+disk) finds
that the bulge luminosity can either either 18.35 ($n=4$) or 19.0 ($n=1.3$) mag.
Jahnke et al. (2004) conclude that the host is a slightly disturbed elliptical
galaxy, using a ground-based image.

{\it HE 1228$-$1637} (Fig. 11.12) --- \ \
This AGN appears to be a merging system. While we fit the companion objects
with four S\'{e}rsic components along with Fourier azimuthal modes, the host
galaxy requires only a single \ser\ component to model. With ground-based 
images, Jahnke et al. (2004) also classify this as an elliptical galaxy but 
they find no clear sign of interaction in the $\sim1$\farcs5 seeing image.

{\it HE 1252+0200} (Fig. 11.12) --- \ \
The residual image shows a significant PSF mismatch at the core.  Aperture
photometry of the host galaxy might be uncertain due to the presence of
companions. The best fit requires only a single \ser\ component to describe the
host galaxy, consistent with findings of Floyd et al. (2004).


{\it HE 1434$-$1600} (Fig. 11.12) --- \ \
The host galaxy is well fit with a single bulge component.  Jahnke et al. (2004)
also conclude that the host is an elliptical galaxy based on the ground-base
image.  The \hst\ image reveals shell-like features which could indicate a 
recent minor merger event.

{\it HE 1503+0228} (Fig. 11.13) --- \ \
The host galaxy is fit with a bulge ($n \sim 0.5$) and a disk with spiral arms.
Using dynamical information obtained from the spatially resolved spectroscopy, 
Courbin et al. (2002) conclude that the host galaxy is a normal spiral.

{\it HE 2345$-$2906} (Fig. 11.13) --- \ \
The host is fit with a very small bulge ($r_e\sim0\farcs06$), a bar, an inner
ring (with the hybrid mode), and a disk with spiral arms. Letawe et al. (2007;
2008) also classify the host as a barred spiral galaxy. 

{\it IC 4329A} (Fig. 11.13) --- \ \
We masked out the dust lane along the disk and fit the host with a 
pseudo-bulge ($n=1.02$) and a disk. While previous studies claim that the
host is a spiral galaxy (Malkan et al. 1998; Crenshaw et al. 2003), 
Kotilainen \& Ward (1994) fail to detect the bulge component in 
ground-based optical images. 

{\it Mrk 10} (Fig. 11.13) --- \ \
We use a mosaic image of WFPC2 for the fit. The image is slightly saturated.
The host galaxy has fitted components that resemble a pseudo-bulge ($n=1.31$)
and a disk with spiral arms.  Previous studies based on the same \hst\ image
also conclude that the host is an unbarred spiral galaxy (Malkan et al. 1998;
Crenshaw et al.  2003).

{\it Mrk 40} (Fig. 11.14) --- \ \
The WFPC2 image is saturated in the center.  While the image does not cover the
entire galaxy, the host galaxy  appears to have a bulge ($n=4.5$) and a disk
with tidal tails. This object is almost certainly an interacting system. 

{\it Mrk 42} (Fig. 11.14) --- \ \
The WFPC2 image is saturated in the center. The host galaxy appears to possess a
pseudo-bulge, a bar, and a disk with spiral arms, modeled with the hybrid
method. An inner ring is apparent in the residual image.  Jiang et al.  (2011)
and Orban de Xivry et al.  (2011) also fit the host with a bulge, a bar, and a
disk components. The bulge to total ratio ($B/T sim 0.12-0.18$) found by  Jiang
et al.  and Orban de Xivry et al. is in good agreement with our estimation
($B/T=0.16$).

{\it Mrk 50} (Fig. 11.14) --- \ \
The WFPC2 image is saturated in the center. The host galaxy appears to have a
bulge and a faint disk ($B/T=0.89$), and might be a S0 galaxy (Malkan et al.
1998).

{\it Mrk 79} (Fig. 11.14) --- \ \
The outer part of the galaxy is outside the detector field of view, thus the
total brightness of the host might be an upper limit.  The host galaxy is fit
with components that resemble a bulge and a disturbed bar and a disk.  Dust
lanes are prominent. Malkan et al.  (1998) and Crenshaw et al. (2003) classify
the host as a barred disk galaxy.  Bennert et al. (2010) find the ratio of
$B/T ~ 0.41$ to be a bit larger than our estimate ($B/T=0.16$). 

{\it Mrk 231} (Fig. 11.15) --- \ \
The WFPC2 mosaic image is saturated in the center.  Because the host galaxy is
highly disturbed ($a_1=0.51$), it is fit with a pseudo-bulge and two disturbed
disks. 


{\it Mrk 352} (Fig. 11.15) --- \ \
The WFPC2 mosaic image is saturated.  The host galaxy is fit with a bulge and a
disk components. The $B/T=0.41$ found in this study matches well with that
($B/T=0.44$) derived in Orban de Xivry et al.  (2011). However, previous studies
(Malkan et al. 1998 and Crenshaw et al. 2003) based on visual inspection conclude
that the host is an elliptical galaxy.

{\it Mrk 359} (Fig. 11.15) --- \ \
The WFPC2 mosaic image is saturated in the center.  We fit the host galaxy with
3 components that resemble pseudo-bulge ($n=0.95$), a bar, and a disk with
spiral arms. Orban de Xivry et al. (2011) derive $B/T=0.11$ which is in
good agreement with our measurement. 

{\it Mrk 372} (Fig. 11.15) --- \ \
The WFPC2 mosaic image is saturated in the center.  The host galaxy appears to
have a bulge and a disk with spiral arms which we fit with a two component model.
There is no sign of a bar in the image (see also Malkan et al. 1998 and Crenshaw
et al. 2003).

{\it Mrk 382} (Fig. 11.16) --- \ \
The WFPC2 mosaic image is saturated in the center.  The host galaxy is fit with
a bulge component ($n=2.06$), a bar, and a disk with spiral arms. Our estimate
on the $B/T$ ($\sim0.2$) ratio is slightly higher than that (0.31) from Orban de
Xivry et al. (2011). This might be because we fit for the bar component
explicitly in the decomposition.

{\it Mrk 423} (Fig. 11.16) --- \ \
The host galaxy in the WFPC2 mosaic image appears to be associated with a
interacting galaxy.  The host is fit with a pseudo-bulge component, a disk with
spiral arms, and tidal tails.  As with Mrk 382, we slightly underestimate $B/T$
($\sim0.24$) compared to Orban de Xivry et al. (2011; $B/T\sim0.42$). 

{\it Mrk 471} (Fig. 11.16) --- \ \
We fit the WFPC2 mosaic image of the host galaxy using a pseudo-bulge ($n=1.3$)
component and a disk with spiral arms. Although Malkan et al. (1998; and
Crenshaw et al. 2003) classify the host as a barred galaxy, the best fit
indicates that an additional bar is unnecessary to model the host.

{\it Mrk 493} (Fig. 11.16) --- \ \
The WFPC2 mosaic image is saturated at the center.  The host galaxy appears to
have  tiny pseudo-bulge ($n=0.25$), a bar, and a disk with spiral arms. The bar
and the disk are fit by the hybrid model.  Orban de Xivry et al. (2011) also
find the host galaxy to contain a pseudo-bulge ($n\sim0.75$). 

{\it Mrk 509} (Fig. 11.17) --- \ \
From ground-based NIR data, the host galaxy appears as an elliptical (e.g.,
Fischer et al.  2006). However, the \hst\ image shows that it s not a typical
elliptical galaxy as it has spiral structures, dusty features, and a ring.  We
model it with a small bulge, a disk, and a ring-like structure (outer ring); the
latter is modeled with the hybrid method. 

{\it Mrk 516} (Fig. 11.17) --- \ \
The host galaxy is fit with a bulge component and a disk with spiral arms in
this WFPC2 mosaic image. From the double nuclei (separation of $\sim0\farcs2$)
it appears that the host is merging system (see also Deo et al. 2006), however
there are no other clear signs of interaction in the WFPC2 image.  Smirnova et
al.  (2010) claim that a wide-field deep image shows shell-like structures,
which might be formed during minor mergers. 

{\it Mrk 543} (Fig. 11.17) --- \ \
The WFPC2 mosaic image is saturated at the center. The host galaxy has a
pseudo-bulge-like ($n=1.51$) component and a disk with spiral arms. Although no
clear signs of interaction are present in the \hst\ image, shells are seen in a
deep ground-based image from Smirnova et al. (2010).

{\it Mrk 590} (Fig. 11.17) --- \ \
The host galaxy has a pseudo-bulge ($n=1.04$) and a disk ($B/T\sim0.26$).  There
is also a faint inner spiral arm-like structure in the residual image. The fit
with a ground-based NIR image (Vika et al. 2012) yields a larger $B/T$
($\sim0.48$) than our results based on an optical image, perhaps due to the
relative prominence of older stellar populations in the NIR. 

{\it Mrk 595} (Fig. 11.18) --- \ \
Since the WFPC2 mosaic image is highly saturated in the center, the fit
parameters may be compromised. The host galaxy is modeled with a bulge and a disk
($B/T\sim0.8$) component. Our result is consistent with Orban de Xivry et al.
(2011; $B/T\sim0.66$), within the errors. 

{\it Mrk 609} (Fig. 11.18) --- \ \
The WFPC2 mosaic image is saturated in the center.  The AGN host appears to have a
pseudo-bulge ($n=0.37$), an inner disk, and another outer disk. We adopt the
hybrid method to model the inner disk.  Our fitting result ($B/T\sim0.11$) is
notably different from that in Orban de Xivry et al. (2011; $B/T\sim0.66$). This
is because we use an inner disk to model the host.  If instead we assume the
inner disk is part of the bulge, $B/T$ becomes dramatically larger ($\sim0.77$)
comparable to that from Orban de Xivry et al.

{\it Mrk 699} (Fig. 11.18) --- \ \
Since cosmic rays are not removed perfectly we mask out pixels affected by
cosmic rays. The WFPC2 mosaic image is saturated in the center.  The host galaxy
is fit with a bulge ($n=2.26$), and a disk component.  Alternatively, a fit with
a classical bulge ($n=4$) and a disk finds a bulge component that is brighter by
0.4 mag. We conclude that the host can be classified as an S0 galaxy.

{\it Mrk 704} (Fig. 11.18) --- \ \
The WFPC2 mosaic image is saturated in the center.  The AGN host is fit with
components: a bulge, a bar, and a disk with spiral arms.  For the bar and disk,
we use the hybrid method. Our result is broadly consistent with Orban de Xivry
et al. (2011).

{\it Mrk 766} (Fig. 11.19) --- \ \
The WFPC2 mosaic image is saturated in the center.  The host galaxy can be
modeled with a bulge, a bar, and a disk with spiral arms, components. The result
is comparable to that from Orban de Xivry et al. (2011).

{\it Mrk 871} (Fig. 11.19) --- \ \
The WFPC2 mosaic image is saturated in the center.  The best fit finds that the
host galaxy has a pseudo-bulge ($n=1.8$), a bar and a disk with spiral arms.
Orban de Xivry et al. (2011) found $B/T\sim0.11$ and $n_{\rm bul}\sim1.3$, which
is in good agreement with our measurements ($B/T\sim0.13$). 

{\it Mrk 896} (Fig. 11.19) --- \ \
The WFPC2 mosaic image is saturated in the center.  The AGN host is fit with a
pseudo-bulge ($n=1.43$), a bar, two (outer and inner) disks with spiral arms
($B/T\sim0.11$), components . Orbad de Xivry et al. (2011) fit the host with a
bulge ($n=2.06$) and a disk ($B/T\sim0.15$).

{\it Mrk 1040} (Fig. 11.19) --- \ \
We use a mosaic image of WFPC2 for the fit, although it appears to cover only a
part of the host galaxy. The image is saturated at the center. We fit the AGN host
with a pseudo-bulge ($n=1.19$) and a disk with spiral arms ($B/T\sim0.15$).
Although a bar is not clearly seen in the \hst\ image (see also Malkan et al.
1998 and Crenshaw et al. 2003), there is gas-dynamical evidence for a bar (Amram
et al. 1992).

{\it Mrk 1044} (Fig. 11.20) --- \ \
The mosaic image of WFPC2 is saturated in the center. The fit finds a small
pseudo-bulge ($n=0.81$), an inner disk, and an outer disk with spiral arms
($B/T\sim0.18$). The results are generally in good agreement with Orbad de Xivry
et al. (2011; $n_{\rm bul}\sim1.45$ and $B/T\sim0.30$). 

{\it Mrk 1048} (Fig. 11.20) --- \ \
The host galaxy appears to be a single bulge from the fit. A companion galaxy
also appears to be a spheroid. There is an extended ring-like structure which is
not easy to see in Figure 11.20. 

{\it Mrk 1095} (Fig. 11.20) --- \ \
There are prominent spiral arms in the image which are not fit.  The best fit
finds that the AGN host can be modeled with a bulge ($m_{\rm bul}\sim16.1$ mag)
component and an exponential disk without spiral arms ($B/T\sim0.25$). While
previous studies claim that the host certainly contains a disk, the $B/T$ ratio
appears to be somewhat uncertain and ranges from 0.35 (Bennert et al. 2010) to
0.49 (Bentz et al. 2009).  

{\it Mrk 1126} (Fig. 11.20) --- \ \
The image is slightly saturated. We use a mosaic image of WFPC2 for the fit.
The AGN host galaxy is fit with a pseudo-bulge and two disks with spiral arms.
The inner disk component has $n=0.5$, which suggests that it can be a ``lens"
(Kormendy 1979). The fitting result is consistent with Orbad de Xivry et al.
(2011; $n_{\rm bul}\sim1.86$ and $B/T\sim0.15$)

{\it Mrk 1330} (Fig. 11.21) --- \ \ 
This SBb galaxy (Sandage 1961) contains dust lanes which make the fitting result
uncertain. The fit uses a pseudo-bulge ($n=0.17$) and a disk component with
spiral arms.  Based on wide-field ground-based images, Kormendy et al.  (2006)
also find the bulge of the host to be an pseudo-bulge.

{\it Mrk 1469} (Fig. 11.21) --- \ \
The WFPC2 image is saturated in the center.  The AGN host is fit with a
pseudo-bulge and a disk, but the result is affected by dust lanes. We omit
Fourier modes in the fit which are even more affected by dust lanes. 

{\it MS 0007.1$-$0231} (Fig. 11.21) --- \ \
The fit is affected by a dust lane which may be perturbing the profile
of the galaxy bulge.


{\it MS 0048.8+2907} (Fig. 11.21) --- \ \
The host galaxy is a very complicated object, and we fit it with a bulge and a
disk with spiral arms.  There are tiny inner arms, large outer arms, and a
highly extended wing which is not shown in this image.  The complicated
structures in the residual image suggest that the bulge parameters may be
uncertain.  We also tried to model the host galaxy with an additional bar but
failed to find an optimal solution.  Contrary to our results, Schade et al.
(2000) claim that the host is a bulge-dominated system ($B/T\sim0.87$) based on
both ground-based and \hst\ images.  However, visual inspection would suggest
that the bulge size and its contribution to the total luminosity should both be
modest to small.

{\it MS 0111.9$-$0132} (Fig. 11.22) --- \ \
The AGN host galaxy seems to have a small bulge ($R_e = 2.6$ pixels) along with
some weak signs of interaction.  Like with MS 0048.8+2907 there is some
disagreement between our model and that from Schade et al. (2000), where they
find the host galaxy to have a single component bulge.  Although we do find that
the fit with a single bulge might work, the residual image from such a fit
clearly shows a more extended structure, possibly indicating a disk in the host.  

{\it MS 0135.4+0256} (Fig. 11.22) --- \ \
While we find that the host galaxy may have a small bulge ($R_e = 0$\farcs23), a
bar, and a disk ($B/T\sim0.13$), Schade et al. (2000) fit the host with a larger
and brighter bulge ($R_e = 1$\farcs66 and $B/T\sim0.54$).

{\it MS 0144.2$-$0055} (Fig. 11.22) --- \ \
The host galaxy appears to have a small bulge ($R_e = $0\farcs32) with a disk.
This is in broad agreement with the fitting results from Schade et al. (2000). 

{\it MS 0321.5$-$6657} (Fig. 11.22) --- \ \
We fit the host galaxy with a small pseudo-bulge ($R_e = $0\farcs18, $n=1.74$),
a bar, and a disk. Our estimate on the $B/T$ ratio ($\sim0.17$) is substantially
smaller than that ($B/T\sim0.56$) from Schade et al. (2000). 

{\it MS 0330.8+0606} (Fig. 11.23) --- \ \
The host galaxy requires a bulge, a bar, and a disk with spiral arms to fit well. 
The estimated $B/T$($\sim0.12$) is slightly smaller than that (0.32) from 
Schade et al. (2000), possibly due to the bar component. 


{\it MS 0412.4$-$0802} (Fig. 11.23) --- \ \ 
Although a single bulge (14.88 mag) seems to suffice to represent the
host galaxy, a fit with a bulge (16.26 mag) and disk (15.54 mag) also works
reasonably well. Schade et al. (2000) classify the host galaxy as an elliptical
galaxy.  

{\it MS 0444.9$-$1000} (Fig. 11.23) --- \ \
The best fit with a bulge, a bar, and a disk finds that the bulge is as bright
as 17.81 mag. A fit without a disk is also acceptable and yields a bulge as
bright as 17.76 mag. Our measurements is consistent with Schade et al. (2000) to
within the measurement uncertainties. 

{\it MS 0457.9+0141} (Fig. 11.23) --- \ \
The host galaxy is fit with a classical bulge ($n=4$, $m_{\rm bul}=16.99$ mag) and 
an exponential disk. But a bulge with $n=5.58$ and $m_{\rm bul}=16.68$ mag
also works very well. Schade et al. (2000) conclude that the host can be 
described by the single bulge ($m_{\rm bul}\sim15.9$ mag). 

{\it MS 0459.5+0327} (Fig. 11.24) --- \ \
The WFPC2 mosaic image is saturated in the center.  The host galaxy has
a bulge, a disk, and a faint tidal tail.  We are unable to fit the tidal tail
with Fourier modes. Crenshaw et al. (2003) classify the host as an elliptical.



{\it MS 0719.9+7100} (Fig. 11.24) --- \ \
While the best fit yields a single bulge for the host galaxy, a fit with two
components also works well. The fitting result is in good agreement with Schade
et al.  (2000).


{\it MS 0754.6+3928} (Fig. 11.24) --- \ \
Either a fit with a single bulge ($15.94$ mag) or a fit with a bulge ($16.18$
mag) + a disk works equally well.  We find a high discrepancy in the host galaxy
magnitude between image fitting and aperture photometry techniques, possibly due
to a significant PSF mismatch.  Schade et al. (2000) argue that the host is an
elliptical galaxy.  However, they estimate the bulge brightness to be $\sim0.7$
mag brighter than our measurement. 



{\it MS 0801.9+2129} (Fig. 11.24) --- \ \
The image is slightly saturated. We mask out pixels where the values are greater
than 1000 ADUs. The bulge magnitude $m_{\rm bul}$ can range from 16.7 mag to
17.1 mag depending on the model. Overall, the best fitting parameters are in
broad agreement with those from Schade et al. (2000). 



{\it MS 0842.7$-$0720} (Fig. 11.25) --- \ \
The host galaxy can be described by a bulge, a disk and a bar.  However, a fit
without a bar appears to be acceptable. The bulge magnitude can be either 17.37
mag with, or 17.50 mag without, a bar. The best fit for the host is broadly
consistent with findings from Schade et al. (2000).

{\it MS 0844.9+1836} (Fig. 11.25) --- \ \
The best fit finds that the host galaxy has a bulge ($n=2.81$)-to-disk ratio as
high as $0.51$. While the bulge luminosity is in good agreement with Schade et
al. (2000), they find a slightly larger $B/T$ ($0.71$) for the host galaxy.


{\it MS 0849.5+0805} (Fig. 11.26) --- \ \
Due to substantial PSF mismatch, it is nontrivial to determine the best fit. 
We choose the double component (bulge+disk) fit as the best model, judging by
the residual image, but a fit with a single component is also acceptable.
Schade et al. (2000) classify the host as an elliptical galaxy with a bulge
magnitude of $14.8$ mag. 

{\it MS 0904.4$-$1505} (Fig. 11.26) --- \ \
There are some weak signs of interaction but the image is too shallow to confirm
it.  The host galaxy shows some very complex features in the center and the best
fit finds the host to require a pseudo-bulge ($n=1.23$), a bar, and a disk,
components.  Schade et al.  (2000) find a substantially larger $B/T\sim0.91$,
which might be due to the bar component being fitted as part of the bulge in
their analysis.  

{\it MS 0905.6$-$0817} (Fig. 11.26) --- \ \
The best fit finds that the host galaxy has a bulge and a slightly disturbed disk.
Our results are in good agreement with the analysis of Schade et al. (2000).

{\it MS 0942.8+0950} (Fig. 11.26) --- \ \
The host galaxy is fit with a bulge, a bar, a disk with spiral arms, yielding 
$m_{\rm bul} \sim 15.6$ mag. But the bulge can be as bright as 
$\sim14.9$ mag for a fit without a bar. The fitting results are consistent 
with those in Schade et al. (2000).



{\it MS 1059.0+7302} (Fig. 11.27) --- \ \
The host galaxy can be fit with a bulge and a disk component. There are weak
signs of an inner ring in the residual. Schade et al. (2000) find a larger $B/T$
($\sim0.74$) than our estimate ($B/T\sim0.25$) and a larger $R_e$ for the
bulge. This discrepancy could be related to the existence of the inner ring. 





{\it MS 1136.5+3413} (Fig. 11.27) --- \ \
We fit the host galaxy with a bulge component, a bar and a disk with spiral
arms. We use the hybrid technique to better model/merge the bar and the disk
components. The fitting results are in good agreement with those in Schade et
al. (2000).

{\it MS 1139.7+1040} (Fig. 11.27) --- \ \
Although the residual image shows weak signs of spiral arms, it is too faint to
be modeled. The host galaxy appears to have a pseudo-bulge ($n=1.33$, $M_{\rm
bul}\sim18.8$ mag) and a disk component. Schade et al. (2000) find a brighter
bulge ($\sim 18$ mag), thus a slightly larger $B/T$ ($\sim0.21$).   

{\it MS 1143.5$-$0411} (Fig. 11.27) --- \ \
The host galaxy has a bulge ($\sim 17.9$ mag), a bar, and a disk with spiral
arms ($B/T \sim 0.15$). The hybrid model is used to merge the bar and disk
components.  The bulge is tiny ($R_e \sim 2.8$ pixels). Like MS 1139.7+1040, the
fitting results from Schade et al. (2000) yield a slightly brighter bulge
brighter ($\sim17.2$ mag) and slightly larger $B/T \sim0.32$ compared to our
results.  

{\it MS 1158.6$-$0323} (Fig. 11.28) --- \ \
A ring-like structure and tidal features are present in the residual image.  The
bulge magnitude can range from 15.95 to 16.86 mag depending on the model.
Schade et al. (2000) obtain a slightly smaller bulge magnitude ($\sim 15.7$
mag).


{\it MS 1205.7+6427} (Fig. 11.28) --- \ \
The host galaxy has a bulge component, and a disturbed disk fitted using two
additional components for the disturbed wings. While we find that the bulge has
$R_e \sim 0.3$\asec, Schade et al. (2000) conclude that $R_e$ of the bulge is
$\sim 2$\asec, which might be the main reason for there being a discrepancy in
the bulge magnitude.

{\it MS 1214.3+3811} (Fig. 11.28) --- \ \
The host galaxy appears to have a tiny bulge ($R_e = $0\farcs15 and $n \sim
0.6$).  The overall fitting results appears to be consistent with those from
Schade et al. (2000).

{\it MS 1217.0+0700} (Fig. 11.28) --- \ \
The host galaxy has a tiny bulge and a disk with spiral arms.
An additional bar is otherwise unhelpful to model the host. Like MS 1214.3+3811, 
our analysis is in good agreement with the study by Schade et al. (2000).

{\it MS 1219.6+7535} (Fig. 11.29) --- \ \
The host galaxy appears to have a pseudo-bulge (16.49 mag and $n=1.3$) while,
the fit with a classical bulge ($n=4$) yields $m_{\rm bul} = 15.92$ mag.  The
lopsidedness parameter, $a_1$, is slightly large (0.15) but this might be caused
by the presence of a companion. In contrast to our findings, Schade et al.
(2000) find a substantially brighter bulge ($\sim 15.1$ mag). 

{\it MS 1220.9+1601} (Fig. 11.29) --- \ \
The bulge magnitude can range from 18.2 ($n=4$) to 18.64 ($n=1.92$) depending on
the adopted \ser\ index. Like MS 1205.7+6427, Schade et al.  (2000) claim that
the size of the bulge ($R_e\sim2.4$\asec) is much larger than our measurement
($R_e\sim0.2$\asec). Also, they find the bulge to be 2.5 mag brighter than our
optimal fit. 

{\it MS 1232.4+1550} (Fig. 11.29) --- \ \
While we fit the host galaxy with a small bulge ($\sim 16.7$ mag), a bar, and
disk with spiral arms, the fit without a bulge appears to work reasonably well.
Schade et al. (2000) find a slightly brighter bulge ($\sim 16.1$ mag) from their
best fit without a bar component. 


{\it MS 1239.2+3219} (Fig. 11.29) --- \ \
There is an inner ring-like structure. The host galaxy has a bulge and a ring
(modeled with the hybrid model), and a disk. The overall result is consistent
with that from Schade et al. (2000). 


{\it MS 1306.1$-$0115} (Fig. 11.30) --- \ \
Although the fit with a bulge ($m{\rm bul} \sim 17.0$ mag) and a disk is deemed
as the best solution, a fit with a single bulge ($m{\rm bul} \sim 16.9$ mag) is
also acceptable.  Schade et al. (2000) conclude that the host can be described
with a single bulge. 

{\it MS 1322.3+2925} (Fig. 11.30) --- \ \
The fitting result is uncertain because the target object is located at the edge
of the image. The host is fit with a single bulge component, which is consistent
with the study by Schade et al. (2000).   

{\it MS 1327.4+3209} (Fig. 11.30) --- \ \
The host galaxy is fit with a bulge, a disk, and an outer ring, components. For
the disk and the ring, we adopt the hybrid model. 

{\it MS 1333.9+5500} (Fig. 11.30) --- \ \
The host galaxy appears to be an edge-on galaxy with a warped disk, and is fit
with a bulge and a disk. The best fitting parameters are in good agreement with
those in Schade et al. (2000). 

{\it MS 1334.6+0351} (Fig. 11.31) --- \ \
The host galaxy has both a bulge ($\sim 17.1$ mag) and a disk. However, the fit 
with a single bulge ($\sim 16.2$ mag) also works reasonably well. 
Schade et al. (2000) conclude that the host is an elliptical galaxy.

{\it MS 1351.6+4005} (Fig. 11.31) --- \ \
The bulge brightness can range from $\sim16.0$ to $\sim16.7$ mag depending on 
the adopted model. A ring-like structure is seen in the residual image. Schade
et al. (2000) find $\sim0.79$ of $B/T$, which is considerably larger than our 
measurement ($B/T\sim0.3$).  

{\it MS 1403.5+5439} (Fig. 11.31) --- \ \
While we fit the host galaxy with a bulge and a disk components, the residuals
show a ring-like or spiral structure in the central region. In the spectrum from
SDSS, the broad \hb\ is very weak, but the broad \hal\ is prominent. The best
fit yields the host to be a bulge-dominated system. Similarly, Schade et al.
(2000) classify the host as an elliptical galaxy. 

{\it MS 1414.0+0130} (Fig. 11.31) --- \ \
Although we fit the host galaxy with a bulge($\sim18.2 mag$, $n=4$) and a disk, 
fits with a single bulge (17.4 mag with $n=4$ or 17.7 mag with $n=2.1$) 
or a pseudo-bulge (18.4 mag) plus a disk also work well. Regardless of adopted
fitting models, the host appears to be a bulge-dominated system. Like 
MS 1403.5+5439, Schade et al. (2000) conclude that the host is an elliptical 
galaxy. 

{\it MS 1420.1+2956} (Fig. 11.32) --- \ \
The host galaxy is decomposed of a small bulge ($R_e \sim $0\farcs2), a bar, and a 
disk with spiral arms. The bar and disk are modeled with the hybrid method.

{\it MS 1455.7+2121} (Fig. 11.32) --- \ \
The host galaxy has a bulge ($\sim 17.35$ mag) and a disk with spiral arms. 
However, based on the \hst\ and ground-based images, Schade et al. (2000) argue
that the host is an elliptical galaxy.  

{\it MS 1456.4+2147} (Fig. 11.32) --- \ \
A single bulge ($\sim 16.1$ mag) is sufficient to describe the host galaxy,
while a fit with a bulge ($16.4$ mag) and a disk also works reasonably well. 

{\it MS 1519.8$-$0633} (Fig. 11.32) --- \ \
We fit the host galaxy with a bulge, a bar, and a disk with spiral arms. The
best fit yields that the $B/T$ ratio is as large as 0.14. Schade et al. (2000)
find significantly larger $B/T$ ($\sim$ 0.56) possibly because they do not model
the bar, which may then be included as part of the bulge component.

{\it MS 1545.3+0305} (Fig. 11.33) --- \ \
The host galaxy is fit with a tiny bulge, a ring like structure (modeled with
the hybrid method) and a disk with spiral arms. Due to the complexity of the
host structure, the fit might be uncertain. Schade et al. (2000) derive a bulge
that is 1 mag brighter than our measurement.   

{\it MS 1846.5$-$7857} (Fig. 11.33) --- \ \
The object extends a bit outside of PC chip field-of-view, which makes the fit
somewhat uncertain.  The best fit finds a pseudo-bulge ($n\sim1.65$), a bar,
and a disk. Like MS 1519.8$-$0633, the decomposition without including bar
yields significantly brighter bulge component (Schade et al. 2000).

{\it MS 2039.5$-$0107} (Fig. 11.33) --- \ \
The host galaxy shows tidal features. While we fit the host with a tiny bulge 
($R_e = $0\farcs1) and a disk, Shade et al. (2000) conclude that the host is 
an elliptical galaxy. 

{\it MS 2128.3+0349} (Fig. 11.33) --- \ \
A few pixels in the center are saturated. The host galaxy is fit with a single
bulge component.  However, a two components model also works reasonably well.
Schade et al. (2000) also classify the host as an elliptical galaxy. 

{\it MS 2144.9$-$2012} (Fig. 11.34) --- \ \
The host galaxy is fit with a bulge, an inner disk and an outer disk with spiral
arms.  We tried to use a bar instead of a exponential profile for the inner
disk, but it was unhelpful. Instead, the inner disk may just be oval in shape.
Schade et al. (2000) use two components (bulge+disk) to fit the host and derive
a brighter bulge ($m_{\rm bul}\sim15.9$ mag) than our measurement ($m_{\rm
bul}\sim17.0$ mag). 

{\it MS 2159.5$-$5713} (Fig. 11.34) --- \ \
The host galaxy is fit with a pseudo-bulge ($n=1.52$) and a disk with spiral
arms.  The central part shows dust lanes. Although Schade et al. derive a
higher $B/T\sim0.95$ than our measurement, both studies agree that the host is
a bulge-dominated system. 

{\it MS 2210.2+1827} (Fig. 11.34) --- \ \
The host galaxy appears to have a small bulge ($R_e=$0\farcs16).  

{\it MS 2254.9$-$3712} (Fig. 11.34) --- \ \
The host galaxy has a pseudo-bulge and a disk with faint spiral arms. The 
fitting results are in good agreement with those from Mathur et al. (2012).

{\it MS 2340.9$-$1511} (Fig. 11.35) --- \ \
The host galaxy is fit with a single bulge (17.47 mag), while a different fit
with two components, where the bulge is as bright as 18.13 mag and $B/T = 0.47$,
works equally well. However, in both cases there are systematically positive
residuals at large radii. Mathur et al. (2012) use the same \hst\ dataset to
estimate the bulge luminosity of the host galaxy and conclude that the host has
two components (bulge+disk), with a relatively larger $B/T \sim0.57$. 

{\it MS 2348.3+3250} (Fig. 11.35) --- \ \
The best fit finds that the host galaxy is well-fit with two components with a
bulge ($\sim17.4$ mag) and a disk. But the fit using a single component bulge
($\sim16.9$ mag) is also acceptable. Schade et al. (2000) classify the host as
an elliptical galaxy. 

{\it MS 2348.6+1956} (Fig. 11.35) --- \ \
The host galaxy has a bulge, a bar, and a disk with spiral arms.

{\it NGC 1019} (Fig. 11.35) --- \ \
The WFPC2 mosaic image is saturated in the center.  The host galaxy has a bulge,
a bar, and an outer disk with spiral arms.  We use hybrid method to model the
bar and the disk. Crenshaw et al. (2003) also classify the host as a barred
spiral galaxy.  There might be a small inner ring in the center. We
simultaneously fit a neighboring galaxy using an exponential profile. 

{\it NGC 3227} (Fig. 11.36) --- \ \
The host galaxy extends outside the FOV of the chip. While the host galaxy is
fit with a disturbed pseudo-bulge and a disturbed disk, the fit might be
uncertain because of the prominent dust lanes and the small field of view of the
image, as was also pointed out in Ho \& Kim (2014).  In contrast, Kormendy \& Ho
(2013) and Virani et al. (2000) derive a lower $B/T$ ($\sim0.1$).

{\it NGC 3516} (Fig. 11.36) --- \ \
The mosaic image of WFPC2 is saturated in the center and the border lines
between chips are masked out in the fit. The host galaxy appears to have a
pseudo-bulge ($n=1.15$), a bar, and a disk. We find that $B/T=0.31$, which is
slightly lower than that ($B/T=0.52-0.86$) derived by Bentz et al. (2009) using
the \hst\ image obtained with a blue filter (F550M). Virani et al.  (2000) find
$B/T\sim0.8$ using the ground-based $R$-band image. 

{\it NGC 3783} (Fig. 11.36) --- \ \
Like NGC 3227, the host galaxy does not fully fit within the FOV of the image.
It might have a large ring-like structure just outside the FOV.  The host galaxy
is fit with a pseudo-bulge ($n=1.87$) and a disk. While we find $B/T\sim0.24$,
Bentz et al. (2009) obtain $B/T\sim0.07-0.12$. 

{\it NGC 3982} (Fig. 11.36) --- \ \
The host galaxy has a pseudo-bulge and a disk with spiral arms, while the
nucleus appears to be very faint.  An optimal PSF is generated by an IDL procedure
(Rhodes et al. 2006, and 2007; http://www.astro.caltech.edu/~rjm/acs/PSF/).
Using a ground-based $R$-band image, Virani et al. (2000) find $B/T\sim0.18$,
which is slightly larger than our measurement ($B/T\sim0.05$).

{\it NGC 4051} (Fig. 11.37) --- \ \
The fit can be highly uncertain due to the presence of a dust lane.  The host is
fit with a pseudo-bulge ($n=1.03$) and a disturbed disk.  We find $m_{\rm
bul}\sim14.3$ mag, which is in good agreement with Virani et al. (2000).
However, Bentz et al. (2009) obtain a significantly brighter bulge ($m_{\rm
bul}\sim12.8$).


{\it NGC 4151} (Fig. 11.37) --- \ \
The host galaxy is fit with a classical bulge ($\sim13.3$ mag; $n=4$) and a disk
($n=1$).  Owing to the prominent dust lanes and the small FoV of the HRC image,
the fitting results may be uncertain.  Based on fitting the same \hst\ image,
Bentz et al. (2009) find a brighter bulge ($\sim 12$ mag) with a much higher
$B/T$ ($\sim0.76$). Using the ground-based images, Virani et al. (2000) find
$B/T\sim0.44$.

{\it NGC 5548} (Fig. 11.37) --- \ \
The host galaxy is fit with a bulge component and a disk with spiral arms.
Bentz et al. (2009) use two components (inner and ordinary bulges) to model the
bulge; our bulge component is consistent with their inner bulge.  The sum of two
bulge components in Bentz et al. is 3.3 times brighter than ours. In
terms of $B/T$ measurements, the fitting result in Virani et al. (2000) appears
to be in between our and Bentz et al. (2009). 
 
{\it NGC 5940} (Fig. 11.38) --- \ \
Because the WFPC2 mosaic image is heavily saturated in the center the fitting
result might be unreliable.  The host galaxy is well-fit with a pseudo-bulge
($n=2.04$), and a disk with spiral arms.  An additional bar does not help. Using
a ground-based image, Virani et al. (2000) find a slightly larger $B/T$
($\sim0.12$) compared to our measurement ($B/T\sim0.04$).  It may be that they
overestimated the bulge brightness by underestimating the nucleus
brightness due to PSF mismatch. 


{\it NGC 6104} (Fig. 11.38) --- \ \
The fit is done with a WFPC2 mosaic image.  The host galaxy has a pseudo-bulge
($n=1.32$), and a disk with spiral arms.  An additional bar for the host is not
helpful. The host appears to be an interacting galaxy (Peletier et al. 1999).

{\it NGC 6212} (Fig. 11.38) --- \ \
The fit is done on a mosaic WFPC2 image. The host galaxy is fit with a 
pseudo-bulge ($n=1.88$) and a disk. While the host is often regarded as an 
elliptical galaxy (e.g., Biermann et al. 1985), the \hst\ image clearly 
reveals flocculent spiral arms (Malkan et al. 1998). 


{\it PG 0003+199} (Fig. 11.39) --- \ \ 
We fit the host galaxy with a bulge ($\sim16.9$ mag) and a disk. Alternatively,
if the host is fit with a single bulge, the magnitude is $\sim15.1$ mag.
However doing so reveals signs of an extended disk in the residual. Bentz et al.
(2009) also find that the host galaxy has an extended disk component.
However, they find a slightly brighter bulge ($\sim16.3$ mag).

{\it PG 0026+129} (Fig. 11.39) --- \ \
While the host galaxy is well-fit with a single bulge, it also can be fit with a
small bulge and a disk (B/T > 0.69). There is a slight PSF mismatch which
may cause a discrepancy between the host magnitude from the best fit and
that from the aperture photometry. Bentz et al. (2009) use two nuclei and a
single \ser\ component to model the central point source and the host,
respectively. Other studies also show that the host is an elliptical galaxy
(McLeod \& McLeod 2001, Guyon et al. 2006; Veilleux et al. 2009).




{\it PG 0844+349} (Fig. 11.40) --- \ \
The host galaxy has a bulge component and a disk with spiral arm. Adding a bar
is not helpful. Our best-fit model yields that the bulge magnitude is $\sim16.1$
mag, while the study of Bentz et al. (2009) finds the bulge ($\sim16.9$ mag) to
be slightly fainter than our measurement for the bulge.

{\it PG 0921+525} (Fig. 11.40) --- \ \ 
The host appears to be a disk galaxy ($B/T \sim 0.2$) and has a relatively small
$\sigma_*\sim90$\kms (Ferrarese et al. 2001).




{\it PG 1001+291} (Fig. 11.41) --- \ \
The host galaxy fit finds a tiny bulge ($R_e = 0.75$ pix), a bar, and a disk
with spiral arms. As there is a significant PSF mismatch as shown in the
residual image, the brightness of nucleus might be underestimated.  Boyce et al.
(1999) claim that the host can either be a barred spiral galaxy or a disturbed
galaxy with signs of tidal interactions. Owing to the large size of the spiral
features ($\sim66$ kpc), Boyce et al. conclude that the latter is more
favorable.  However, the tidal features appear to be symmetric, like well
defined spiral arms. The scale-length of the disk ($\sim10$ kpc) is consistent
with that expected for massive early spiral galaxies (de Jong et al. 2004).



{\it PG 1202+281} (Fig. 11.42) --- \ \
The image is saturated in the center. Bahcall et al. (1997) classify
the host as an elliptical galaxy.



{\it PG 1229+204} (Fig. 11.43) --- \ \
There is a faint tidal tail which might be evidence of interaction or minor
merging. The host galaxy is fit with a pseudo-bulge ($n=1.51$) and a disk. We
find $B/T\approx0.25$. Based on the \hst\ NIR image, Veilleux et al. (2009)
reach a similar conclusion ($B/T\approx0.2$).

{\it PG 1302$-$102} (Fig. 11.43) --- \ \
The image is highly saturated in the core. While we fit the host galaxy with a
single bulge, significant structures remain in the residual image (Kim et al.
2008b).  Guyon et al. (2006) also classify the host as an elliptical galaxy, but
Veilleux et al. (2009) conclude that the morphology is ambiguous. 




{\it PG 1351+695} (Fig. 11.44) --- \ \ 
The host galaxy has a disturbed bulge ($a_1=0.46$) and a disturbed disk
($a_1=0.09$), which might be caused either by interaction or dusty lanes.  While
the best-fit model yields $m_{\rm bul}\approx 16.4$ and $B/T\approx 0.21$,
previous studies (Bentz et al. 2009; Bennert et al. 2010; Orban de Xivry et al.
2011) suggest a slightly higher $B/T$ ($\approx 0.3-0.4$).  

{\it PG 1402+261} (Fig. 11.44) --- \ \
The image is highly saturated, thus the fitting result may be unreliable.  The
host galaxy is fit with a pseudo-bulge ($n=0.22$), a bar, and a disk with spiral
arms. Bahcall et al. (1997) also classify the morphology of the host galaxy to
be an SBb. 



{\it PG 1416$-$129} (Fig. 11.45) --- \ \ 
The image is saturated at the center. A single bulge component is sufficient to
model the host galaxy. Although Schade et al. (2000) find that the host is
well fit with an additional disk, the ratio of $B/T \approx 0.6$ means that
it is a bulge-dominated (early-type) galaxy. Using ground-based images, Jahnke
et al. (2004) argue that the host galaxy is an elliptical galaxy.  

{\it PG 1426+015} (Fig. 11.45) --- \ \
Although the host galaxy has a highly disturbed appearance it can be fit with a
single \ser\ component, suggesting that it is a disturbed elliptical galaxy.  We
simultaneously fit the merging galaxy with three components.  Guyon et al.
(2006) also classify the host an an elliptical galaxy. However, Schade et al.
(2000) use two components to fit the host ($B/T \sim 0.54$). 

{\it PG 1434+590} (Fig. 11.45) --- \ \ 
The FoV of the \hst\ image is not large enough to cover the outer part of the
host galaxy, making the analysis uncertain.  The host galaxy apparently has a
small bulge ($n=0.22$, $m_{\rm bul}\approx18.13$ mag), a bar and a disk with
spiral arms. Crenshaw et al. (2003) also classify the host as being a barred
galaxy.  Without using a bar component, Bentz et al. (2009) and Bennert et al.
(2010) find a slightly fainter bulge ($m_{\rm bul}\approx17.4-17.7$ mag).

{\it PG 1440+356} (Fig. 11.46) --- \ \ 
The host galaxy has a pseudo-bulge (17.07 mag, $n=0.64$) and a disk. 
Guyon et al. (2006) and Mathur et al. (2012) also find that the disk component
is required to model the host galaxy adequately.
 


{\it PG 1534+580} (Fig. 11.46) --- \ \ 
The WFPC2 mosaic image is saturated in the center.  The host galaxy is fit with
a bulge and a disk components. However, a single component bulge fit also
appears to be reliable. Earlier studies (Orban de Xivry et al. 2011; Bentz et
al.  2013) also suggest that the host galaxy is well-fit with two components
(bulge+disk).

{\it PG 1545+210} (Fig. 11.46) --- \ \
Also known as 3C323.3, this is a radio-loud QSO. The image is slightly saturated
in the center. The host galaxy is fit with a single bulge, but another fit with
two components also works reasonably well. There are positive residuals in  the
1-D profile due to a companion galaxy. Bahcall et al. (1997) also classify the
host as an elliptical. 

{\it PG 1613+658} (Fig. 11.47) --- \ \
The host galaxy is highly disturbed and has double nuclei. It appears that the
host galaxy can be well-fit with a single component bulge modified by large
Fourier amplitudes ($a_1\sim0.21$).  Guyon et al. (2006) and Bentz et al. (2009)
also classify the host as an elliptical galaxy. 

{\it PG 1617+175} (Fig. 11.47) --- \ \
While multiple \hst\ images taken with different instrument configuration are
available for this object, we use the deeper ACS/WFC images in our analysis.
The best-fit model finds the host galaxy is a single component bulge.  Other
studies also categorize this object as an elliptical galaxy (Guyon et al.
2006; Bentz et al. 2009; Veilleux et al. 2009). 

{\it PG 1700+518} (Fig. 11.47) --- \ \
Although the host galaxy is fit with three components to account for the tidally
disturbed features, the galaxy is bulge dominated ($B/T \approx 0.87$). Previous
studies also findthat the host is either a disturbed elliptical or
bulge-dominated galaxy (Guyon et al. 2006; Bentz et al. 2009; Veilleux et al.
2009).

{\it PG 2130+099} (Fig. 11.47) --- \ \
The host galaxy has a pseudo-bulge ($n \approx 0.45$) and a disturbed disk.  The
outer disk has a low \ser\ index $n \approx 0.28$ which might be due to the
outer ring. The bulge magnitude spans from 17.8 to 18.9 mag according to
previous studies (Bentz et al. 2009; Bennert et al. 2010), while the best-fit
model from this study yields $m_{\rm bul}\approx 17.3$ mag.  




{\it PKS 0312$-$77} (Fig. 11.48) --- \ \
The image is saturated in the center, thus the fit might be uncertain.  The host
galaxy is a single component bulge, but a fit with two components also works
reasonably well. Boyce et al. (1998) and Veron-Cetty \& Woltjer (1990) classify
the host as an elliptical type.

{\it PKS 0518$-$45} (Fig. 11.49) --- \ \
The WFPC2 mosaic image is saturated in the center.  The host galaxy has a bulge
and a disk. Using the ground-based images, Zirbel et al. (1996) claim that the
host has a de Vaucouleurs profile ($n=4$). However, Inskip et al. (2010) find
that a disky model ($n\approx2$) is more appropriate to represent the host,
which is consistent with this study. 









{\it POX 52} (Fig. 11.50) --- \ \
The host galaxy is fit with a single bulge with $n = 4$. There is a blob in the
central region. Thornton et al. (2008) also analyze the same \hst\ image with
GALFIT.  While our galaxy parameters are in good agreement to within the errors,
their nuclear luminosity is brighter by 0.4 mag.  While the origin of the 
discrepancy is unclear, it is possible that our method to broaden the images 
could alleviate the PSF mismatch significantly.  

{\it RX J1117.1+6522} (Fig. 11.51) --- \ \
The host galaxy is modeled with a bulge and a disk with spiral arms. While the
best-fit model finds that the host is a bulge-dominated spiral galaxy ($B/T
\approx 0.43$), Mathur et al. (2012) suggest the bulge fraction is fairly modest
($B/T \approx 0.17$).

{\it RX J1209.8+3217} (Fig. 11.51) --- \ \
The host galaxy model is of a single bulge component. The residual image reveals
a ring-like feature in the center. Based on the same \hst\ images, Mathur et al.
(2012) also find that the host is well represented by a single component with a
small \ser\ index ($n=1$). 

{\it RX J1702.5+3247} (Fig. 11.51) --- \ \
The host galaxy has a tiny single bulge ($R_e \approx$ 0\farcs07).  Its
effective radius is comparable to FWHM of the PSF. However, the study by Mathur
et al. (2012) show that the host is well represented by a spiral galaxy with a
bulge. 


{\it RX J2216.8$-$4451} (Fig. 11.51) --- \ \
The host galaxy is modeled with a pseudo-bulge, a disk, and a ring like
structure.  Owing to the ring, there are sharp discontinuities in the 1-D
profile. Mathur et al. (2012) find a slightly larger $B/T\approx 0.5$.

{\it RX J2217.9$-$5941} (Fig. 11.52) --- \ \
The host galaxy is fit with a bulge and a disk with spiral arms. Clearly the
host is interacting with a companion galaxy. The best-fit parameters are in good
agreement with those in Mathur et al. (2012).  

{\it SDSS 003043.59$-$103517.6} (Fig. 11.52) --- \ \
The host galaxy model has a pseudo-bulge ($n=0.54$), a bar, and a disk with
spiral arm, components. The optimal psf is generated by IDL procedure (Rhodes et
al. 2006, and 2007; http://www.astro.caltech.edu/~rjm/acs/PSF/). Cales et al.
(2011) find that the bulge is relatively large ($R_e \sim$2\farcs3) compared to
our measurement ($R_e \sim$0\farcs3). While it is unclear what caused this
discrepancy, visual inspection shows that the size of the bulge ought to be
be smaller than 1\arcsec.  

{\it SDSS 005739.19+010044.9} (Fig. 11.52) --- \ \
The host galaxy model has a pseudo-bulge ($n=0.36$) and a disk with spiral arm
components.  Cales et al. (2011) also classify the host galaxy as a
grand-design spiral galaxy. 


{\it SDSS 020258.94$-$002807.5} (Fig. 11.52) --- \ \
The host galaxy has very complicated features (tidal tails, spiral arms and SF
arc; Cales et al. 2011).  We fit it with a pseudo-bulge, a disk and another disk
with spiral arms. 

{\it SDSS 021447.00$-$003250.6} (Fig. 11.53) --- \ \
Since the host galaxy has complicated features, the fit might be rather
uncertain.  There is a companion galaxy but whether it is interacting with the
host galaxy is unknown. While we model the host with a bulge and a disk ($B/T
\sim 0.18$), Cales et al. (2011) argue that the host is a bulge-dominated galaxy
($B/T \sim 0.77$).

{\it SDSS 023700.30$-$010130.5} (Fig. 11.53) --- \ \
The host galaxy model consists a pseudo-bulge, a disk with spiral arms, and a
ring-like structure. We also fit three unresolved objects simultaneously, which
are not considered part the host. Cales et al. (2011) use a bulge and a disk to
model the host galaxy, finding that the host galaxy is bulge-dominated ($B/T
\sim 0.83$).

{\it SDSS 040210.90$-$054630.3} (Fig. 11.53) --- \ \
The host galaxy fit consists a bulge, an outer disk, and another inner disk with
spiral arms. Our finding that the host galaxy is disk-dominated ($B/T \sim 0.1$)
is in good agreement with Cales et al. (2011).

{\it SDSS 074621.06+335040.7} (Fig. 11.53) --- \ \
The host galaxy is fit with either a single bulge ($19.09$ mag; $n\sim 2.7$) or
a small blob with a disk (19.61 mag). 

{\it SDSS 075521.30+295039.2} (Fig. 11.54) --- \ \
The host galaxy model has a pseudo-bulge ($n=0.93$), a disk, and a another disk
with spiral arms. The bulge luminosity does not change significantly even 
though the fit is done with a bulge and a disk. Cales et al. (2011) also argue 
that the host light is disk-dominated ($B/T\sim0.27$).

{\it SDSS 081018.67+250921.2} (Fig. 11.54) --- \ \
The image is slightly saturated in the center. The host galaxy appears to be
highly disturbed but is well-fit with a bulge ($n=1.93$), a disturbed disk, and
two fine structures. It is unclear whether the bulge can be classified as a
pseudo-bulge or classical bulge based on the \ser\ index and $B/T$ ($\sim0.5$).
Cales et al. (2011) find that the host is a bulge-dominated galaxy ($B/T
\sim 0.8$). 

{\it SDSS 105816.81+102414.5} (Fig. 11.54) --- \ \
The host galaxy has a single bulge component, but a fit with two components also
works reasonably well. The bulge luminosity can either be 18.56 mag for a single
component bulge, or 18.63 mag from B/D decomposition. Cales et al. (2011) also
classify the host as a bulge-dominated galaxy. 

{\it SDSS 115159.59+673604.8} (Fig. 11.54) --- \ \
The host galaxy model consists a tiny pseudo-bulge ($R_e = $0\farcs12; $n=1.66$), 
a bar, and a disk. 

{\it SDSS 115355.58+582442.3} (Fig. 11.55) --- \ \
A single bulge component is sufficient to fit the host galaxy. 

{\it SDSS 123043.41+614821.8} (Fig. 11.55) --- \ \
The host galaxy is fit with a bulge ($n\sim0.85$) and a disk component. The
$B/T$ ratio is relatively large ($\sim 0.47$), and Cales et al. (2011) find that
the host is likely to be an early-type galaxy. There are several nearby
companions, but it is unclear whether they are associated with the host galaxy.

{\it SDSS 124833.52+563507.4} (Fig. 11.55) --- \ \
There are fine structures in the host galaxy, including shells and tidal
features.  The host galaxy is modeled using a classical bulge component and a
disk with spiral arms. 

{\it SDSS 145640.99+524727.2} (Fig. 11.55) --- \ \
The host galaxy has a pseudo-bulge ($n=0.88$), a bar, and a disk with spiral
arms in the model. The result shows a $B/T\sim0.4$, which is consistent
with Cales et al. (2011). 

{\it SDSS 145658.15+593202.3} (Fig. 11.56) --- \ \
This is a merging system with $2-3$ companion galaxies so the fit may be
uncertain. The host galaxy has a pseudo-bulge ($n=0.43$) and a disk with disturbed
spiral arms. The closest companion model consists a unresolved nucleus and a
disk with spiral arms. The second companion is fit with a disk with disturbed
spiral arms. Finally, the other companion that is slightly apart from the host
is fit with a disk. 

{\it SDSS 154534.55+573625.1} (Fig. 11.56) --- \ \
The host galaxy is fit with a pseudo-bulge ($n=0.5$), a disk, and a ring-like,
components. 

{\it SDSS 164444.92+423304.5} (Fig. 11.56) --- \ \
The image shows some weak signatures of interaction. The host is well-fit using
a single bulge component. A fit with two components produces a bulge magnitude of
20.18, but it is not obvious that this model is more appropriate than the one
component model. Cales et al. (2011) also classify this host galaxy as an
early-type. 

{\it SDSS 170046.95+622056.4} (Fig. 11.56) --- \ \
This object appears to be a merging system. The host galaxy is fit with a
pseudo-bulge ($n~1.1$) component and a disk. We simultaneously fit the neighboring
companion.  Cales et al. (2011) also conclude that the host is a disk-dominated
galaxy.

{\it SDSS 210200.42+000501.8} (Fig. 11.57) --- \ \
This object is a merging system. The host galaxy model consists a pseudo-bulge
and a disk with spiral arms. While our model finds that the host is disk
dominated ($B/T \sim 0.25$), Cales et al. (2011) argue that it is more likely to
be bulge dominated ($B/T \sim 0.7$).  

{\it SDSS 212843.42+002435.6} (Fig. 11.57) --- \ \
The object resides in merging system, possibly associated with 3 companion 
galaxies. We fit the host galaxy with a single \ser\ profile (19.72 mag; 
$n=1.48$). Cales et al. (2011) also find that the host is an early-type 
galaxy. 

{\it SDSS 230614.18$-$010024.4} (Fig. 11.57) --- \ \
The host galaxy has a bulge with a small \ser\ index ($n=0.68$) and a disk with
spiral arms. However, the $B/T$ ratio is relatively large ($\sim0.5$).  Cales et al. (2011) conclude also that the host galaxy is 
bulge-dominated ($B/T\sim0.7$).

{\it SDSS 234403.55+154214.0} (Fig. 11.57) --- \ \
We fit the host galaxy with a pseudo-bulge component ($n=0.49$) and a disk with
spiral arms.

{\it TOL 1059+105} (Fig. 11.58) --- \ \
The image is slightly saturated at the center. There is a neighboring galaxy
whose association with the host galaxy is unclear. The host is fit with a bulge
and a disk component. Crenshaw et al. (2003) classify the host as an S0 galaxy.

{\it TOL 1347+023} (Fig. 11.58) --- \ \ 
The image is saturated in the center. The fit consists a bulge and two 
disk components. 

{\it TOL 2327$-$027} (Fig. 11.58) --- \ \
The fit uses a WFPC2 mosaic image. The host galaxy model consists a bulge, a
bar, an inner disk, and an outer disk  with spiral arms.  We use the hybrid
method to merge the bar and two disk components. There are dust lanes in the
center. 

{\it UGC 3223} (Fig. 11.58) --- \ \
The WFPC2 mosaic image is saturated at the center.  Model fitting finds that the
host galaxy has a bulge, a bar, and a disk with spiral arms.  The host is
classified as a barred spiral galaxy (SBa) from de Vaucouleurs et al. (1991).

{\it UGC 10683B} (Fig. 11.59) --- \ \
The WFPC2 mosaic image is saturated at the center.  While it is not present in
the \hst\ image, there is a companion galaxy located 1 kpc away from UGC 10683B.
The host galaxy is fit with a bulge, an inner disk, and another outer disk
component with spiral arms. Although the host appears to be a disk-dominated
galaxy ($B/T\sim0.1$) the bulge has a relatively large \ser\ index
($\sim3.49$).



{\it WAS 45} (Fig. 11.59) --- \ \
The WFPC2 mosaic image is saturated in the galaxy center, which is located on
the WF4 chip.  The host galaxy is fit with a pseudo-bulge ($n=0.77$) and two
disks components with spiral arm.

\clearpage


\begin{thebibliography}{}

\bibitem[]{}
Alam, S., Albareti, F. D., Allende Prieto, C., et al. 2015, ApJS, 219, 12

\bibitem[]{}
Amram, P., Marcelin, M., Bonnarel, F., et al. 1992, \aa, 263, 69

\bibitem[]{}
Bagchi, J., Vivek, M., Vikram, V., et al. 2014, ApJ, 788, 174

\bibitem[]{}
Bahcall, J. N., Kirhakos, S., Saxe, D. H., \& Schneider, D. P. 1997, \aj, 479,
642

\bibitem[]{}
Barth, A. J., Ho, L. C., Rutledge, R. E., \& Sargent, W. L. W. 2004,
\apj, 607, 90

\bibitem[]{}
Barvainis, R., Lonsdale, C., \& Antonucci, R. 1996, \aj, 111, 1431

\bibitem[]{}
Becker, R. H., White, R. L., \& Edwards, A. L. 1991, \apjs, 75, 1

\bibitem[]{}
Becker, R. H., White, R. L., \& Helfand, D. J. 1995, \apj, 450, 559

\bibitem[]{}
Bender, R., Burstein, D., \& Faber, S. M. 1992, ApJ, 399, 462

\bibitem[]{}
Bennert, N., Canalizo, G., Jungwiert, B., et al. 2008, ApJ, 677, 846

\bibitem[]{}
Bennert, N., Jungwiert, B., Komossa, S., Haas, M., \& Chini, R.
2006, \aa, 459, 55

\bibitem[]{}
Bennert, V. N., Treu, T., Woo, J., et al. 2010, \apj, 708, 1507

\bibitem[]{}
Bentz, M. C., Denney, K. D., Grier, C. J., et al. 2013, ApJ, 767, 149

\bibitem[]{}
Bessell, M. S. 2005, ARA\&A, 43, 293

\bibitem[]{}
Best, P. N., Kauffmann, G., Heckman, T. M., et al. 2005, MNRAS, 362, 25 

\bibitem[]{}
Biermann, P., Strom, R., \& Bartel, N. 1985, \aa, 147, L27

\bibitem[]{}
B\"{o}hm, A., Wisotzki, L., Bell, E. F., et al. 2013, A\&A, 549, A46

\bibitem[]{}
Boller, Th. 2004, Progress of Theoretical Physics Supplement, 155, 217

\bibitem[]{}
Botte, V., Ciroi, S., Rafanelli, P., \& Di Mille, F. 2004, \aj, 127, 3168

\bibitem[]{}
Boyce, P. J., Disney, M. J., Blades, J. C., et al. 1998, MNRAS, 298, 121

\bibitem[]{}
Boyce, P. J., Disney, M. J., \& Bleaken, D. G. 1999, MNRAS, 302, L39

\bibitem[]{}
Brotherton, M. S. 1996, \apjs, 102, 1

\bibitem[]{}
Bruzual, G., \& Charlot, S. 2003, MNRAS, 344, 1000

\bibitem[]{}
Cales, S. L., Brotherton, M. S., Shang, Z., et al. 2011, ApJ, 741, 106

\bibitem[]{}
Calzetti, D., Kinney, A. L., \& Storchi-Bergmann, T. 1994, \apj 429, 582

\bibitem[]{}
Cameron, E. 2011, PASA, 28, 128

\bibitem[]{}
Cardelli, J.~A., Clayton, G.~C., \& Mathis, J.~S. 1989, \apj, 345, 245

\bibitem[]{}
Cassata, P., Guzzo, L., Franceschini, A., et al. 2007, ApJS, 172, 270 

\bibitem[]{}
Chen, Y.-M., Wang, J.-M., Yan, C.-S., Hu, C., \& Zhang, S. 2009, ApJL, 695, L130

\bibitem[]{}
Cisternas, M., Jahnke, K., \& Inskip, K. J. et al. 2011, \apj, 726, 57

\bibitem[]{}
Collin, S., Kawaguchi, T., Peterson, B. M., \& Vestergaard, M. 2006, \aa, 
456, 75

\bibitem[]{}
Condon, J. J., Cotton, W. D., \& Broderick, J. J. 2002, \aj, 124, 675

\bibitem[]{}
Condon, J. J., Cotton, W. D., Greisen, E. W., et al.  1998, \aj, 115, 1693 

\bibitem[]{}
Courbin, F., Letawe, G., Magain, P., et al. 2002, \aa, 394, 863

\bibitem[]{}
Courteau, S., Dutton, A. A., van den Bosch, F. C., et al. 2007, ApJ, 671, 203

\bibitem[]{}
Crenshaw, D. M., Kraemer, S. B., \& Gabel, J. R. 2003, AJ, 126, 1690

\bibitem[]{}
Darg, D. W., Kaviraj, S., Lintott, C. J., et al. 2010, MNRAS, 401, 1043

\bibitem[]{}
de Jong, R. S. 1996, A\&AS, 118, 557

\bibitem[]{}
de Jong, R. S., Simard, L., Davies, R. L., et al. 2004, MNRAS, 355, 1155

\bibitem[]{}
de Koff, S., Baum, S. A., Sparks, W. B., et al. 1996, ApJS, 107, 621


\bibitem[]{}
Dennefeld, M., Boller, T., Rigopoulou, D., \&Spoon, H. W. W. 20
03, A\&A, 406, 527

\bibitem[]{}
Denney, K. D., Peterson, B. M., Dietrich, M., Vestergaard, M., \&
Bentz, M. C. 2009, \apj, 692, 246

\bibitem[]{}
Deo, R. P., Crenshaw, D. M., \& Kraemer, S. B. 2006, AJ, 132, 321

\bibitem[]{}
de Vaucouleurs, G., de Vaucouleurs, A., Corwin Jr., H. G., et al. 1991, Third
Reference Catalogue of Bright Galaxies, Version 3.9 (New York: Springer)


\bibitem[]{}
Donzelli, C. J., Chiaberge, M., Macchetto, F. D., et al. 2007, ApJ, 667, 780


\bibitem[]{}
Dunlop, J. S., McLure, R. J., Kukula, M. J., et al. 2003, MNRAS, 340, 1095

\bibitem[]{}
Edelson, R. A. 1987, \apj, 313, 651

\bibitem[]{}
Eracleous, M., \& Halpern, J. P. 1994, \apjs, 90, 1

\bibitem[]{}
Eracleous, M., \& Halpern, J. P. 2003, \apj, 599, 886

\bibitem[]{}
Ferrarese, L., \& Merritt, D. 2000, \apjl, 539, L9

\bibitem[]{}
Ferrarese, L., Pogge, R. W., Peterson, B. M., et al. 2001, ApJL, 555, L79

\bibitem[]{}
Fischer, S., Iserlohe, C., Zuther, J., et al. 2006, \aa, 452, 827

\bibitem[]{}
Fisher, D. B., \& Drory, N. 2008, AJ, 136, 77

\bibitem[]{}
Floyd, D. J. E., Axon, D., Baum, S., et al. 2008, \apjs, 177, 148

\bibitem[]{}
Floyd, D. J. E., Axon, D., Baum, S., et al. 2010, \apj, 713, 66

\bibitem[]{}
Floyd, D. J. E., Kukula, M. J., Dunlop, J. S., et al. 2004, \mnras, 355, 196

\bibitem[]{}
Freeman, K. C. 1966, MNRAS, 133, 47

\bibitem[]{}
Fukugita, M., Shimasaku, K., \& Ichikawa, T. 1995, PASP, 107, 945

\bibitem[]{}
Gadotti, D. A. 2009, \mnras, 393, 1531

\bibitem[]{}
Gallimore, J. F., Axon, D. J., O'Dea, C. P., Baum, S. A., \& Pedlar, A. 2006,
\aj, 132, 546

\bibitem[]{}
Gao, H., \& Ho, L. C. 2017, \apj, submitted

\bibitem[]{}
Gear, W. K., Stevens, J. A., Hughes, D. H., et al. 1994, \mnras, 267, 167

\bibitem[]{}
Gebhardt, K., Bender, R., Bower, G., et al. 2000, \apjl, 539, L13

\bibitem[]{}
Ghigo, F. D., Wycko, S., Wardle, J. F. C., \& Cohen, N. L. 1982, \aj, 87,
1438

\bibitem[]{}
Giavalisco, M., Ferguson, H. C., Koekemoer, A. M., et al. 2004, ApJL, 600, L93

\bibitem[]{}
Grandi, S. A., \& Osterbrock, D. E. 1978, \apj, 220, 783

\bibitem[]{}
Greene, J. E., \& Ho, L. C. 2005, ApJ, 630, 122

\bibitem[]{}
Greene, J. E., \& Ho, L. C. 2006a, \apjl, 641, L21

\bibitem[]{}
Greene, J. E., \& Ho, L. C. 2006b, \apj, 641, 117

\bibitem[]{}
Greene, J. E., Ho, L. C., \& Barth, A. J. 2008, \apj, 688, 159

\bibitem[]{}
Greene, J. E., Peng, C. Y., Kim, M., et al. 2010, ApJ, 721, 26

\bibitem[]{}
Gregory, S. A., Tifft, W. G., \& Cocke, W. J. 1991, \aj, 102, 1977

\bibitem[]{}
Grupe, D., Beuermann, K., Mannheim, K., \& Thomas, H.-C. 1999, \aa, 350, 805

\bibitem[]{}
Grupe, D., \& Mathur, S. 2004, \apjl, 606, L41

\bibitem[]{}
G\"{u}ltekin, K., Richstone, D. O., Gebhardt, K., et al. 2009, \apj, 698, 198

\bibitem[]{}
Guyon, O., Sanders, D. B., \& Stockton, A. 2006, ApJS, 166, 89

\bibitem[]{}
Halpern, J. P., \& Filippenko, A. 1986, \aj, 91, 1019

\bibitem[]{}
Hamilton, T. S., Casertano, S., \& Turnshek, D. A. 2002, ApJ, 576, 61

\bibitem[]{}
Hamilton, T. S., Casertano, S., \& Turnshek, D. A. 2008, ApJ, 678, 22 

\bibitem[]{}
Heckman, T. M. 1983, \apj, 268, 628

\bibitem[]{}
Helmboldt, J. F., Taylor, G. B., Tremblay, S., et al. 2007, \apj, 658, 203

\bibitem[]{}
Hill, G. J., Goodrich, R. W., \& Depoy, D. L. 1996, \apj, 462, 163

\bibitem[]{}
Ho, L. C., Filippenko, A. V., Sargent, W. L. W., \& Peng, C. Y. 1997,
\apjs, 112, 391

\bibitem[]{}
Ho, L. C., \& Kim, M. 2009, \apjs, 184, 398

\bibitem[]{}
Ho, L. C., \& Kim, M. 2014, \apj, 789, 17

\bibitem[]{}
Ho, L. C., \& Kim, M. 2015, \apj, 809, 123

\bibitem[]{}
Ho, L. C., \& Peng, C. Y. 2001, ApJ, 555, 650

\bibitem[]{}
Ho, L. C., \& Ulvestad, J. S. 2001, \apjs, 133, 77

\bibitem[]{}
Hong, J., Im, M., Kim, M., \& Ho, L. C. 2015, ApJ, 804, 34

\bibitem[]{}
Hopkins, P. F., Hernquist, L., Cox, T. J., et al. 2006, \apjs, 163, 1

\bibitem[]{}
Hutchings, J. B., Holtzman, J., Sparks, W. B., et al. 1994, ApJ, 429, L1

\bibitem[]{}
Hutchings, J. B., \& Neff, S. G. 1992, AJ, 104, 1

\bibitem[]{}
Im, M., Simard, L., Faber, S. M., et al. 2002, ApJ, 571, 136

\bibitem[]{}
Inskip, K. J., Tadhunter, C. N., Morganti, R., et al. 2010, MNRAS, 407, 1739

\bibitem[]{}
Ivezi\'{c}, \u{Z}., Menou, K., Knapp, G.R., et al. 2002, AJ, 124, 2364

\bibitem[]{}
Ivezi\'{c}, \u{Z}., Smith, J. A., Miknaitis, G., et al. 2007, AJ, 134, 973



\bibitem[]{}
Jiang, Y.-F., Greene, J. E., Ho, L. C., Xiao, T., \& Barth, A. J. 2011, \apj, 742, 68

\bibitem[]{}
Jones, D. H., Saunders, W., Colless, M., et al. 2004, MNRAS, 355, 747 

\bibitem[]{}
Kaspi, S., Smith, P. S., Netzer, H., et al. 2000, \apj, 533, 631

\bibitem[]{}
Kauffmann, G., \& Haehnelt, M. 2000, \mnras, 311, 576


\bibitem[]{}
Keel, W. C. 1985, \aj, 90, 1449

\bibitem[]{}
Kellermann, K. I., Pauliny-Toth, I. I. K., \& Williams, P. J. S. 1969, \apj, 157, 1

\bibitem[]{}
Kellermann, K.~I., Sramek, R.~A., Schmidt, M., Shaffer, D.~B., \& Green, R.~F. 1989, \aj, 98, 1195

\bibitem[]{}
Kim, M., Ho, L. C., \& Im, M. 2006, \apj, 642, 702

\bibitem[]{}
Kim, M., Ho, L. C., Peng, C. Y., Barth, A. J., \& Im, M. 2008a, \apjs, 179, 283

\bibitem[]{}
Kim, M., Ho, L. C., Peng, C. Y., et al. 2008b, \apj, 687, 767

\bibitem[]{}
Kim, M., Ho, L. C., Peng, C. Y., \& Im, M. 2007, \apj, 658, 107

\bibitem[]{}
Kinney, A. L., Calzetti, D., \& Bohlin, R. C. 1996, \apj, 467, 38

\bibitem[]{}
Koekemoer, A. M., Faber, S. M., Ferguson, H. C., et al. 2011, ApJS, 197, 36 

\bibitem[]{}
Koekemoer, A. M., Fruchter, A. S., Hook, R. N., \& Hack, W. 2002, The 2002 HST
Calibration Workshop: Hubble After the Installation of the ACS and the NICMOS
Cooling System, ed. S. Arribas, A. M. Koekemoer, \& B. Whitmore
(Baltimore, MD: STScI), 337 

\bibitem[]{}
Kojoian, G., Tovmassian, H. M., Dickinson, D. F., \& Dinger, A. S.
1980, \aj, 85, 1462


\bibitem[]{}
Kormendy, J. 1979, ApJ, 227, 714

\bibitem[]{}
Kormendy, J., Cornell, M. E., Block, D. L., Knapen, J. H., \& Allard, E. L.
2006, ApJ, 642, 765

\bibitem[]{}
Kormendy, J., Fisher, D. B., Cornell, M. E., \& Bender, R. 2009, ApJS, 182, 216

\bibitem[]{}
Kormendy, J., \& Ho 2013, ARA\&A, 51, 511

\bibitem[]{}
Kormendy, J., \& Kennicutt Jr., R. C. 2004, ARA\&A, 42, 603

\bibitem[]{}
Kormendy, J., \& Richstone, D.~O. 1995, \annrev, 33, 581


\bibitem[]{}
Kotilainen J. K., Tavares J. L., Olgu\'{i}n-Iglesias, et al. 2016, ApJ, 832, 157 
\bibitem[]{}
Kotilainen, J. K., \& Ward, M. J. 1994, MNRAS, 266, 953

\bibitem[]{}
Kotilainen, J. K., Ward, M. J., \& Williger, G. M. 1993, MNRAS, 263, 655

\bibitem[]{}
Krist, J. 1995, in Astronomical Data Analysis Software and
Systems IV, ed. R. A. Shaw, H. E. Payne, \& J. J. E. Hayes
(San Francisco: ASP), 349

\bibitem[]{}
Kuraszkiewicz, J. K., Green, P. J., Crenshaw, D. M., et al. 2004, ApJS, 150, 165

\bibitem[]{}
K\"{u}hr, H., Witzel, A., Pauliny-Toth, I. I. K., \& Nauber, U. 1981,
A\&AS, 45, 367

\bibitem[]{}
La Barbera, F., Busarello, G., Merluzzi, P., Massarotti, M., \& Capaccioli, M. 
2003, ApJ, 595, 127

\bibitem[]{}
Laing, R., \& Peacock, J. A. 1980, \mnras, 190, 903

\bibitem[]{}
Lal, D. V., \& Ho, L. C. 2010, AJ, 139, 1089

\bibitem[]{}
LaMassa, S. M., Heckman, T. M., Ptak, A., \& Urry, C. M. 2013, ApJL, 765, L33

\bibitem[]{}
Laurikainen, E., Salo, H., Buta, R., Knapen, J. H., \& Comer\'{o}n, S. 2010, 
MNRAS, 405, 1089

\bibitem[]{}
Lawrence, C. R., Zucker, J. R., \& Readhead, A. C. S. 1996, \apjs, 107, 541

\bibitem[]{}
Leipski, C., Falcke, H., Bennert, N., \& H\"{u}ttemeister, S. 2006,
\aa, 455, 161

\bibitem[]{}
Letawe, G., Magain, P., \& Courbin, F. 2007, \mnras, 378, 83

\bibitem[]{}
Letawe, Y., Magain, P., Letawe, G., Courbin, F., \& Hutsem\'{e}kers, D. 2008, 
ApJ, 679, 967

\bibitem[]{}
Letawe, Y., Letawe, G., \& Magain, P. 2010, \mnras, 403, 2088	


\bibitem[]{}
Lotz, J. M., Jonsson, P., Cox, T. J., \& Primack, J. 2008, MNRAS, 391, 1137

\bibitem[]{}
Lotz, J. M., Jonsson, P., Cox, T. J., \& Primack, J. 2010, MNRAS, 404, 590

\bibitem[]{}
Madrid, J. P., Chiaberge, M., Floyd, D., et al. 2006, ApJS, 164, 307

\bibitem[]{}
Magorrian, J., Tremaine, S., Richstone, D., et al. 1998, \aj, 115, 2285

\bibitem[]{}
Malkan, M. A., Gorjian, V., \& Tam, R. 1998, ApJS, 117, 25

\bibitem[]{}
Marble, A. R., Hines, D. C., Schmidt, G. D., et al. 2003, ApJ, 590, 707

\bibitem[]{}
Marshall, H. L., Schwartz, D. A., Lovell, J. E. J., et al. 2005, \apjs, 156, 13

\bibitem[]{}
Martel, A. R., Baum, S. A., Sparks, W. B., et al. 1999, ApJS, 122, 81

\bibitem[]{}
Marziani, P., Sulentic, J. W., \& Zamanov, R. 2003, \apjs, 145, 199

\bibitem[]{}
Masters, K. L., Nichol, R. C., Hoyle, B., et al. 2011, MNRAS, 411, 202

\bibitem[]{}
Mathur, S., Fields, D., Peterson, B. M., \& Grupe, D. 2012, ApJ, 754, 146


\bibitem[]{}
Matthews, T. A., Morgan, W. W., \& Schmidt, M. 1964, ApJ, 140, 35

\bibitem[]{}
Mauch, T., Murphy, T., \& Buttery, H. J. 2003, \mnras, 342, 1117

\bibitem[]{}
McCarthy, P. J., Baum, S. A., \& Spinrad, H. 1996, \apjs, 106, 281

\bibitem[]{}
McLeod, K. K., \& McLeod, B. A. 2001, ApJ, 546, 782

\bibitem[]{}
McLeod, K. K., \& Rieke, G. H. 1995a, \apj, 441, 96

\bibitem[]{}
McLeod, K. K., \& Rieke, G. H. 1995b, ApJL, 454, L77

\bibitem[]{}
McLure, R. J., \& Dunlop, J. S. 2001, \mnras, 327, 199

\bibitem[]{}
McLure, R.~J., \& Dunlop, J.~S. 2004, \mnras, 352, 1390

\bibitem[]{}
McLure, R.~J., Kukula, M. J., Dunlop, J. S., et al. 1999, MNRAS, 308, 377

\bibitem[]{}
Nagar, N. M., Falcke, H., \& Wilson, A. S. 2005, \aa, 435, 521


\bibitem[]{}
Ohta, K., Aoki, K., Kawaguchi, T., \& Kiuchi, G. 2007, ApJS, 169, 1

\bibitem[]{}
Onken, C.~A., Ferrarese, L., Merritt, D., et al. 2004, \apj, 615, 645

\bibitem[]{}
Orban de Xivry, G., Davies, R., \& Schartmann, M. 2011, MNRAS, 417, 2721

\bibitem[]{}
Osterbrock, D. E. 1977, \apj, 215, 733

\bibitem[]{}
Osterbrock, D. E. 1981, \apj, 249, 462

\bibitem[]{}
Park, D., Woo, J.-H., Treu, T., et al. 2012, ApJ, 747, 30

\bibitem[]{}
Peletier, R. F., Knapen, J. H., Shlosman, I., et al. 1999, ApJS, 125, 363

\bibitem[]{}
Peng, C. Y., Ho, L. C., Impey, C. D., \& Rix, H.-W. 2002, \aj, 124, 266

\bibitem[]{}
Peng, C. Y., Ho, L. C., Impey, C. D., \& Rix, H.-W. 2010, \aj, 139, 2097

\bibitem[]{}
Peterson, B. M., Ferrarese, L., Gilbert, K. M., et al. 2004, \apj, 613, 682

\bibitem[]{}
Planck Collaboration, Ade, P. A. R., Aghanim, N., et al. 2016, \aa, 594, 13

\bibitem[]{}
Polletta, M., Tajer, M., Maraschi, L., et al. 2007, ApJ, 663, 81

\bibitem[]{}
Pounds, K. A., Done, C., \& Osborne, J. P. 1995, MNRAS, 277, L5

\bibitem[]{}
Ramos Almeida, C., Bessiere, P. S., Tadhunter, C. N., et al. 2012, MNRAS, 419, 687

\bibitem[]{}
Rhodes, J. D., Massey, R., \& Albert, J. 2006, in The 2005 Calibration 
Workshop, ed. A. Koekemoer, P. Goudfrooij, \& L. Dressel (Baltimore: STScI), 21

\bibitem[]{}
Rhodes, J. D., Massey, R. J., Albert, J., et al. 2007, ApJS, 172, 203

\bibitem[]{}
Rix, H.-W., Barden, M., Beckwith, S. V. W., et al. 2004, ApJS, 152, 163 

\bibitem[]{}
Rodr\'{i}guez-Ardila, A., Binette, L., Pastoriza, M. G., \&
Donzelli, C. J. 2000, \apj, 538, 581

\bibitem[]{}
Rothberg, B., Fischer, J., Rodrigues, M., \& Sanders, D.~B. 2013, \apj, 767, 72

\bibitem[]{}
Salpeter, E. E. 1955, \apj, 121, 161

\bibitem[]{}
Salpeter, E. E., \& Hoffman, G. L. 1996, ApJ, 465, 595



\bibitem[]{}
Sani, E., Lutz, D., Risaliti, G., et al. 2010, MNRAS, 403, 1246

\bibitem[]{}
Schade, D. J., Boyle, B. J., \& Letawsky, M. 2000, MNRAS, 315, 498

\bibitem[]{}
Schlafly, E. F., \& Finkbeiner, D. P. 2011, \apj, 737, 103

\bibitem[]{}
Scoville, N., Abraham, R. G., Aussel, H., et al. 2007, ApJS, 172, 38 

\bibitem[]{}
S\'{e}rsic, J. L. 1968, Atlas de Galaxias Australes (C\'{o}rdoba: Observatorio Astron\'{o}mico, Univ. C\'{o}rdoba)

\bibitem[]{}
Sikora, M., Stawarz, \l{L}., \& Lasota, J.-P. 2007, ApJ, 658, 815

\bibitem[]{}
Simien, F., \& de Vaucouleurs, G. 1986, ApJ, 302, 564

\bibitem[]{}
Smirnova, A. A., Moiseev, A. V., \& Afanasiev, V. L. 2010, MNRAS, 408, 400

\bibitem[]{}
Smith, E. P., Heckman, T. M., Bothun, G. D., Romanishin, W., \& Balick, B. 1986, ApJ, 306, 64

\bibitem[]{}
Stephens, S. A. 1989, \aj, 97, 10

\bibitem[]{}
Stirpe, G. M. 1990, A\&AS, 85, 1049

\bibitem[]{}
Stockton, A. 1982, ApJ, 257, 33

\bibitem[]{}
Storchi-Bergmann, T., Bica, E., \& Pastoriza, M. G. 1990, \mnras, 245, 749

\bibitem[]{}
Strateva, I. V., Brandt, W. N., Eracleous, M., Schneider, D. P.,
\& Chartas, G. 2006, \apj, 651, 749

\bibitem[]{}
Taylor, G. L., Dunlop, J. S., Hughes, D. H., \& Robson, E. I. 1996, MNRAS,
283, 930

\bibitem[]{}
Tempel, E., Saar, E., Liivam\"{a}gi, L. J., et al. 2011, \aa, 529, 53

\bibitem[]{}
Thornton, C. E., Barth, A. J., Ho, L. C., Rutledge, R. E., \& Greene, J. E. 
2008, ApJ, 686, 892

\bibitem[]{}
Tovmassian, H. M. 1972, \aj, 77, 705

\bibitem[]{}
Treu, T., Stiavelli, M., Casertano, S., M{\o}ller, P., \& Bertin, G. 2002, ApJL, 564, L13

\bibitem[]{}
Treu, T., Woo., J.-H., Malkan, M. A., \& Blandford, R. D. 2007, \apj, 667, 117

\bibitem[]{}
Treister, E., Schawinski, K., Urry, C. M., \& Simmons, B. D. 2012, ApJ,
758, L39

\bibitem[]{}
Ulvestad, J. S., Antonucci, R. R. J., \& Barvainis, R. 2005, \apj, 621, 123

\bibitem[]{}
Ulvestad, J. S., \& Wilson, A. S. 1984, \apj, 285, 439

\bibitem[]{}
Vanden Berk, D. E., Richards, G. T., Bauer, A., et al. 2001, \aj, 122, 549

\bibitem[]{}
van Bemmel, I. M., Barthel, P. D., \& de Graauw, T. 2000, \aa, 359, 523

\bibitem[]{}
van Dokkum, P. G. 2001, PASP, 113, 1420

\bibitem[]{}
Veilleux, S., Rupke, D. S. N., Kim, D.-C., et al. 2009, ApJS, 182, 628

\bibitem[]{}
V\'{e}ron-Cetty, M.-P., V\'{e}ron, P., \& Gon\c{c}alves, A. C. 2001,
\aa, 372, 73

\bibitem[]{}
V\'{e}ron-Cetty, M.-P., \& Woltjer, L. 1990, \aa, 236, 69

\bibitem[]{}
Vestergaard, M., \& Peterson, B. M. 2006, ApJ, 641, 689 

\bibitem[]{}
Vika, M., Driver, S. P., Cameron, E., Kelvin, L., \& Robotham, A. 2012,
MNRAS, 419, 2264

\bibitem[]{}
Virani, S., De Robertis, M. M., \& VanDalfsen, M. L. 2000, AJ, 120, 1739

\bibitem[]{}
Wang, J., Wei, J. Y., \& He, X. T. 2006, \apj, 638, 106

\bibitem[]{}
White, R. L., \& Becker, R. H. 1992, \apjs, 79, 331

\bibitem[]{}
White, R. L., Becker, R. H., Gregg, M. D., et al. 2000, ApJS, 126, 133

\bibitem[]{}
Wilhite, B. C., Vanden Berk, D. E., Brunner, R. J., \& Brinkmann, J. V.
2006, \apj, 641, 78

\bibitem[]{}
Winkler, H. 1992, \mnras, 257, 677

\bibitem[]{}
Woo, J.-H., Treu, T., Barth, A. J., et al. 2010, \apj, 716, 269

\bibitem[]{}
Wright, A., \& Otrupcek, R. 1990, PKS Catalog, 0

\bibitem[]{}
Xu, D. W., Komossa, S., Wei, J. Y., Qian, Y., \& Zhen, X. Z. 2003, \apj, 590, 73

\bibitem[]{}
Zheng, X. Z., Xia, X. Y., Mao, S., Wu, H., \& Deng, Z. G. 2002, \aj, 124, 18

\bibitem[]{}
Zhou, H.-Y., Wang, T.-G., \& Yuan, W. 2006, \apjs, 166, 12

\bibitem[]{}
Zirbel, E. L. 1996, ApJ, 473, 713

\end{thebibliography}
\end{document}